\documentclass[%
 reprint,
 amsmath,amssymb,
 prd,
]{revtex4-2}

\usepackage{amsfonts}
\usepackage{epsfig}
\usepackage{color}
\usepackage{amstext}
\usepackage{amssymb}
\usepackage{verbatim}
\usepackage{graphicx}
\usepackage{mathrsfs}
\usepackage{wrapfig}
\usepackage[T1]{fontenc}
\usepackage[percent]{overpic}
\usepackage{xcolor}
\usepackage[]{hyperref}
\usepackage{booktabs}
\hypersetup{
	colorlinks,
	linkcolor={blue},
	linktocpage=true,
	citecolor={red},
	urlcolor={olive}
}

\newcommand{\taupi}{$\tau^-\rightarrow\pi^-\nut$~}
\newcommand{\taue}{$\tau^-\rightarrow e^-\nueb\nut$~}
\newcommand{\taumu}{$\tau^-\rightarrow \mu^-\numb\nut$~}
\newcommand{\taurho}{$\tau^-\rightarrow\rho^-\nut\rightarrow\pi^-\pi^0\nut$~}
\newcommand{\taua}{$\tau\rightarrow(a_1\rightarrow\pi^\pm\pi^0\pi^0)\nutb$~}

\newcommand{\taurhoinv}{$\tau^+\rightarrow\rho^+\nutb\rightarrow\pi^+\pi^0\nutb$~}
\newcommand{\ifb}{fb$^{-1}$}
\newcommand{\epemtauptaum}{e$^+$e$^-$ $\rightarrow \tau^+\tau^-$}
\newcommand{\polresult}{0.0035}
\newcommand{\polstat}{0.0024}
\newcommand{\polsys}{0.0029}

\makeatletter
\renewcommand\subsubsection{\@startsection{subsubsection}{3}{\z@}%
                                     {-3.25ex\@plus -1ex \@minus -.2ex}%
                                     {-1.5ex \@plus -.2ex}
                                     {\normalfont\normalsize\bfseries}}
\makeatother

\newcommand{\BaBarYear}       {23}
\newcommand{\BaBarNumber}     {004}
\newcommand{\SLACPubNumber} {17732}

\newcommand{\BaBarType}      {PUB}  

\input babarsym

\begin{document}

{\pagestyle{empty}
		
\onecolumngrid{
\flushleft{
SLAC-PUB-\SLACPubNumber \hfill \\
\babar-\BaBarType-\BaBarYear/\BaBarNumber\hfill 
}
}
		
\title{Precision \texorpdfstring{$e^-$}{e-} Beam Polarimetry  at an \texorpdfstring{\epem}{e+e-} B Factory using Tau-Pair Events}

\author{J.~P.~Lees}
\author{V.~Poireau}
\author{V.~Tisserand}
\author{E.~Grauges}
\author{A.~Palano}
\author{G.~Eigen}
\author{D.~N.~Brown}
\author{Yu.~G.~Kolomensky}
\author{M.~Fritsch}
\author{H.~Koch}
% ...
\collaboration{The \babar\ Collaboration}
\noaffiliation


\begin{abstract}

We present a new technique, `Tau Polarimetry', for measuring the longitudinal beam polarization present in an \epem collider through the analysis of \epemtauptaum~events. By exploiting the sensitivity of $\tau$ decay kinematics to the longitudinal polarization of the beams, we demonstrate that the longitudinal polarization can be measured with a 3 per mil systematic uncertainty at the interaction point using a technique that is independent of spin and beam transport modeling. Using 424.2$\pm$1.8~\ifb~of \babar~data at $\sqrt{s}=10.58$~GeV, the average longitudinal polarization of the \pep2 \epem collider has been measured to be $\langle P\rangle=\polresult \pm \polstat_{\textrm{stat}}\pm \polsys_{\textrm{sys}}$. The systematic uncertainty studies are described in detail, which can serve as a guide for future applications of Tau Polarimetry. A proposed $e^-$ beam longitudinal polarization upgrade to the SuperKEKB \epem collider would benefit from this technique. \\
\begin{flushleft}
\hspace{55pt}PACS numbers: 13.88.+e, 14.60.Fg, 29.27.Hj
\end{flushleft}
\end{abstract}		

\maketitle

\vfill

\newpage
		
}


\setcounter{footnote}{0}

\section{Introduction}
We present in this paper a novel method for measuring the average longitudinal beam polarization in an $e^+e^-$ collider, referred to as `Tau Polarimetry'. Tau Polarimetry uses \epemtauptaum events measured in the detector and determines the average  longitudinal beam polarization using the sensitivity of the $\tau$ decay kinematics to the beam polarization. The technique is developed using data from the \babar~experiment at the PEP-II collider which operated with a center-of-mass energy of 10.58~GeV and is expected to have no beam polarization.  Using \babar~data, this paper reports on the statistical sensitivity of the technique and the determination of the dominant systematic uncertainties in the beam polarization.  
The motivation for this study is to benchmark the precision to which the beam polarization can be measured using Tau Polarimetery with Belle~II at a future polarization upgrade of the SuperKEKB collider.

Precision measurements of the weak mixing angle can be performed with experimental determinations of the left-right asymmetry, $A_{\textrm{LR}}$, for each of the $\epem\rightarrow f {\overline f}$ processes, where $f$ is a charged lepton or quark. The asymmetry is defined as the normalized difference between the production cross-sections for a left and right handed process:
\begin{equation}
	A_{\textrm {LR}}=\frac{\sigma_{\textrm{L}}-\sigma_{\textrm{R}}}{\sigma_{\textrm{L}}+\sigma_{\textrm{R}}},
	\label{eqn:alrdef}
\end{equation}
where the L and R subscripts refer to the chirality of the initial state electron in the $\epem\rightarrow f\overline f$ process. In the past, the SLAC Large Detector (SLD) experiment, operating at the $Z$-pole, used measurements of $A_{\textrm {LR}}$ to make the most precise determination of sin$^2\theta_W$~\cite{SLD1,SLD2}. At electron-positron colliders, a non-zero value of this asymmetry arises from $\gamma-Z$ interference~\cite{alrcalcs} and the measured value of $A_{\textrm{LR}}$ scales linearly with the average longitudinal polarization of the beams~\cite{SuperB,ALRee}:
\begin{equation}
	A^{f}_{\textrm{LR}}\propto\left(\frac{sG_F}{\alpha}\right)g^e_Ag^f_V\langle P\rangle,
	\label{eqn:ALR}
\end{equation}
where $G_F$ is the Fermi constant, $s$ is the square of the \epem center-of-mass (c.m.) energy, $\alpha$ is the fine structure constant, $g^e_A$ is the neutral current axial coupling of the electron, $g^f_V = T_3^f  -2 Q_f \textrm {sin}^2\theta_W$ is the neutral current vector coupling of fermion $f$, where $T_3^f$ is the third component of isospin, $Q_f$ is the electric charge, and $\sin^2\theta_W$ is the weak mixing angle. $\langle P\rangle$ is the average longitudinal polarization of the mediator in the \epem collision, defined as:
\begin{equation} \label{eqn:pol_def}
    \langle P\rangle=\frac{R^+L^--L^+R^-}{R^+L^-+L^+R^-},
\end{equation}
where $L^{\pm}$ ($R^\pm$) is the fraction of positrons ($+$) or electrons ($-$) in their respective beams that have left-handed (right-handed) spin, so that ($L^\pm+R^\pm\equiv1$). 

Tau Polarimetry relies on two convenient properties. The first is the linear relationship between the longitudinal polarization present in the beams and the polarization of the $\tau$ leptons produced in the \epem\ra\tautau process, where at $\sqrt{s}=10.58$~GeV~\cite{tau_beam_Polarization}:
\begin{equation}
\begin{split}
	P_\tau= & P\frac{\cos\theta}{1+\cos^2\theta} \\
 & -\frac{8G_Fs}{4\sqrt{2}\pi\alpha}g^\tau_V\left(g^\tau_A\frac{|\vec{p}|}{p^0}+2g^e_A\frac{\cos\theta}{1+\cos^2\theta}\right).\\
\end{split}
\label{eqn:ePoltoTauPol}
\end{equation}
$P_{\tau}$ is the polarization of the $\tau$, $P$ is the longitudinal polarization of the beams, $\theta$ is the angle between the emitted $\tau^-$ and the electron beam in the c.m.\ frame, and $\vec{p}$ and $p^0$ are the 3-momentum and energy of the $\tau$, respectively. At $\sqrt{s}=10.58$ GeV the size of the electroweak correction is small and known with a high precision $(8G_Fs)/(4\sqrt{2}\pi\alpha) g^\tau_V =-0.0029\pm0.0001.$
The majority of the uncertainty in the electroweak correction arises from the world average for $g_V^\tau$ at $m_Z$~\cite{tauZpole}. The electroweak correction is accounted for in the analysis and the associated uncertainties are negligible compared to the systematic uncertainties in the beam polarization measurement. 

The second property arises from the chirality of neutrinos and the correlation of the chirality and kinematic distributions of the $\tau$ decay~\cite{BernUpsilon,SuperB}. This correlation has been exploited by LEP to extract precision measurements of the weak mixing angle~\cite{tauZpole,ALEPHtau,DELPHItau,L3tau,OPALTauPolar}. By combining Eqn. \ref{eqn:ePoltoTauPol} with the kinematic dependence on polarization, a precision measurement of $\langle P\rangle$ can be made.

The longitudinal beam polarization in \babar~data is expected to be near zero due to the beam rings at \pep2 being unsuited to the build up of polarization through the Sokolov-Ternov effect~\cite{Sokolov,babarTrans}. Any polarization that would build up under this effect would be transversely polarized and only a longitudinal component would be visible to this analysis. In addition the \pep2 design expects a depolarization time of 1.5 minutes for fully transversely polarized beams and a residual transverse polarization of less than 0.8\%~\cite{babarTrans}. By measuring the near-zero average longitudinal polarization in \pep2, \babar~is able to determine the dominant systematic uncertainties in the Tau Polarimetry method.

Due to the similarities between the \babar~and Belle~II detectors, and the fact that both involve \epem collisions at $\sqrt{s}=10.58$~GeV  (Belle~II at SuperKEKB and \babar~ at PEP-II), \babar~can demonstrate the feasibility of the Tau Polarimetry technique,  and indicate the expected level of both statistical and systematic sensitivity that Belle~II might achieve in a polarization-upgraded SuperKEKB collider. This polarization upgrade is being considered for SuperKEKB in an upgrade referred to as `Chiral Belle'~\cite{b2polarizationWhitePaper}. This upgrade would introduce polarization to the $e^-$ beam only, which simplifies Eqn. \ref{eqn:pol_def} to $\langle P\rangle=L^--R^-$, as $L^+=R^+=0.5$. This definition of $\langle P\rangle$ is equivalent to the average longitudinal polarization of the $e^-$ beam. 

With the addition of $e^-$ beam polarization Belle II intends to significantly improve the precision with which the neutral current vector couplings, and hence sin$^2\theta_W$, can be determined separately for electrons, muons, $\tau$s, $c$ quarks, and $b$ quarks; enabling not only precision measurements of sin$^2\theta_W$ in a region away from the $Z$-Pole, but also the world's highest precision measurements of universality. Chiral Belle intends to also measure other fundamental parameters, such as the anomalous magnetic moment of the $\tau$~\cite{b2polarizationWhitePaper,BernUpsilon,Crivellin:2021spu}. The largest systematic uncertainty on these proposed measurements is expected to be the precision to which $\langle P\rangle$ is known. 

The Chiral Belle upgrade includes a Compton polarimeter on the electron beam to provide continuous monitoring of the beam polarization. The Compton polarimeter must be physically located outside of the Belle II detector and as such is expected to have uncertainties related to the modeling of the spin transport when extrapolating to the polarization present at the interaction point (IP). Tau Polarimetry provides a second, and complementary, way to determine the average longitudinal polarization at the IP, although on a much longer time scale. The primary advantage of the Tau Polarimetry measurement is its independence of any spin transport modeling and an increased precision for large data sets, $\mathcal{O}$(100 \ifb).

\section{\pep2 and The \texorpdfstring{\babar}{BaBar} Detector}\label{sec:babar}
The \babar~detector~\cite{BaBarDet,BaBarUpgrade} operated from 1999 to 2008 at the \pep2 asymmetric \epem collider, which collided 9.0 GeV electrons with 3.1 GeV positrons. 

Particles in the \babar~detector were identified by combining information from its sub-detectors. Charged-particle momenta were determined using tracks measured both in a five-layer silicon vertex tracker and in a 40-layer drift chamber (DCH) operated in a 1.5 T solenoidal magnetic field. Photons and electrons had their energy and angle measured in the electromagnetic calorimeter (EMC) consisting of 6580 CsI(Tl) crystals.
Muons were identified by resistive-plate chambers and streamer tubes in the instrumented magnetic-flux-return iron (IFR). Charged-particle identification (PID) was based on energy-loss measurements in the silicon vertex tracker and DCH, and on information from a ring-imaging Cherenkov detector, the EMC, and the IFR. 
The \babar~coordinate system features the z axis aligned with the principal axis of the solenoid field, which was offset by 20 mrad from the beam axis. The y axis was orientated upwards and the x axis was directed outwards from the center of \pep2.
The studies reported in this paper use the data collected by \babar~at a c.m.\ energy of 10.58~GeV, the $\Upsilon$(4S) resonance, with an integrated luminosity of 424.2$\pm$1.8~fb$^{-1}$~\cite{LumiPaper}. 

A total of 700 million polarized \tautau Monte Carlo (MC) simulated events, equivalent to 643~fb$^{-1}$ , were produced for both a fully left and right handed beam polarization with the \kkmc generator~\cite{kkmc}. A number of MC generators were used to produce unpolarized samples of various processes of interest: the continuum \mumu and \tautau were produced with \kkmc, which invoked \tauola~\cite{tauola} to simulate the decays of final-state $\tau$ leptons; the $\epem\ra\epem$ Bhabha process was simulated using the \bhwide~\cite{bhwide} generator; and the \evtgen~\cite{evtgen} generator provided the hadronic continuum MC.  \photos~\cite{photos} was employed to calculate the final-state radiation effects. These simulated processes then underwent a detector response simulation implemented with \geant~\cite{geant,geant2}. Roughly twice as much \mumu and \ccbar MC, and roughly four times as many $\uubar, \ddbar, \ssbar, \bbbar$ and \tautau MC events were produced compared to the number expected in 424.4~\ifb. As \babar~relies heavily on data-driven approaches to study and control Bhabha backgrounds, a smaller sample of Bhabha MC events was exploited for low-statistics studies.

\section{Polarization Sensitivity}\label{sec:PolarVar}
While all $\tau$ decay modes are sensitive to beam polarization, the hadronic decays are the most sensitive as there is only one neutrino carrying away angular momentum. In the case of the \taurho~decay (and charge conjugate (c.c.)), which has the largest $\tau$ decay branching fraction (25.49\%)~\cite{pdg2022}, three angular variables (including $\cos\theta$, with $\theta$ being the angle between the $\tau^-$ momentum and the electron beam direction in the c.m.\ frame) are required to extract the beam polarization, and capture all the angular momentum information from the spin-1 $\rho$ decay. The other two polarization sensitive variables are defined as~\cite{HAGIWARArho}:\\
\begin{equation}
	\cos\theta^\star=\frac{2z-1-m^2_{\rho}/m^2_{\tau}}{1-m^2_{\rho}/m^2_{\tau}} \hspace{2cm} z\equiv\frac{E_\rho}{E_{\textrm{beam}}}
	\label{eqn:zct}
\end{equation}
\begin{equation}
	\cos\psi=\frac{2x-1}{\sqrt{1-m^2_{\pi}/m^2_{\rho}}} \hspace{2cm} x\equiv\frac{E_\pi}{E_{\rho}}
	\label{eqn:xct}
\end{equation}
where $E_\pi$ and $E_\rho$ are, respectively, the reconstructed energies of the charged pion and $\rho$ in the c.m.\ frame, and $E_{\textrm{beam}}\equiv\sqrt{s}/2$. For the mass of the charged pion and the $\tau$ we use the world-average values~\cite{pdg2022}, while for the mass of the $\rho$, due to its large width, we use the event-by-event reconstructed $\pi^\pm\pi^0$ mass. The observable $\theta^\star$ is defined as the polar angle of the $\rho$ momentum in the $\tau$ rest frame, where the polar axis is the boost direction of the $\tau$ in the c.m.\ frame. Similarly, $\psi$ is the polar angle of the charged pion momentum in the $\rho$ rest frame, where the polar axis is the boost direction of the $\rho$ in the c.m.\ frame. Both $\cos\theta^\star$ and $\cos\psi$ exhibit mirrored polarization sensitivity depending on whether the $\tau^-$ decays in the forward ($\cos\theta$>0) or backward ($\cos\theta$<0) hemisphere. Figure \ref{fig:cartoons} illustrates the angle definitions. The distributions of these variables are depicted in Figs. \ref{fig:ct_pol_sense} to \ref{fig:xctpolar} for both the left and right chiral states of the electron beam.
\begin{figure}
	\includegraphics[width=0.3\linewidth]{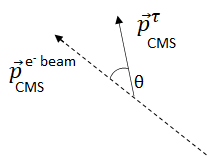}
	\includegraphics[width=0.3\linewidth]{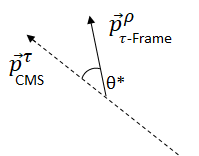}
	\includegraphics[width=0.3\linewidth]{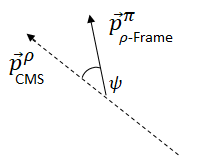}
	\caption{Diagrams illustrating $\theta$ (left) where $f$ represents a final-state particle, $\theta^\star$ (center), and $\psi$ (right).}
	\label{fig:cartoons}
\end{figure}
\begin{figure*}
	\centering
    \begin{tabular}{cc}
        \begin{overpic}[width=0.45\linewidth]{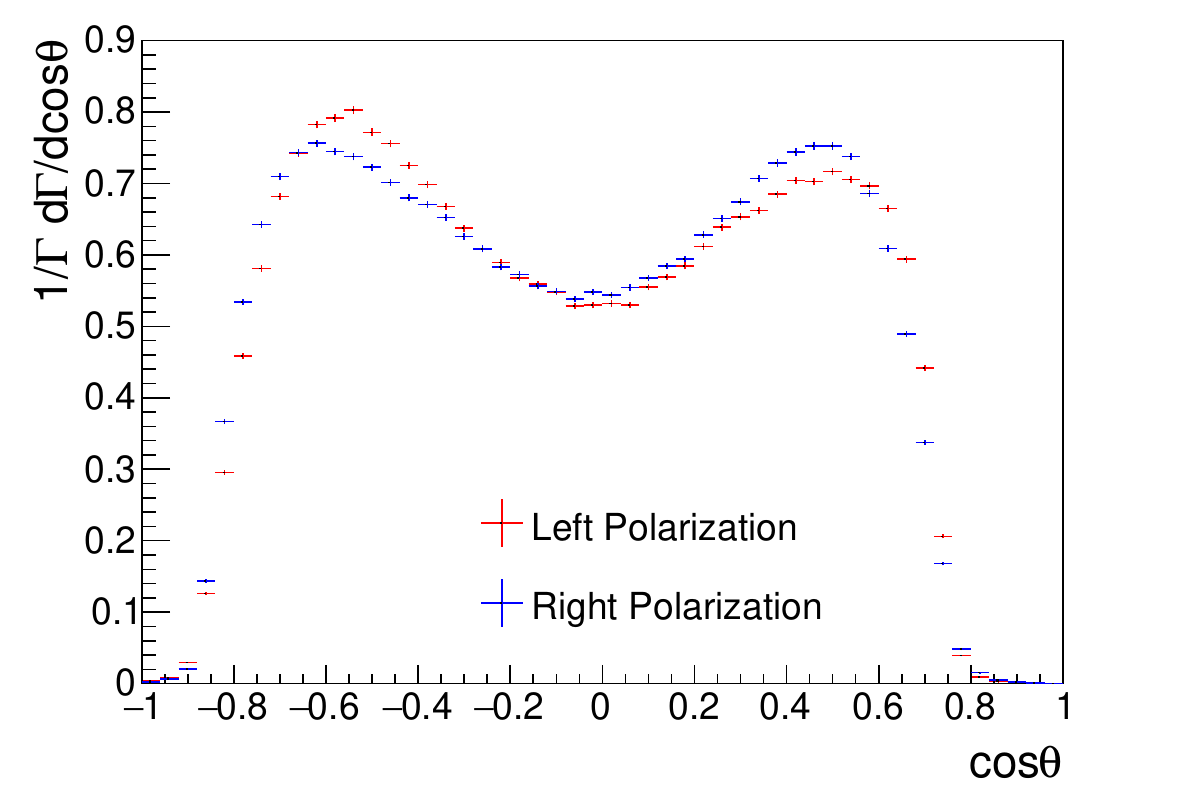} 
            \put(43,60){\textbf{a)} \taurhoinv}
        \end{overpic}
	    &
        \begin{overpic}[width=0.45\linewidth]{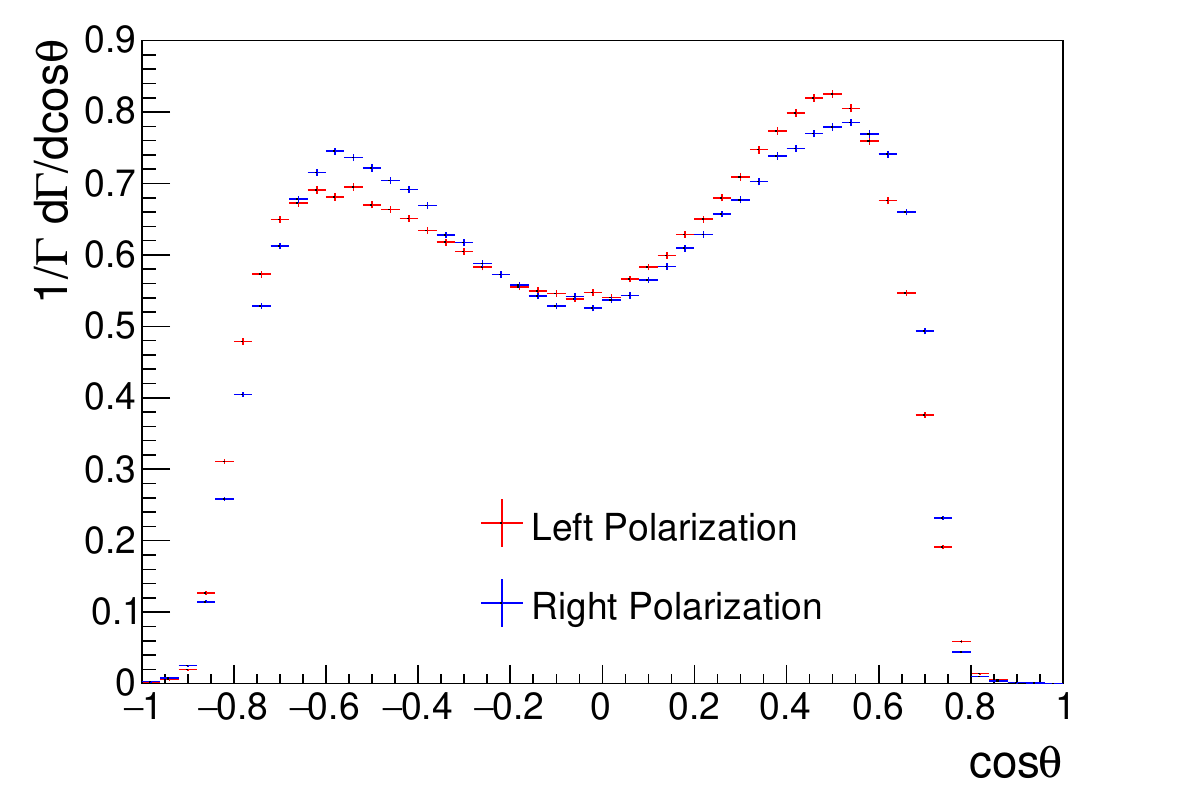}
            \put(43,60){\textbf{b)} \taurho}
        \end{overpic}
    \end{tabular}
	\caption{Distribution of $\cos\theta$ for simulated events after detector response reconstruction for \textbf{a)} positively and \textbf{b)} negatively charged $\tau$ decays to $\rho^\pm \rightarrow \pi^\pm \pi^0$. $\Gamma$ represents the total number of entries and d$\cos\theta$ represents the bin width of 0.04.} 
	\label{fig:ct_pol_sense}
\end{figure*}
\begin{figure*}
	\centering
    \begin{tabular}{cc}
        \begin{overpic}[width=0.45\textwidth]{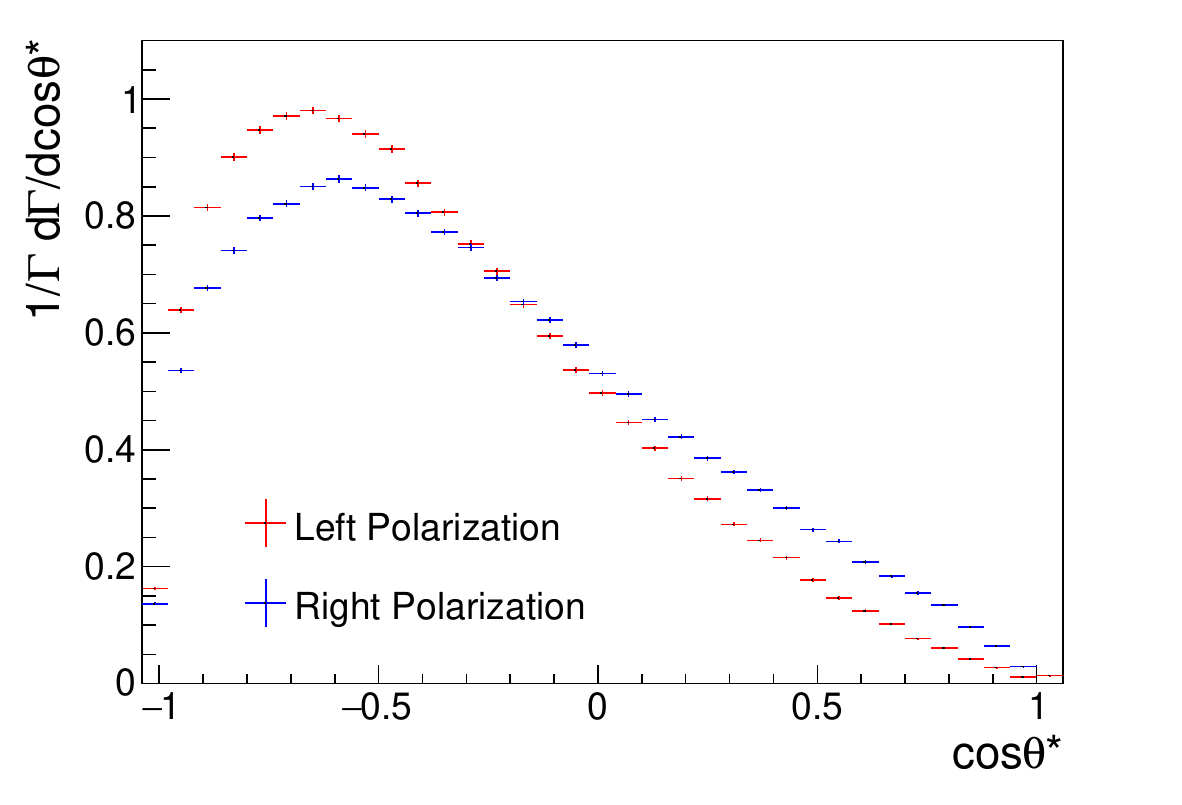}
            \put (60,60) {\textbf{a)} $\tau^+, \cos\theta<0$}
        \end{overpic} &
        \begin{overpic}[width=0.45\textwidth]{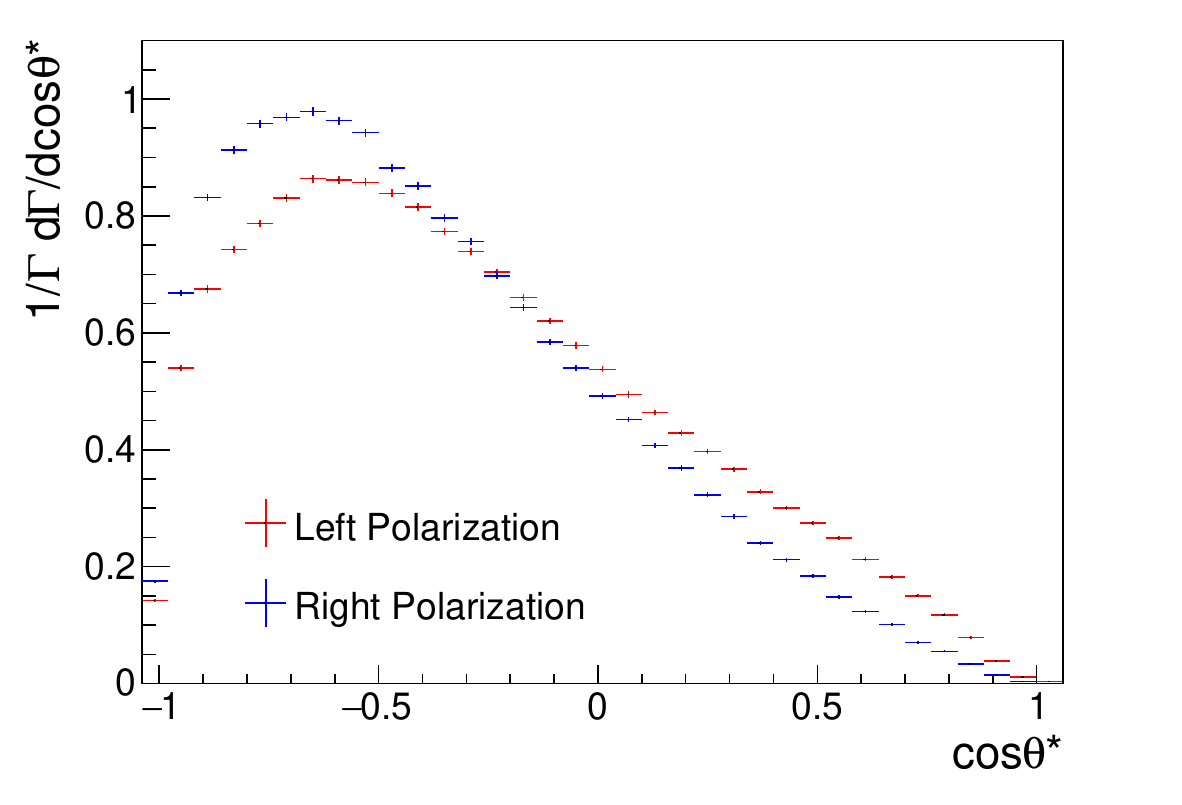}
            \put (60,60) {\textbf{b)} $\tau^+, \cos\theta>0$}
        \end{overpic} \\
        \begin{overpic}[width=0.45\textwidth]{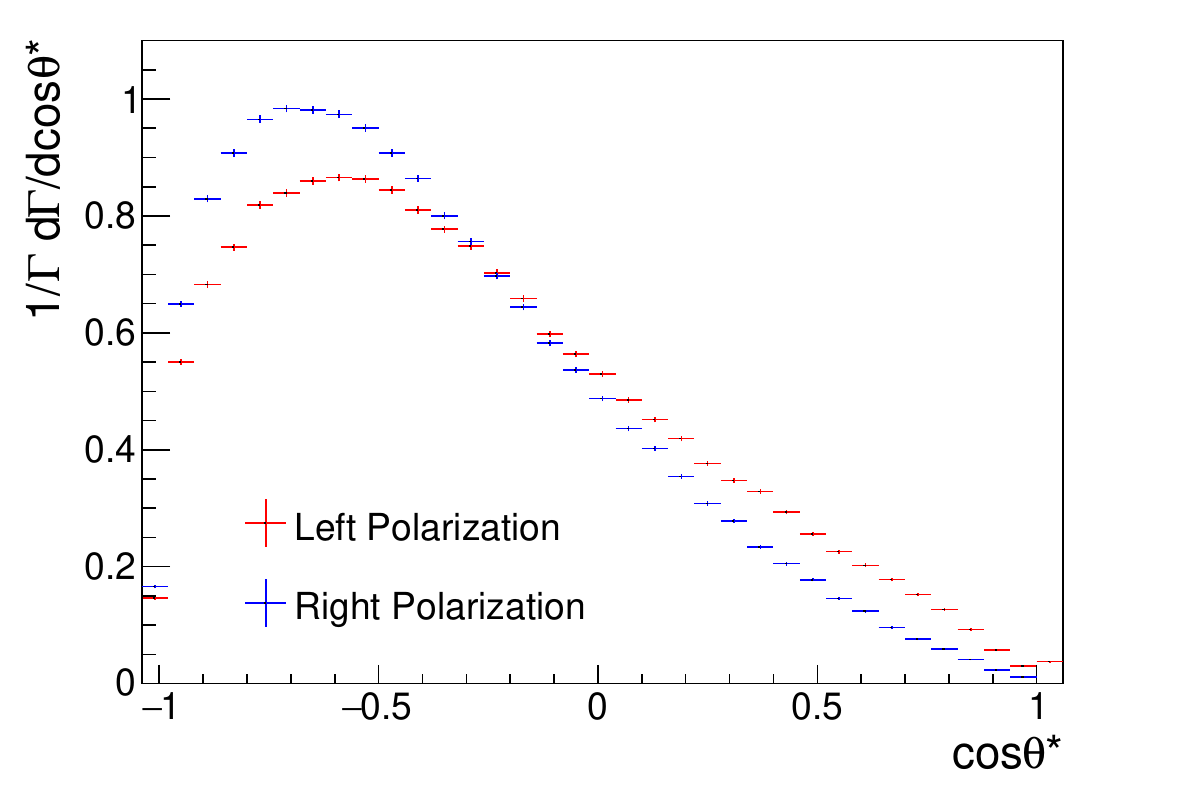}
            \put (60,60) {\textbf{b)} $\tau^-, \cos\theta<0$}
        \end{overpic} &
        \begin{overpic}[width=0.45\textwidth]{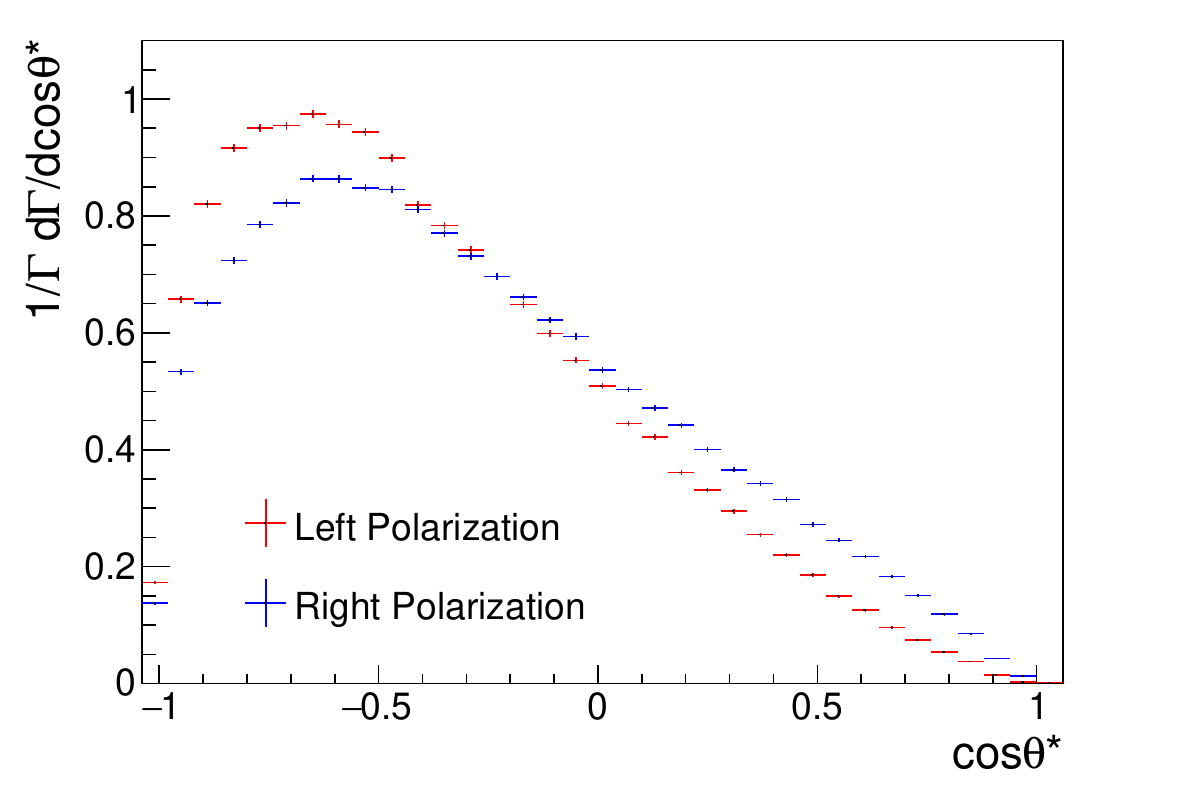}
            \put (60,60) {\textbf{d)} $\tau^-, \cos\theta>0$}
        \end{overpic} \\ 
    \end{tabular}
	\caption{Distribution of $\cos\theta^*$ for simulated events after detector response reconstruction for: $\tau^+$ decays for \textbf{a)} $\cos\theta<0$ and \textbf{b)} $\cos\theta>0$, and for $\tau^-$ decays for \textbf{c)} $\cos\theta<0$ and \textbf{d)} $\cos\theta>0$.$\Gamma$ represents the total number of entries and d$\cos\theta^*$ represents the bin width of 0.04.}
	\label{fig:zctpolar} 
\end{figure*}
\begin{figure*}
	\centering
    \begin{tabular}{cc}
        \begin{overpic}[width=0.45\textwidth]{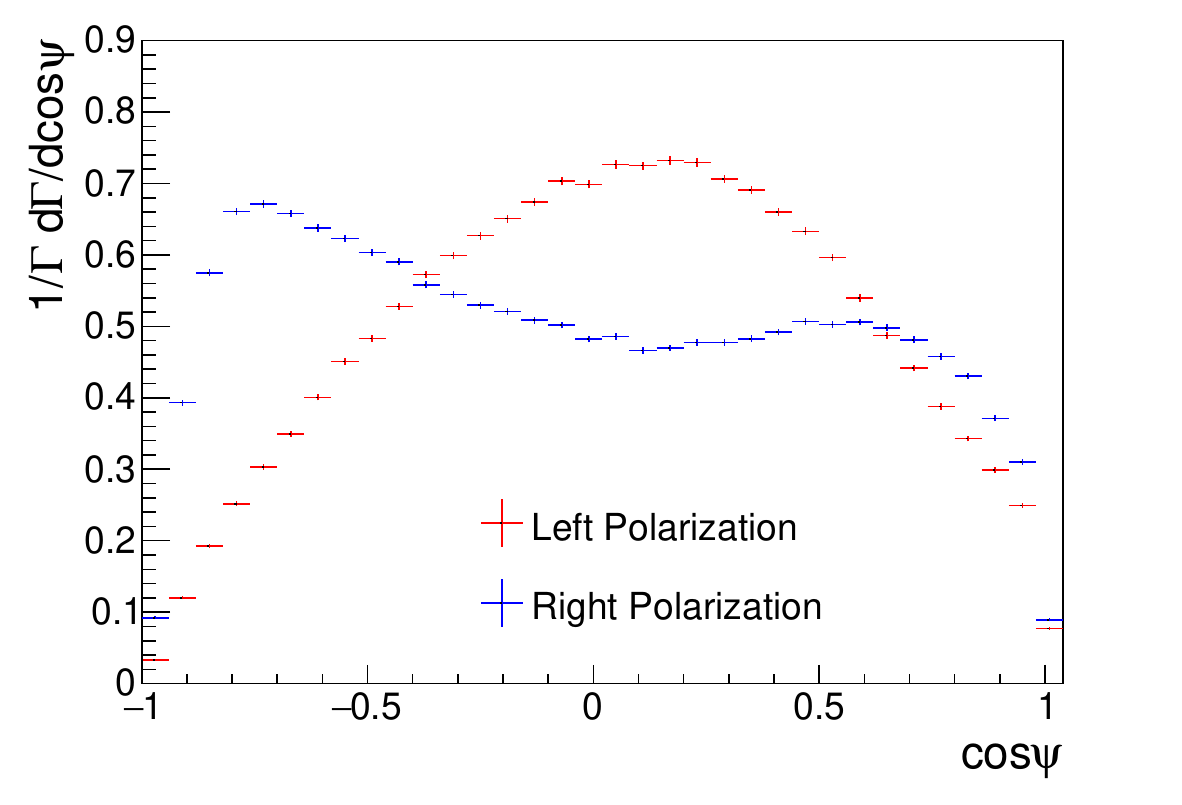}
            \put (60,60) {\textbf{a)} $\tau^+, \cos\theta<0$}
        \end{overpic} &
        \begin{overpic}[width=0.45\textwidth]{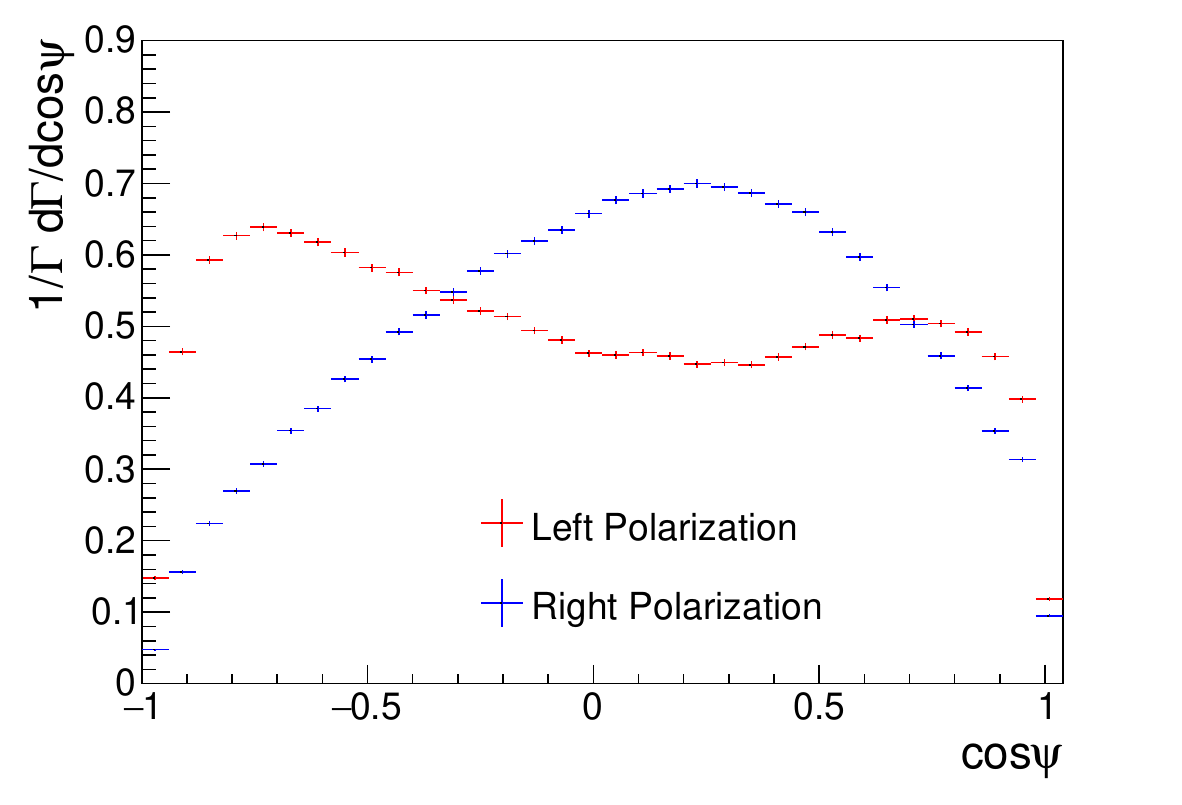}
            \put (60,60) {\textbf{b)} $\tau^+, \cos\theta>0$}
        \end{overpic} \\
        \begin{overpic}[width=0.45\textwidth]{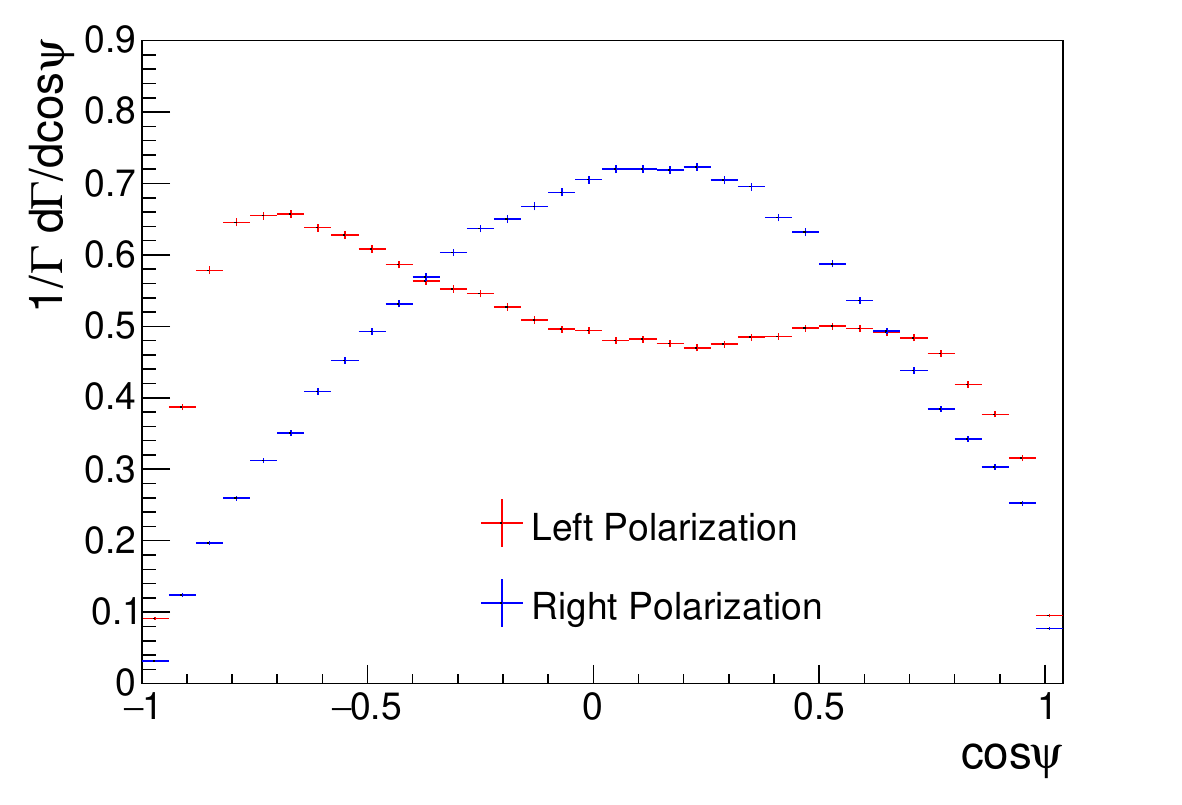}
            \put (60,60) {\textbf{c)} $\tau^-, \cos\theta<0$}
        \end{overpic} &
        \begin{overpic}[width=0.45\textwidth]{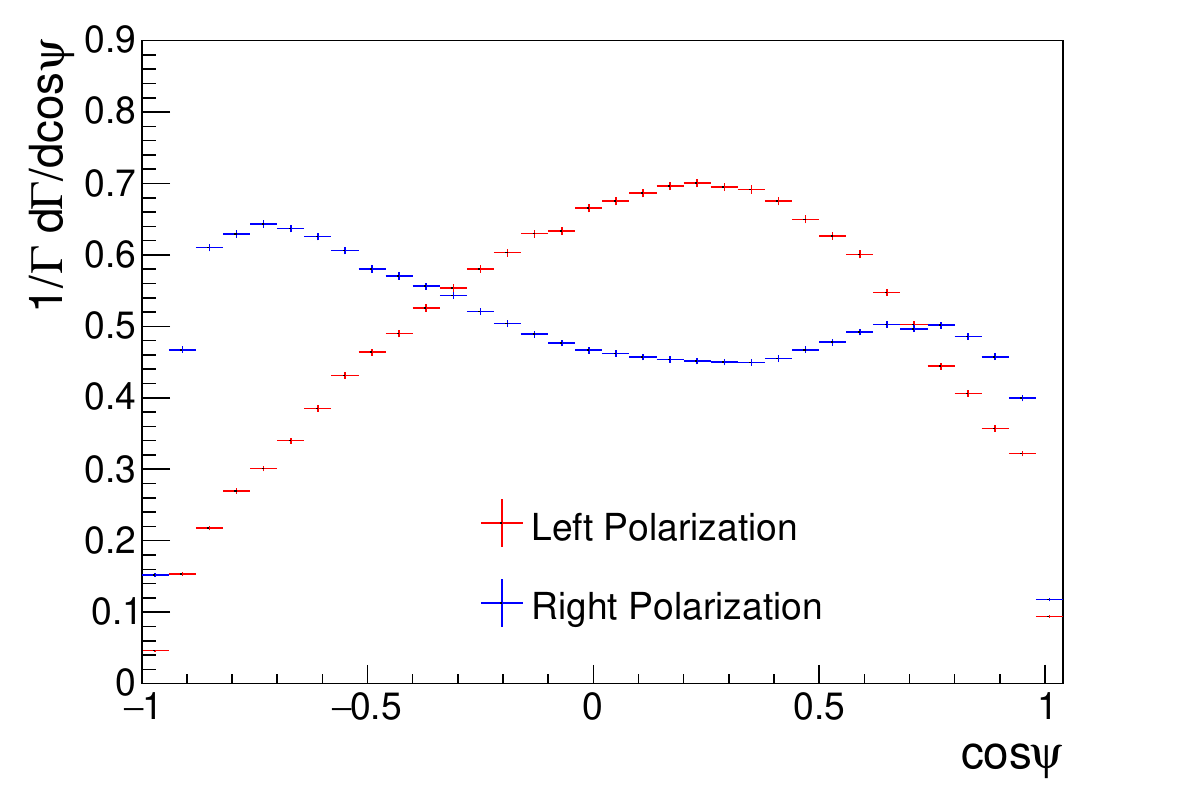}
            \put (60,60) {\textbf{d)} $\tau^-, \cos\theta>0$}
        \end{overpic} \\ 
    \end{tabular}
	\caption{Distribution of $\cos\psi$ for simulated events after detector response reconstruction for: $\tau^+$ decays for \textbf{a)} $\cos\theta<0$ and \textbf{b)} $\cos\theta>0$, and for $\tau^-$ decays for \textbf{c)} $\cos\theta<0$ and \textbf{d)} $\cos\theta>0$.$\Gamma$ represents the total number of entries and d$\cos\psi$ represents the bin width of 0.04.}
	\label{fig:xctpolar}
\end{figure*}

\section{Fitting Methodology}\label{sec:fitting}
To extract the average beam polarization we perform a binned likelihood fit on the normalized distribution using the Barlow and Beeston method as implemented in ROOT~\cite{barlow,rootv6}. We fill three-dimensional histograms of $\cos\theta^\star$, $\cos\psi$, and $\cos\theta$ for each of the data, the $\epem\rightarrow\tautau$ MC for a left polarized beam, the $\epem\rightarrow\tautau$ MC for a right polarized beam, the $\epem\rightarrow\epem$ MC, the $\epem\rightarrow\mumu$ MC, and the $\epem\rightarrow\qqbar$ MC, where $q=u,d,s,c$. The \bbbar final-states were found to contribute no events to the final selection in MC studies.

A linear combination of the MC sample 3D histograms is then fit to the data distribution.
\begin{equation}
    \small
    H_{\textrm{data}}= a_{\textrm{L}}H_{\textrm{L}}+
    a_{\textrm{R}}H_{\textrm{R}}+
    a_{e}H_{e}+
    a_{\mu}H_{\mu}+
    a_{uds}H_{uds}+
    a_{c}H_{c}
\end{equation}
Where $H$ refers to the histograms for data or the MC samples, and $a$ refers to the weights in the fit. The relative weights of the non-$\tau$ backgrounds ($a_e,a_\mu,a_{uds},a_c$) are fixed based on MC efficiency studies. The contributions from the $\tautau$ MC for a left and right polarized $e^-$ beam ($a_L$ and $a_R$) are extracted from the fit and the average beam polarization is calculated from the difference, $\langle P\rangle=a_L-a_R$.

\subsection{MC Validation}\label{sec:PolarSense}
In order to validate the Tau Polarimetry technique at non-zero beam polarization states, the polarized \tautau MC is used to produce and measure different beam polarization states. This was done by splitting each of the left and right polarized $\tautau$ MC in half, one half is used to fill the templates used to perform the polarization fit and the other for mixing beam polarization states. Specific beam polarization states can be created by  combining left and right beam polarization MC samples with appropriate weights, e.g., 70\% polarized is made with 85\% left polarized MC and 15\% right polarized MC. Using this technique we tested polarization states from $-1$ to 1 in steps of 0.1, the results of which are presented in Fig. \ref{fig:polarSense}. The results from fits to the MC samples are in good agreement with the input MC beam polarization states, which demonstrates the measurement technique will yield the correct polarization for any beam polarization within uncertainties.
\begin{figure}
	\centering
	\includegraphics[width=\linewidth]{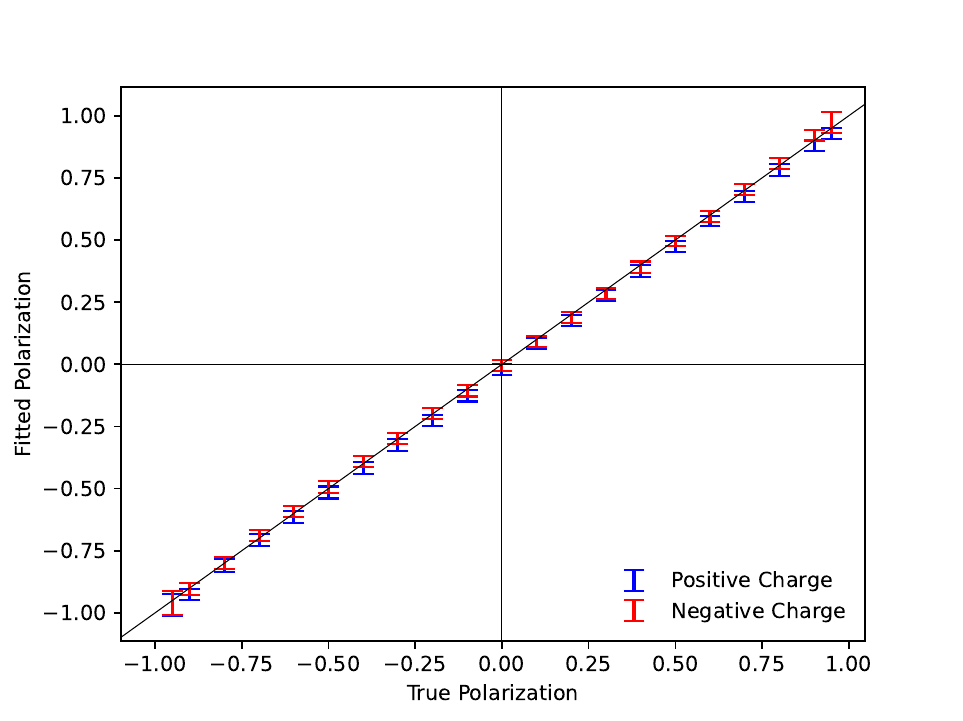}
	\caption{Fit validation MC study: beam polarization outputs of fits as a function of input polarization, produced by mixing polarized \tautau MC as described in the text. The red points correspond to the measurements for positively charged signal candidates while the blue points correspond to the negatively charged candidates. Diagonal line plotted to show optimal correlation.}
	\label{fig:polarSense}
\end{figure}
\section{Event Selection}\label{sec:events}
In order to obtain a pure sample of \taurho events, we tag the second $\tau$ lepton in the event by a decay to \taue or \taumu (or c.c.). Figure \ref{fig:topo} shows the event topology for a signal event tagged with an electron. 
\begin{figure}
	\centering
	\includegraphics[width=0.8\linewidth]{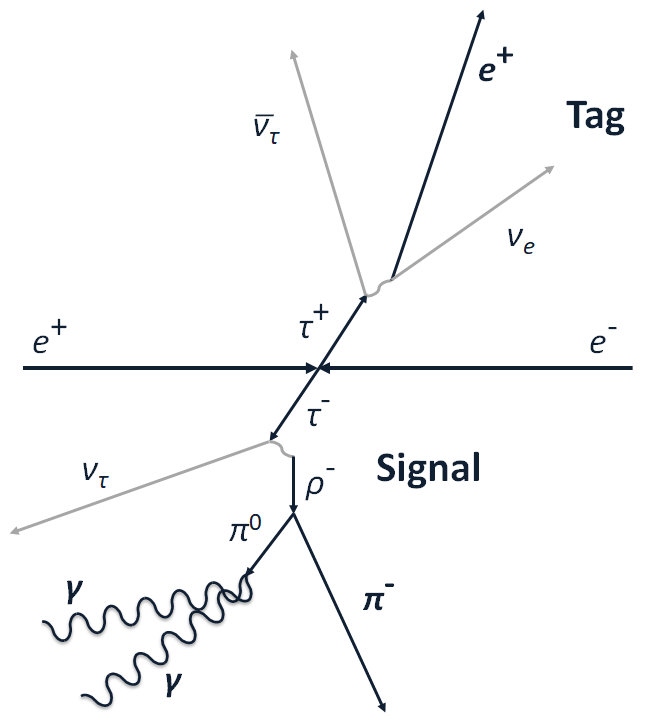}
	\caption{Schematic of a signal type event in the c.m.\ frame with a signal \taurho~decay tagged with a \taue~decay.}
	\label{fig:topo}
\end{figure}
We select this topology by requiring the event to contain two charged particles, one of which is identified as a lepton, and a neutral pion.

The two charged particles are required to originate from within 3~cm of the collision point, as measured along the beam axis, and within 1.5~cm in the transverse plane. We split the event into two hemispheres based on the thrust axis~\cite{thrust1,thrust2} of the event, the signal side and the tag side. The signal side is required to contain the \piz, and the tag side the lepton. The lepton is required to be consistent with either a muon or electron via PID requirements on the track. Both the muon and electron selectors have been trained with machine learning techniques; a boosted decision tree for muons, and an error correcting output code utilizing bootstrap aggregate decision trees for the electrons~\cite{BaBarUpgrade}.

Neutral particles candidates are required to have energy depositions in the EMC exceeding 50 MeV with no associated charged particle identified nearby. Neutral particles within 40 cm (at the EMC) of a charged particle are combined with the charged particle to reduce sensitivity to MC modeling of split-offs arising from hadronic interactions of charged hadrons in the EMC. After this merging, the tag side of the event is required to be free of any neutral particles. Neutral pions are reconstructed from neutral clusters which exceed 100 MeV of deposited energy in one of two ways. First, \babar~is able to identify neutral pions where both photons are detected within the same EMC cluster (a `merged $\pi^0$')~\cite{BaBarDet,BaBarUpgrade}. If no \piz is identified this way, a search for a suitable candidate is performed by evaluating the invariant masses of pairs of neutral clusters. The invariant mass of the reconstructed neutral pion is required to be within a mass window of 115 MeV to 155 MeV for the event to be selected. If multiple candidates exist, the one closest to the $\piz$ mass is accepted as the candidate.

The $\epem\rightarrow\ellell$ ($\ell=e,\mu$), and two-photon ($\epem\rightarrow\epem X$) events, where $X$ is any allowed final state, are primarily rejected by requiring the transverse momentum of each charged particle to exceed 350 MeV, as well as the total event transverse moment (summed over all charged and neutral particles) to exceed 350 MeV. The surviving Bhabha events are reduced by approximately a factor of two by requiring the EMC energy in the lab frame not to exceed 10 GeV, at the cost of 0.028\% of signal events.

The acceptance in $\theta$ for charged particle tracks is slightly reduced such that each track is within the calorimeter acceptance, $0.430<\theta_{\textrm{lab}}<2.350$ rads. This fiducial requirement improves PID performance and data/MC agreement, and reduces the contamination from Bhabha events. The Bhabha contamination is further reduced by a factor of three by requiring  $-1<\cos\theta^\star<0.9$ and $-0.9<\cos\psi<1$. 

The event selection is further refined by reconstructing the $\rho$ on the signal side, and requiring the reconstructed mass to exceed 300~MeV. This ensures $\cos\psi$ remains physical. The reconstructed $\rho$ is also required to exhibit an angle between it's decay products in the c.m.\ frame of $\cos\alpha<0.9$, where $\alpha$ is the angle between the charged and neutral pion. This reduces sensitivity to MC modeling of events where the hadronic shower of the charged pion can overlap with the electromagnetic showers associated with the \piz. As the true $\tau$ direction is not reconstructed because of the missing neutrino, the reconstructed $\rho$ direction is used to determine $\cos\theta$. This approach was found to supply the least biased estimate for the true $\tau$ direction through MC studies.

These requirements result in a final \epem\ra\tautau selection that is 99.9\% pure and selects 1.4\% of all $\tau^+\tau^-$ events. This corresponds to a 7.8\% overall efficiency for selecting $\tau^\pm\tau^\mp\rightarrow\rho^\pm\nu_\tau+\ell^\mp\nu_\ell\overline{\nu_\tau}$ events. The largest non-$\tau$ background sources are Bhabha and $\mumu$ events, each of which make up 0.05\% of the final sample. The final event selection break-down as predicted by the MC simulations is shown in Table \ref{tab:events}. There is a small but statistically significant difference between the efficiency for selecting left and right polarized events; $\Delta\varepsilon=0.011\%\pm0.001\%$. This is small enough that it will have a negligible effect on the extracted polarization.
\begin{table}
\centering
\caption{Fraction of event types expected in data in the final event selection based on MC efficiencies. The $\tau$ pair events are further broken down to show the decay mode composition of the events selected on the signal side.}
\label{tab:events}
\begin{tabular}{lr} \toprule\toprule
MC source & \quad Fraction \\ \midrule
Bhabha & 0.046\%\\
\mumu & 0.046\%\\
\uubar,\ddbar,\ssbar & 0.030\%\\
\ccbar & 0.006\%\\
\bbbar & 0.000\%\\
\tautau & 99.871\%\\
\end{tabular}
\hspace{2cm}
\begin{tabular}{lr} \midrule\midrule
	Tau Signal & Fraction \\ \midrule
	\taue	&	0.018\%	\\
	\taumu	&	0.031\%	\\
	\taupi	&	0.035\%	\\
	\taurho	&	87.858\%	\\
	\taua	&	9.785\%	\\
	$\tau\rightarrow$ else	&	2.145\%	\\ \bottomrule\bottomrule
\end{tabular}
\end{table} 
\section{Fit Results}\label{sec:results}
As is evident in Figures \ref{fig:ct_pol_sense} to \ref{fig:xctpolar}, the distributions for a left-handed electron beam generates distributions for $\tau^-$ leptons that are the same as those for right-handed beams and $\tau^+$ leptons. Consequently, we fit the positive and negative charged distributions separately. As the \babar~data is split into chronological periods, runs 1 through 6, each run is treated independently. As such we obtain six measurements of the beam polarization and corresponding statistical and systematic uncertainties. Table \ref{tab:fitresults} shows the fit results for each run and the associated statistical uncertainty only. 
Taking the weighted mean of these fit results gives the overall average beam polarization for \pep2 to be $\langle P\rangle=\polresult\pm \polstat_{\textrm{stat}}$. The two dimensional projections of $\cos\theta^*$ and $\cos\psi$ for positively charged events are shown in Fig. \ref{fig:2d_proj_pos} and the negatively charged events in Fig. \ref{fig:2d_proj_neg}. The one dimensional projections are included in Appendix \ref{app:plots}.
\begin{table*}
	\centering
    \caption{Average beam polarization measured for each run period of the \babar~data set. The average for each run is obtained from the weighted mean of the positive and negative fit results. The reported uncertainties are statistical only.}
	\label{tab:fitresults}
	\begin{tabular}{lrrr} \toprule\toprule
		Data Set (fb$^{-1}$)&Positive Charge  & \hspace{10pt}Negative Charge & \hspace{10pt}Average Polarization \\ \midrule
		Run 1 (20.4) & 0.0018$\pm$0.014\phantom{0} & -0.0047$\pm$0.014\phantom{0} &  -0.0014$\pm$0.010\phantom{0}\\ 
        Run 2 (61.3) &	0.0075$\pm$0.0083 & 0.0007$\pm$0.0083 & 0.0041$\pm$0.0059	\\ 
        Run 3 (32.3) &	0.0151$\pm$0.012\phantom{0} &-0.0047$\pm$0.012\phantom{0}	&  0.0048$\pm$0.0083	\\ 
        Run 4 (99.6) &-0.0035$\pm$0.0072 & 0.0010$\pm$0.0067	&-0.0011$\pm$0.0049	\\ 
        Run 5 (132.3) &-0.0028$\pm$0.0062 & 0.0136$\pm$0.0064	& 0.0052$\pm$0.0045	\\ 
        Run 6 (78.3) &	0.0036$\pm$0.0089 & 0.0133$\pm$0.0088	& 0.0084$\pm$0.0062	\\   \midrule
	    424.18$\pm$1.8& 0.0015$\pm$0.0034& 0.0055$\pm$0.0034& 0.0035$\pm$0.0024 \\ \bottomrule\bottomrule
	\end{tabular}
\end{table*}
\begin{figure*}
    \centering
    \begin{tabular}{cc}
        \begin{overpic}[width=0.4\textwidth]{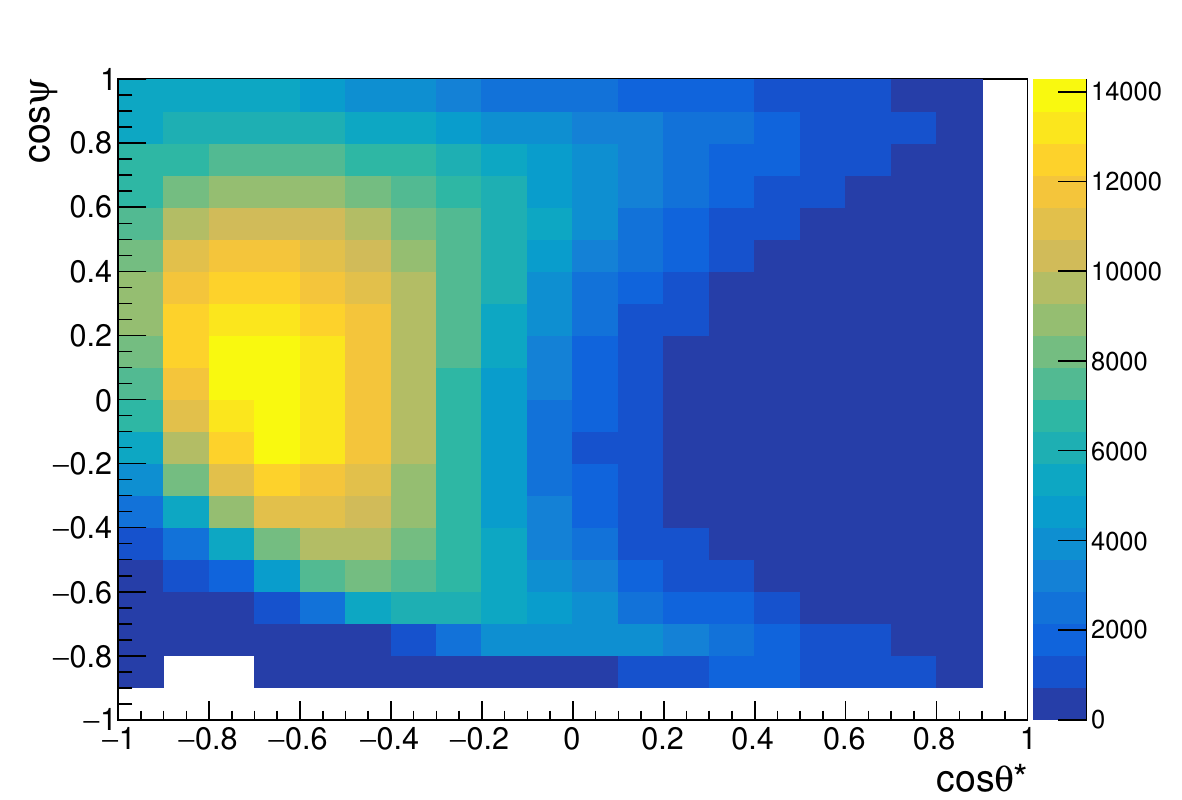}
            \put (10,62) {Data, $\cos\theta<0$}
            \put (75,48) {\color{white}\textbf{a)}}
            \put (75,62) {\babar}
        \end{overpic} &
        \begin{overpic}[width=0.4\textwidth]{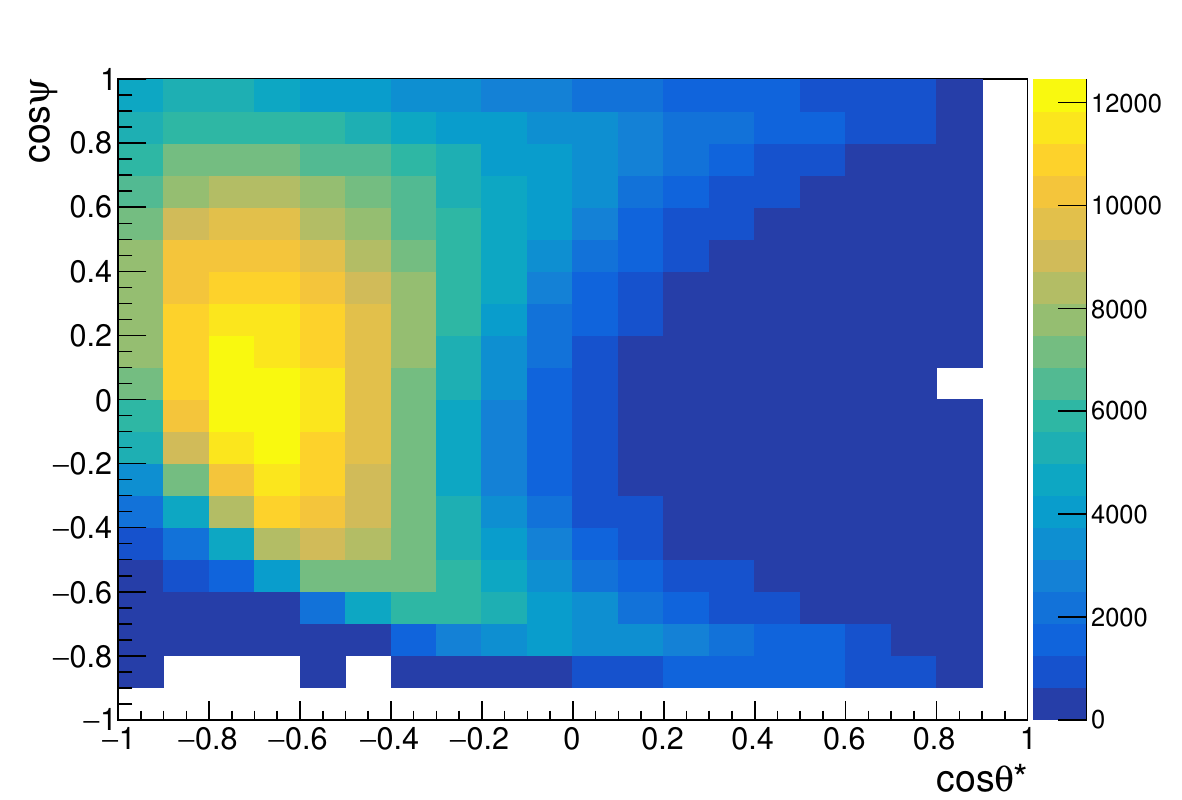}
            \put (10,62) {Data, $\cos\theta>0$}
            \put (75,48) {\color{white}\textbf{b)}}
            \put (75,62) {\babar}
        \end{overpic} \\
        \begin{overpic}[width=0.4\textwidth]{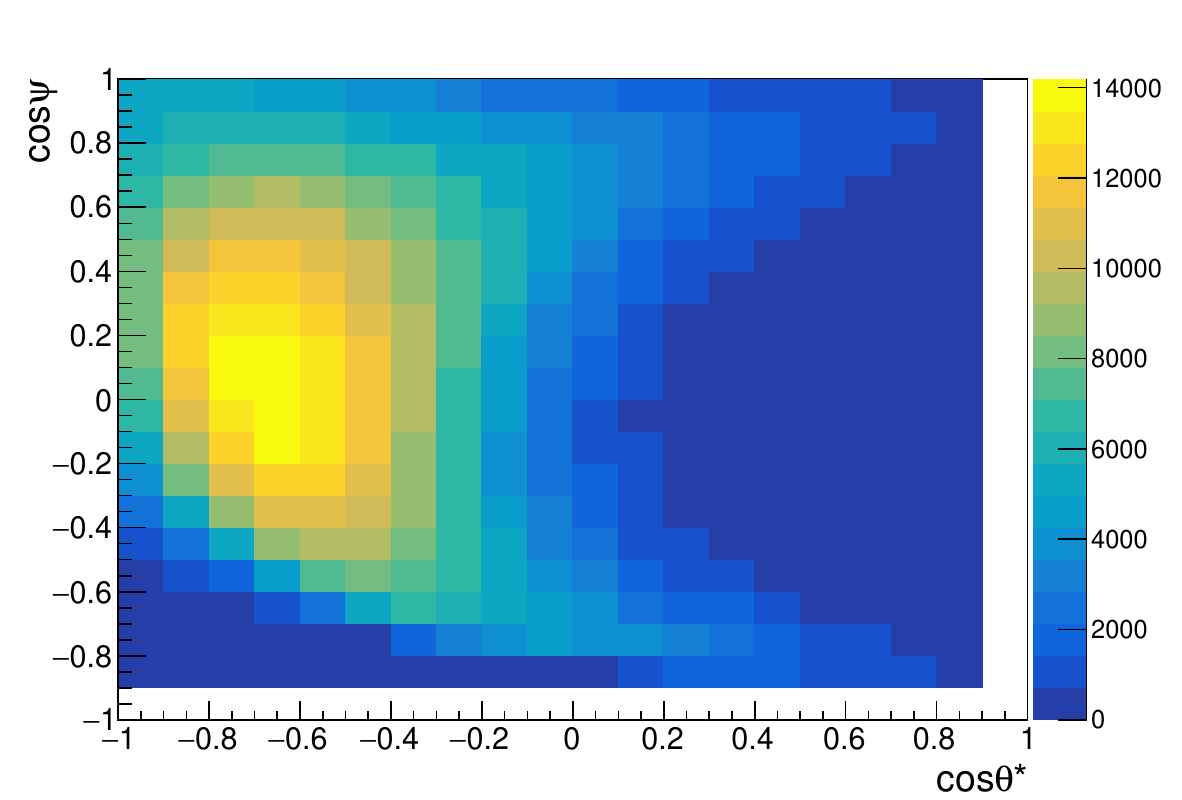}
            \put (10,62) {Fit Result, $\cos\theta<0$}
            \put (75,35) {\color{white}\textbf{c)}}
            \put (60,62) {\babar}
        \end{overpic} &
        \begin{overpic}[width=0.4\textwidth]{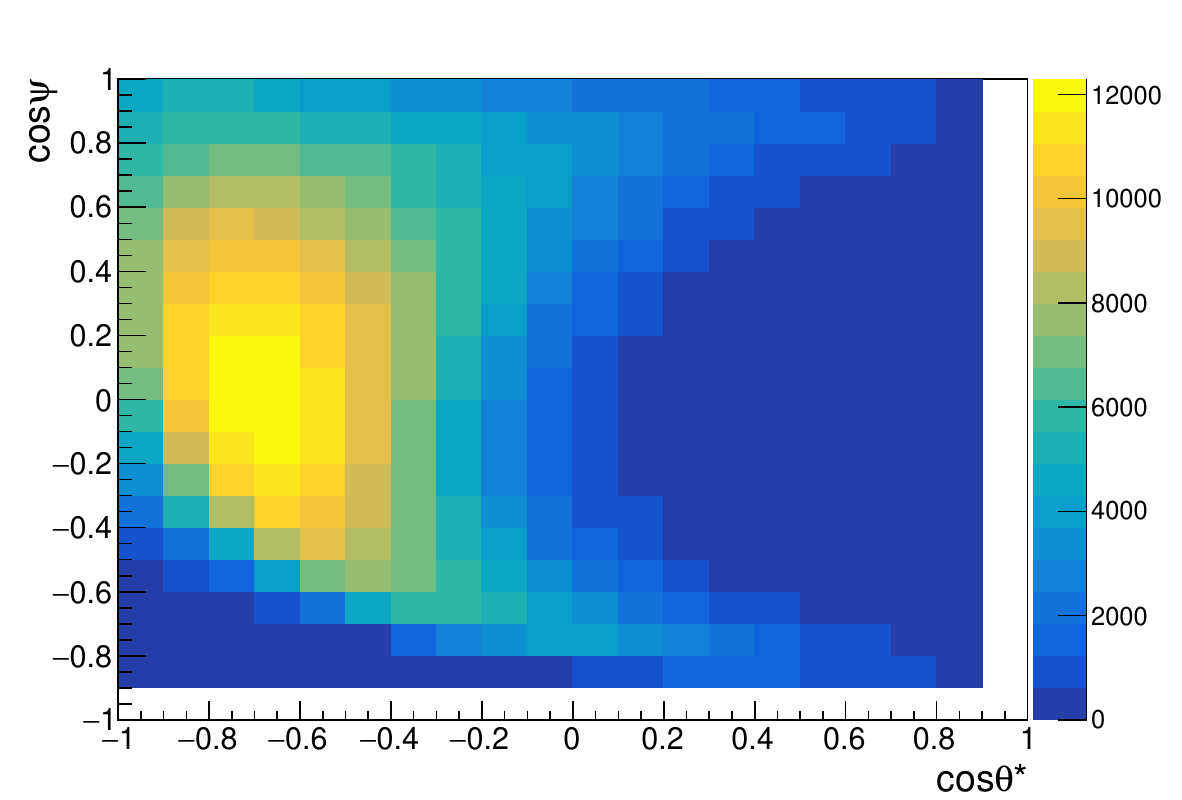}
            \put (10,62) {Fit Result, $\cos\theta<0$}
            \put (75,35) {\color{white}\textbf{d)}}
            \put (60,62) {\babar}
        \end{overpic} \\
        \begin{overpic}[width=0.4\textwidth]{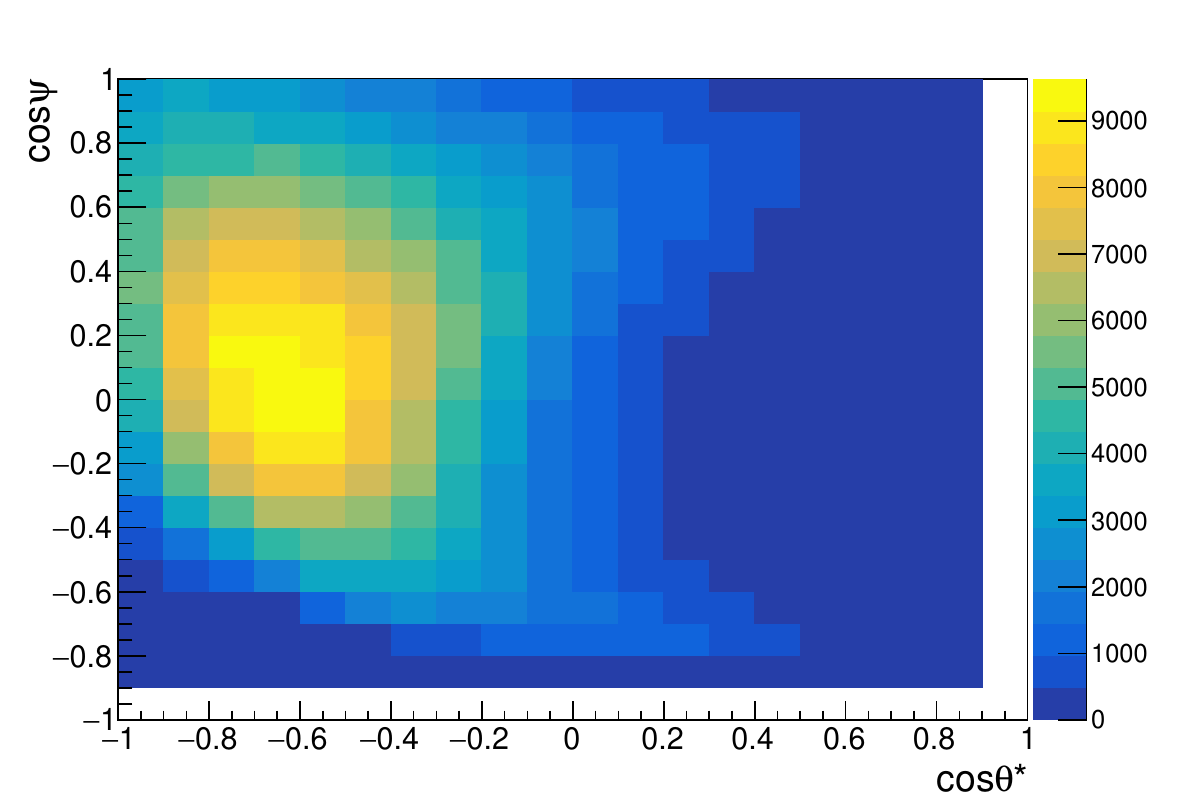}
            \put (10,62) {Left Polarized MC, $\cos\theta<0$}
            \put (75,48) {\color{white}\textbf{e)}}
            \put (75,62) {\babar}
        \end{overpic} &
        \begin{overpic}[width=0.4\textwidth]{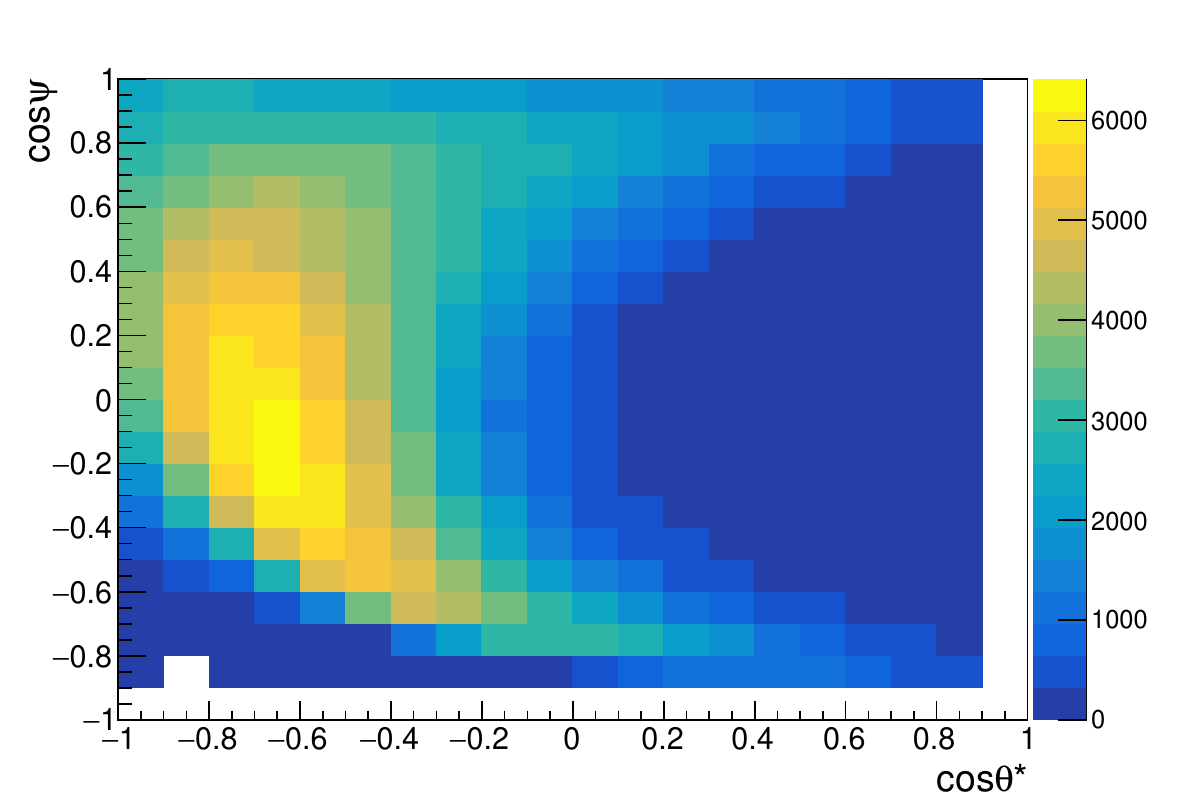}
            \put (10,62) {Left Polarized MC, $\cos\theta>0$}
            \put (75,48) {\color{white}\textbf{f)}}
            \put (75,62) {\babar}
        \end{overpic} \\
        \begin{overpic}[width=0.4\textwidth]{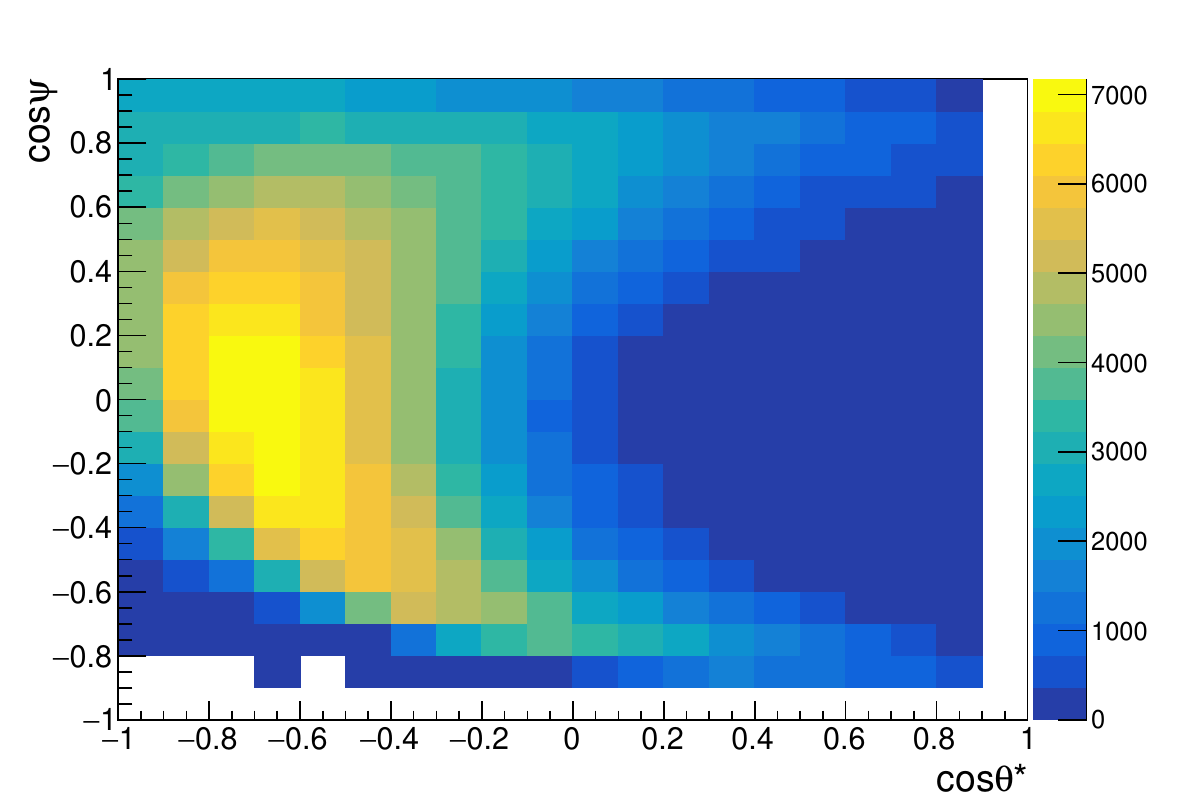}
            \put (10,62) {Right Polarized MC, $\cos\theta<0$}
            \put (75,48) {\color{white}\textbf{g)}}
            \put (75,62) {\babar}
        \end{overpic} &
        \begin{overpic}[width=0.4\textwidth]{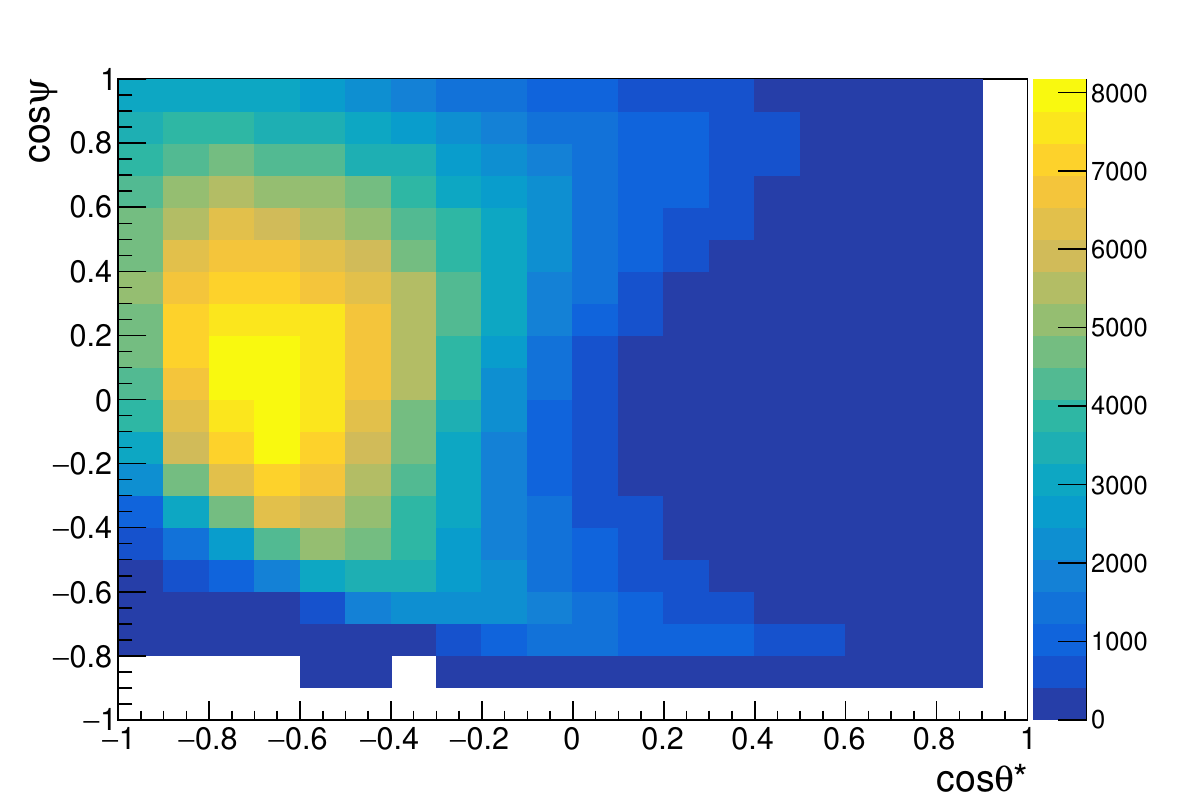}
            \put (10,62) {Right Polarized MC, $\cos\theta>0$}
            \put (75,48) {\color{white}\textbf{h)}}
            \put (75,62) {\babar}
        \end{overpic} \\
    \end{tabular}
    \caption{Projection of the 3-dimensional event distributions onto the $\cos\theta^*$ vs $\cos\psi$ plane for positively charged signal events. \textbf{a)} Data, $\cos\theta<0$, \textbf{b)} data, $\cos\theta>0$, \textbf{c)} fit result, $\cos\theta<0$, \textbf{d)} fit result, $\cos\theta>0$, \textbf{e)} left-polarized MC, $\cos\theta<0$, \textbf{f)} left-polarized MC, $\cos\theta>0$, \textbf{g)} right-polarized MC, $\cos\theta<0$, \textbf{h)} right-polarized MC, $\cos\theta>0$.}
    \label{fig:2d_proj_pos}
\end{figure*}
\begin{figure*}
    \centering
    \begin{tabular}{cc}      
        \begin{overpic}[width=0.4\textwidth]{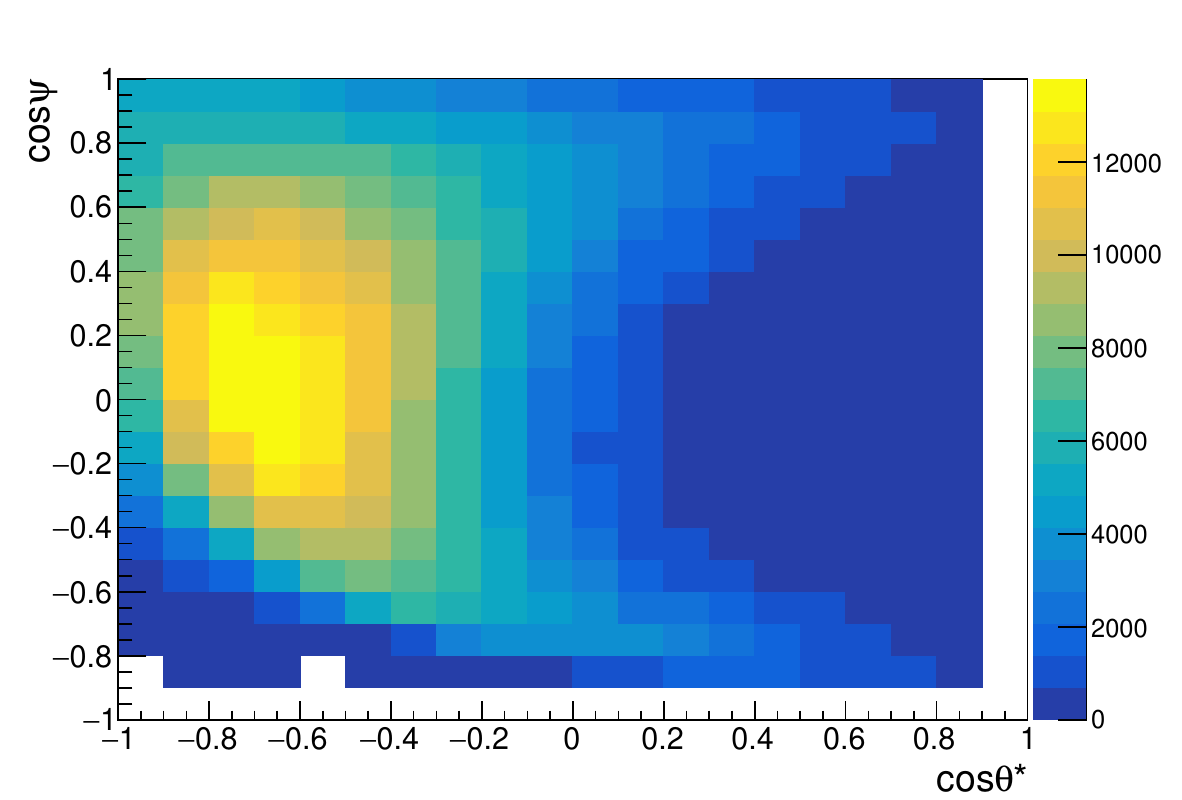}
            \put (10,62) {Data, $\cos\theta<0$}
            \put (75,48) {\color{white}\textbf{a)}}
            \put (75,62) {\babar}
        \end{overpic} &
        \begin{overpic}[width=0.4\textwidth]{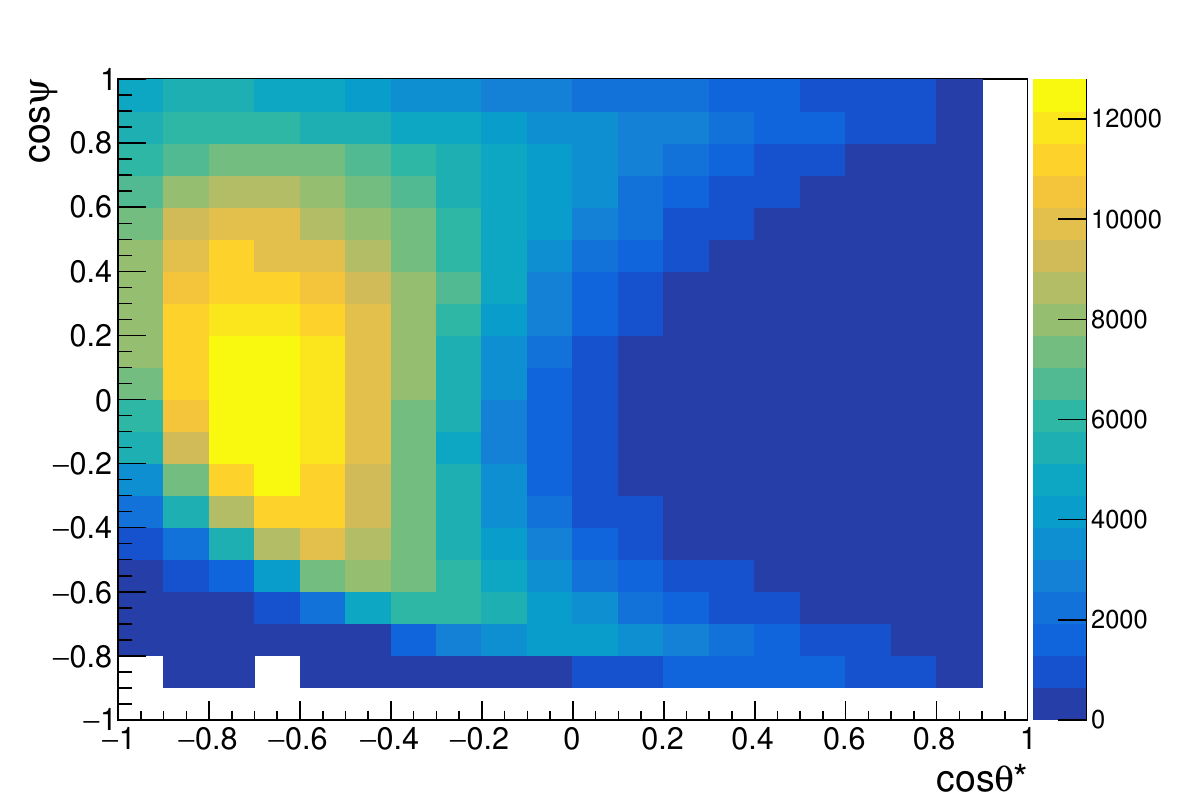}
            \put (10,62) {Data, $\cos\theta>0$}
            \put (75,48) {\color{white}\textbf{b)}}
            \put (75,62) {\babar}
        \end{overpic} \\
        \begin{overpic}[width=0.4\textwidth]{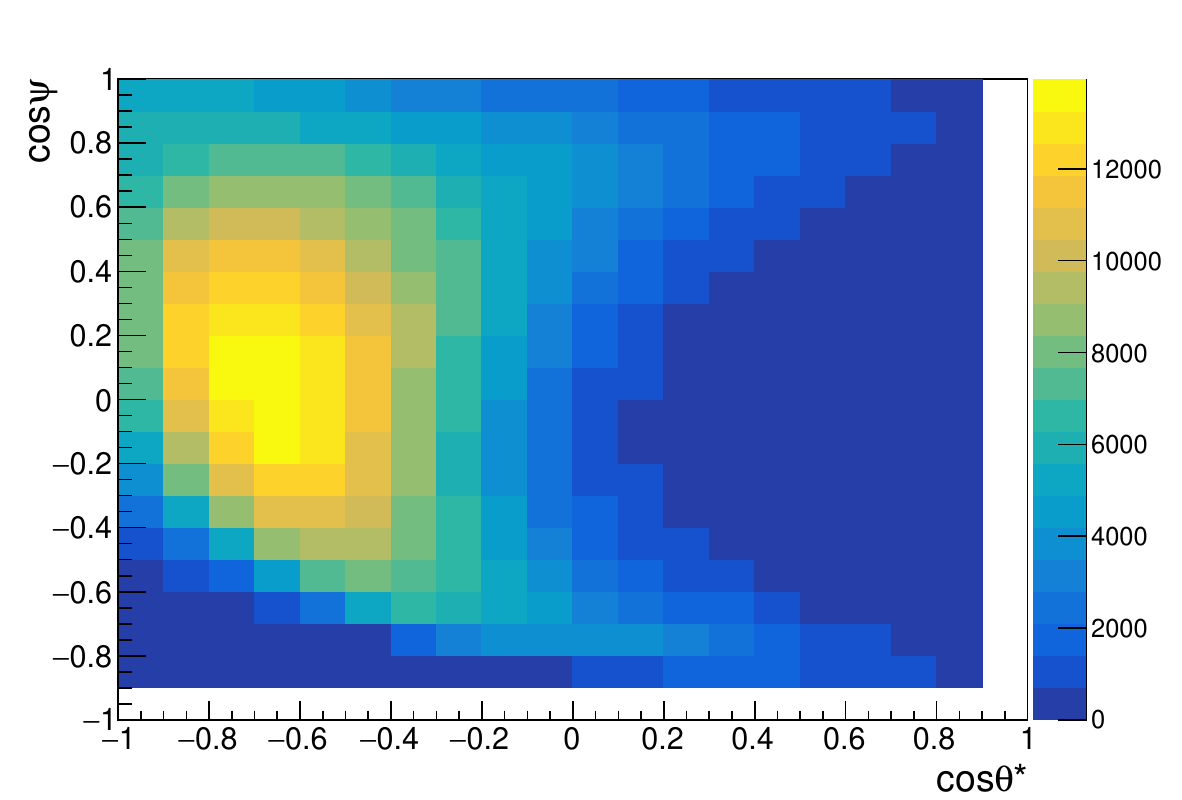}
            \put (10,62) {Fit Result, $\cos\theta<0$}
            \put (75,35) {\color{white}\textbf{c)}}
            \put (60,62) {\babar}
        \end{overpic} &
        \begin{overpic}[width=0.4\textwidth]{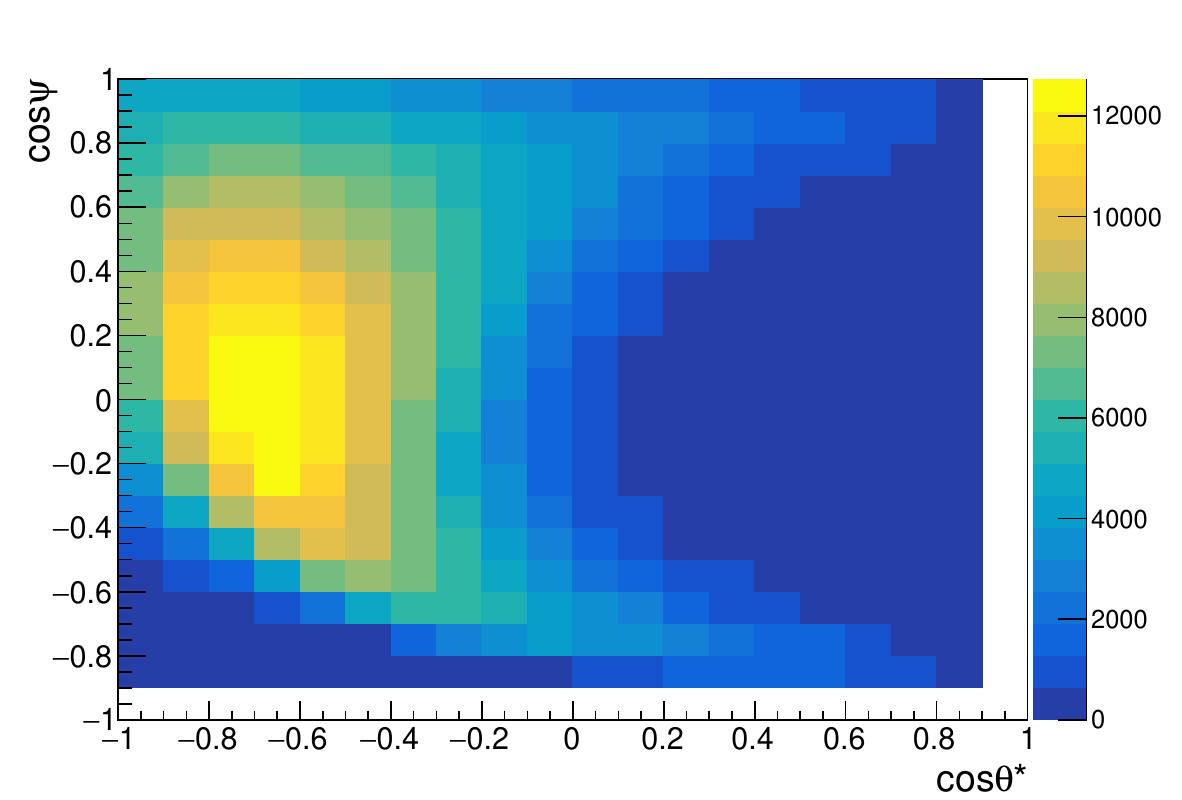}
            \put (10,62) {Fit Result, $\cos\theta<0$}
            \put (75,35) {\color{white}\textbf{d)}}
            \put (60,62) {\babar}
        \end{overpic} \\
        \begin{overpic}[width=0.4\textwidth]{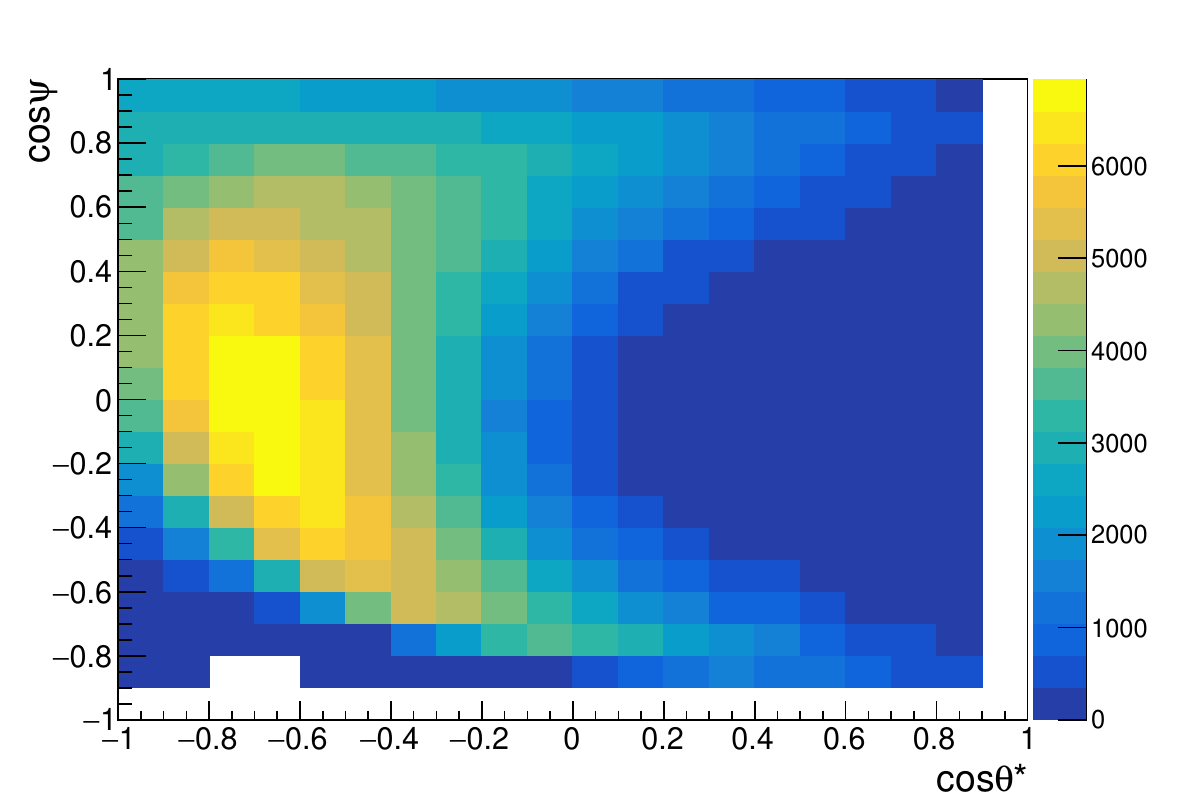}
            \put (10,62) {Left Polarized MC, $\cos\theta<0$}
            \put (75,48) {\color{white}\textbf{e)}}
            \put (75,62) {\babar}
        \end{overpic} &
        \begin{overpic}[width=0.4\textwidth]{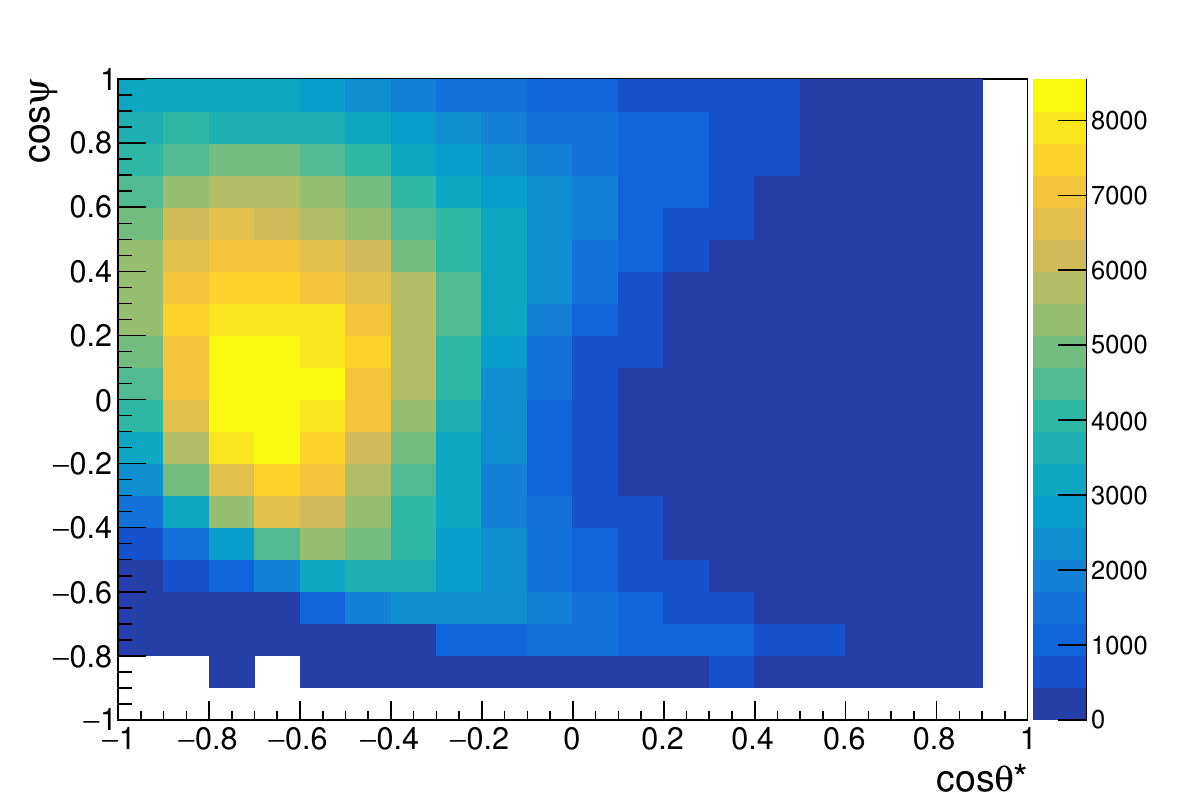}
            \put (10,62) {Left Polarized MC, $\cos\theta>0$}
            \put (75,48) {\color{white}\textbf{f)}}
            \put (75,62) {\babar}
        \end{overpic} \\
        \begin{overpic}[width=0.4\textwidth]{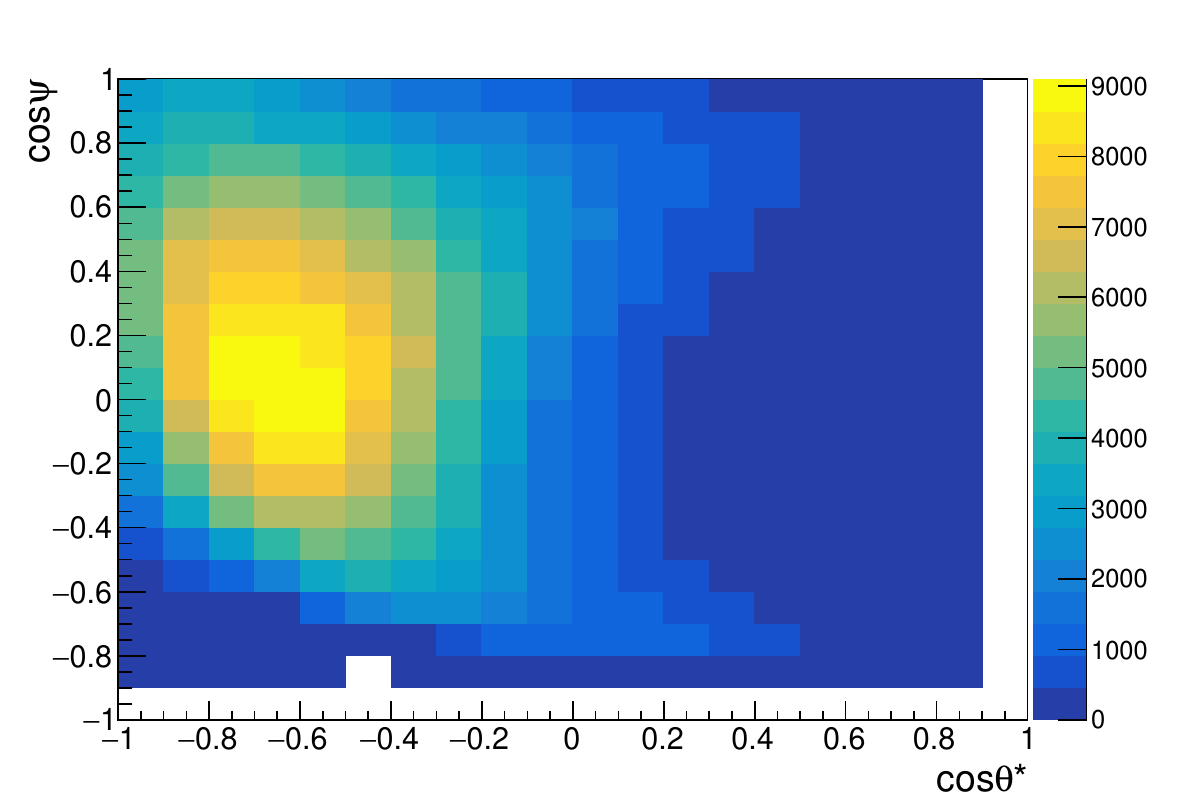}
            \put (10,62) {Right Polarized MC, $\cos\theta<0$}
            \put (75,48) {\color{white}\textbf{g)}}
            \put (75,62) {\babar}
        \end{overpic} &
        \begin{overpic}[width=0.4\textwidth]{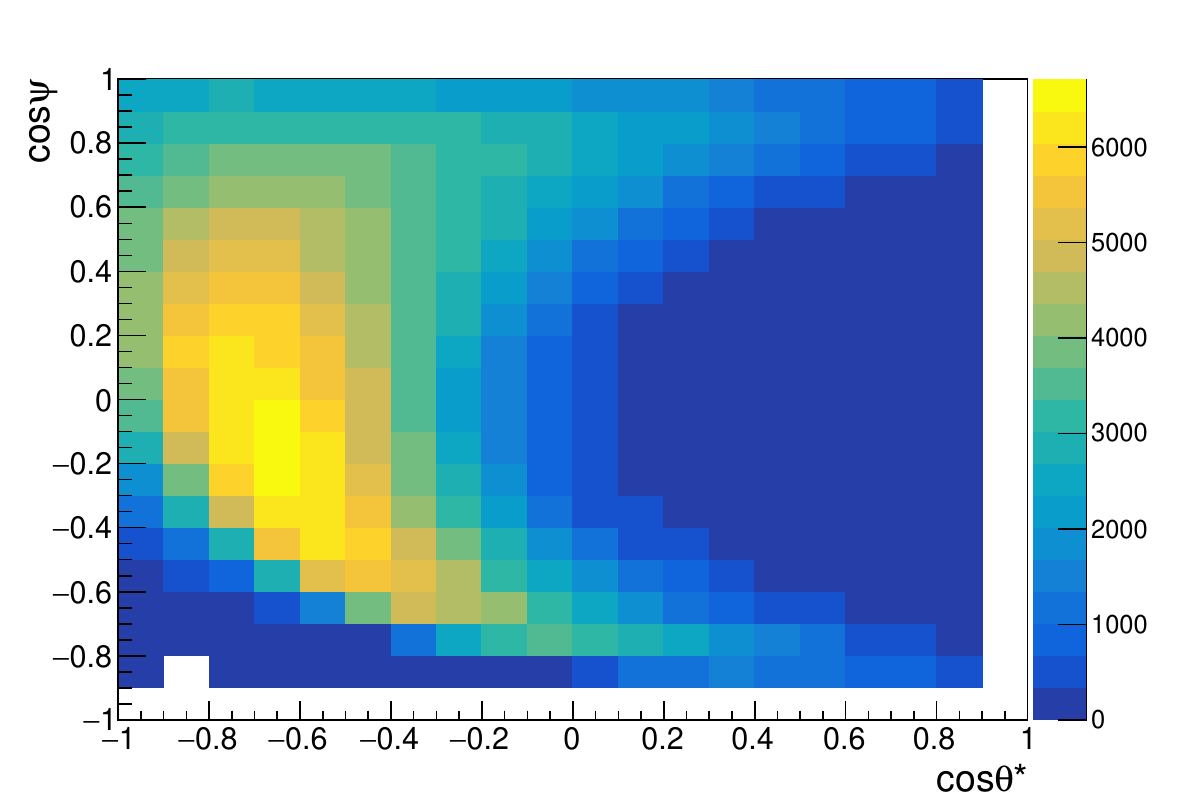}
            \put (10,62) {Right Polarized MC, $\cos\theta>0$}
            \put (75,48) {\color{white}\textbf{h)}}
            \put (75,62) {\babar}
        \end{overpic} \\
    \end{tabular}
    \caption{Projection of the 3-dimensional event distributions onto the $\cos\theta^*$ vs $\cos\psi$ plane for negatively charged signal events. \textbf{a)} Data, $\cos\theta<0$, \textbf{b)} data, $\cos\theta>0$, \textbf{c)} fit result, $\cos\theta<0$, \textbf{d)} fit result, $\cos\theta>0$, \textbf{e)} left-polarized MC, $\cos\theta<0$, \textbf{f)} left-polarized MC, $\cos\theta>0$, \textbf{g)} right-polarized MC, $\cos\theta<0$, \textbf{h)} right-polarized MC, $\cos\theta>0$.}
    \label{fig:2d_proj_neg}
\end{figure*}

\section{Systematic Uncertainty Studies}\label{sec:systematics}
Each of the systematic uncertainties have been evaluated with a method that best suits the particular source of systematic uncertainty. The first method used is a controlled variation of MC distributions in order to adjust the fit templates and determine the effect on the beam polarization measurements. The second method is a variation of the selection applied to the variable. This is primarily used in regions where the selection is designed to remove uncontrolled sources of backgrounds or poor MC modeling. The final method is used to evaluate the PID requirements, where different selectors are employed and the different effects on the data and MC are used as an estimator of the bias introduced by the selectors. 
This section discusses these methods in more detail and how they apply to each variable. For all of the approaches, the intent is to capture an approximate 68\% interval on the systematic variations. The systematic uncertainties are combined in a way that accounts for correlations between runs and summed in quadrature to deliver a total uncertainty. Table \ref{tab:systematicSummary} shows a summary of all the systematic uncertainties associated with the polarization measurement.
\subsection{Controlled variation of MC templates}
\subsubsection{\texorpdfstring{\piz}{Neutral pion} efficiency correction} \label{sec:piz_eff}
The \piz selection efficiency is notably different in data and MC and so is corrected in the polarized $\tau$ MC fit templates using the lab-frame momentum and lab-frame $\cos\theta$ distributions in the unpolarized $\tau$ MC and data.
This is done by binning the \piz lab-frame momentum and $\cos\theta$ data/unpolarized-MC ratios, for both $\tau$ charges combined, to obtain a set of correction factors. These corrections are then applied to the polarized MC. A systematic uncertainty is evaluated by varying the correction factors up and down by the statistical uncertainty in each bin. This process results in a systematic uncertainty of $\sigma=0.0013$. 
By combining both charged states in the correction procedure, the efficiency correction is independent of polarization effects. To verify the correction does not introduce a bias to the polarization measurement, the procedure was performed on a $70\%$ polarized MC sample, which demonstrated a negligible effect on the polarization measurement. 
\subsubsection{Neutral particle energy calibration} \label{sec:neu_scale}
The energy calibration of photons in the \babar~detector is known to within 0.3\%~\cite{BaBarUpgrade}. Increasing and decreasing the energy calibration of all photons in the MC results in a systematic uncertainty of $\sigma=0.0010$.  
\subsubsection{Boost correction} \label{sec:boost}
As the beam energies in \pep2 are asymmetric a boost is required to move between the lab and c.m.\ frames. As a mismodeling of the boost vector can affect the polarization measurement, a sample of $\epem\rightarrow\mumu$ events were studied to quantify the effect. A small offset in the acollinearity of the muon pairs between the data and MC indicated a 4 MeV discrepancy in the z component of the boost vector. Correcting this offset in the MC templates shifts the data polarization fit by the assessed systematic uncertainty of $\sigma=0.0004$.\\
\subsubsection{Momentum calibration and resolution} \label{sec:p_scale}
The same selection of $\epem\ra\mumu$ events is used to correct and quantify the momentum calibration and momentum resolution of the charged particles. This is done by first fitting the $p_{CM}/p_{CM}^{\textrm{Max}}$ distribution with a Crystal Ball function~\cite{Skwarnicki}, where $p_{CM}^{\textrm{Max}}$ is the beam constrained maximum muon momentum ($\sqrt{s}/2-2m_\mu$). 
The fit is performed on both data and MC for each run and a scaling factor, $S_p$, and resolution factor, $R_p$, are extracted. 
$S_p$ is the ratio of the mean values of the Gaussian components of the fits ($S_p\equiv\overline{\mu}_{\textrm{data}}/\overline{\mu}_{\textrm{MC}}$), and $R_p$ is similarly the ratio of the widths ($R_p\equiv\sigma_{\textrm{data}}/\sigma_{\textrm{MC}}$). 
From these two factors, the momentum is corrected as:
\begin{equation}
	p^{\textrm{corr}}_{\textrm{recon}}=(p_{\textrm{truth}}-R_{p}(p_{\textrm{truth}}-p_{\textrm{recon}}))S_p,
\end{equation}
where recon and truth refer respectively to MC that has undergone a detector response simulation or not. Typical values of $S_p$ differ from 1 by $\pm$0.1\% and the statistical uncertainties are $\sim$0.01\%. The resolution factor is more significant, $R_p\sim$0.92, and has a statistical uncertainty of 0.1\%. In order to evaluate a systematic uncertainty associated with the correction, the two factors are varied by the respective uncertainties found in the Crystal Ball fit. The shifts in the corrected momentum due to these variations result in a systematic uncertainty of $\sigma=0.0004$ for the momentum calibration, and $\sigma=0.0003$ for the momentum resolution.
\subsubsection{\texorpdfstring{$\tau$}{Tau} direction definition} \label{sec:tau_dir}
In order to evaluate the level of bias in $\cos\theta$ due to the choice of the $\rho$ direction as the estimator, we evaluate the acollinearity between the $\rho$ and tagging lepton direction in data and MC. Adjusting the $\rho$ direction in each event by $\Delta\cos\theta=\pm0.001$, as indicated by the study, results in a systematic uncertainty of $\sigma=0.0003$.
\subsubsection{Angular resolution} \label{sec:theta_res}
The angular resolution in $\theta$ is 0.897 mrad~\cite{nugent}. Varying the angles by this factor and evaluating the effect on the polarization measurement results in a systematic uncertainty of $\sigma=0.0003$.
\subsubsection{Background contributions}\label{sec:backgrounds}
The effects of the background contributions, primarily Bhabha and $\epem\ra\mumu$ events, are evaluated conservatively by varying the weights of their respective templates in the polarization fit by a factor of 2. This method results in a systematic uncertainty of $\sigma=0.0003$.
\subsubsection{\texorpdfstring{$\tau$}{Tau} branching fraction} \label{tau_bf}
The $\tau$ branching fraction uncertainties are evaluated by varying the weights of the $\tau$ decay templates in the fit. The uncertainties in the world-average branching fractions~\cite{pdg2022} are used, obtaining a systematic uncertainty of $\sigma=0.0002$.
\subsection{Variation of selection value}
\subsubsection{Split-off modeling} \label{sec:split_off}
In order to reduce sensitivity to the modeling of low energy neutrals emitted by charged particles interacting hadronically in the EMC,  all energy depositions in the EMC within 40 cm of the charged particle at the EMC surface are recombined with the charged energy deposition. The distance in MC modeling agrees with the data distribution to within 0.72 cm, so a $\pm$1 cm variation is conservatively used for the systematic study. This results in a systematic uncertainty of $\sigma=0.0011$.
\subsubsection{\texorpdfstring{\piz}{Neutral pion} mass acceptance window} \label{piz_mass}
The systematic uncertainty associated with the 115-155 MeV window for the reconstructed \piz mass is expected to be partially related to the overall photon energy calibration. However, the presence of two photons in the reconstruction also brings in correlations and angular dependencies. At the risk of partially double-counting systematic uncertainties, a separate systematic uncertainty is conservatively assigned to the acceptance window as well. This is done by varying the acceptance window by $\pm$1 MeV, based on the agreement between the average data and MC reconstructed mass. This variation is performed on each side of the acceptance, which results in a systematic uncertainty of $\sigma=0.0008$. 
\subsubsection{\texorpdfstring{$\rho$}{Rho} decay product collinearity}\label{sec:pipiz_angle}
The opening angle between the charged and neutral pion in the $\rho$ decay in the c.m.\ frame is a particularly sensitive variable in this analysis. Removing events with approximately collinear charged pions and $\pi^0$'s by requiring $\cos\alpha>0.9$, improved the data/MC agreement and reduced the fit discrepancies between the separate charged fits. A study of the modeling and the selection threshold was carried out and the systematic uncertainty of $\sigma=0.0007$ was determined by varying the $\cos\alpha>0.9$ requirement by $\pm$0.001. This uncertainty in $\cos\alpha$ was established by studying the difference between the mean of the reconstructed $\rho$ mass in data and MC, which is related to the uncertainty in $\cos\alpha$. This uncertainty in $\cos\alpha$ was further validated by comparing the shifts in data and MC means in the $\cos\alpha$ distribution after the cut was applied.

\subsubsection{Merged \texorpdfstring{\piz}{neutral pion} likelihood}\label{sec:piz_like}
The merged \piz candidates are associated with a likelihood score. At low likelihood values, a significant amount of \mumu and Bhabha events can mimic the presence of a \piz in the final state. An acceptance value for the likelihood was established at the point where nearly all di-lepton events are excluded. A systematic variation in the acceptance was determined from the level of agreement in data/MC and this variation results in a systematic uncertainty of $\sigma=0.0007$.
\subsubsection{Track event transverse momentum} \label{sec:track_pt}
The transverse momentum associated with the charged particle tracks is closely associated with the overall momentum scale and resolution factors. However at low values of $p_T$, Bhabha and unmodeled two-photon final-states can contaminate the data set. Based on the comparison of data and MC, the 350 MeV minimum energy selection criteria was varied by $\pm$2 MeV. This results in a systematic uncertainty of $\sigma=0.0008$. 
\subsubsection{Total transverse momentum} \label{sec:tot_pt}
An additional systematic uncertainty is required for the total $p_T$ to account for any unmodeled effects contributing to the data set. For the total $p_T$, a $\pm$1 MeV variation was used to estimate a $\sigma=0.0003$ systematic uncertainty.
\subsubsection{Maximum calorimeter response} \label{sec:e_max}
The requirement for events to deposit less than 10 GeV in the EMC removes about half of the remaining Bhabha backgrounds. A $\pm$8 MeV variation is used to assess the systematic uncertainty of $\sigma=0.0003$. The $\pm$8 MeV variation is determined from the level of data and MC agreement in the means of the EMC energy distributions for events exceeding 10 GeV.
\subsubsection{\texorpdfstring{$\rho$}{Rho} mass acceptance} \label{sec:rho_mass}
The requirement for the reconstructed $\rho$ mass to exceed 300 MeV is needed to ensure $\cos\psi$ remains physical. The level of data to MC agreement in the mass distribution shows agreement at a $\pm$2 MeV level. Varying the selection by this amount results in a systematic uncertainty of $\sigma=0.0003$.
\subsubsection{\texorpdfstring{$\cos\theta^*$ and $\cos\psi$}{Angular} acceptance} \label{sec:cts_ctp_angles}
The acceptance for $\cos\theta^\star$ and $\cos\psi$ is constrained in order to remove Bhabha events. As the Bhabha distributions are not well modeled a variation on the selection value is used to evaluate the systematic uncertainty. 
MC comparisons with data found variations of $\pm$0.002 and $\pm$0.01 respectively in the level of agreement. Performing the polarization fits with these variations yields systematic uncertainties of: $\sigma_{\theta^*}=0.0002, \sigma_{\psi}=0.0002$. 
\subsection{Lepton identification} \label{sec:pid}
The uncertainty associated with the different criteria in the lepton identification procedures was evaluated by switching between \babar~predefined selection algorithms. For both the muon and electron selection algorithms, the fit response was evaluated with the use of lepton selectors with more stringent requirements for classifying particles as leptons. This reduces the selection efficiency by $\sim$5\% for the muons, and $\sim$1\% for the electrons. Systematic uncertainties of $\sigma=0.0012$ and $\sigma=0.0005$ are assigned for the muon and electron identification respectively. This approach was limited by the statistical uncertainties associated with the change in selection efficiency rather than a systematic bias in the polarization fit. 
\subsection{Other effects}
In addition to the primary systematic sources, the efficiency of the $\tau$ trigger decision, luminosity weightings, particle quality definitions, and effects of histogram rebinning are all evaluated. All of these effects are negligible compared to the uncertainties already discussed.
\subsection{Total systematic uncertainty}
The total systematic uncertainty in the polarization measurement is found by summing the uncertainties in quadrature and results in $\sigma_{\textrm{sys}}=0.0029$. This result, and the breakdown of the uncertainties across all runs is presented in Table \ref{tab:systematicSummary}.
\begin{table*}[ht]
	\centering
    \caption{Summary of systematic uncertainties associated with the Tau Polarimetry polarization measurement. The combined column accounts for correlations between runs in the combination.}
	\label{tab:systematicSummary}
	\begin{tabular}{lrrrrrrr} \toprule\toprule
Source	&	\hspace{8pt} Run 1	&	Run 2	&	Run 3	&	Run 4	&	Run 5	&	Run 6	&		\hspace{5pt} Combined		\\	\midrule
$\piz$ efficiency~(\ref{sec:piz_eff})	&	0.0025	&	0.0016	&	0.0013	&	0.0018	&	0.0006	&	0.0017	&	\textbf{	0.0013	}	\\	
Muon PID~(\ref{sec:pid})	&	0.0018	&	0.0018	&	0.0029	&	0.0011	&	0.0006	&	0.0016	&	\textbf{	0.0012	}	\\	
Split-off modeling~(\ref{sec:split_off})	&	0.0015	&	0.0017	&	0.0016	&	0.0006	&	0.0016	&	0.0020	&	\textbf{	0.0011	}	\\	
Neutral energy calibration~(\ref{sec:neu_scale})	&	0.0027	&	0.0012	&	0.0023	&	0.0009	&	0.0014	&	0.0008	&	\textbf{	0.0010	}	\\	
$\piz$ mass~(\ref{piz_mass})	&	0.0018	&	0.0028	&	0.0010	&	0.0005	&	0.0004	&	0.0004	&	\textbf{	0.0008	}	\\	
$\cos\alpha$~(\ref{sec:pipiz_angle}) &	0.0015	&	0.0009	&	0.0016	&	0.0007	&	0.0005	&	0.0005	&	\textbf{	0.0007	}	\\	
$\piz$ likelihood~(\ref{sec:piz_like})	&	0.0015	&	0.0009	&	0.0015	&	0.0006	&	0.0003	&	0.0010	&	\textbf{	0.0006	}	\\	
Electron PID~(\ref{sec:pid}) &	0.0011	&	0.0020	&	0.0008	&	0.0006	&	0.0005	&	0.0001	&	\textbf{	0.0005	}	\\	
Particle transverse momentum~(\ref{sec:track_pt})	&	0.0012	&	0.0007	&	0.0009	&	0.0002	&	0.0003	&	0.0006	&	\textbf{	0.0004	}	\\	
Boost modeling~(\ref{sec:boost})	&	0.0004	&	0.0019	&	0.0003	&	0.0004	&	0.0004	&	0.0004	&	\textbf{	0.0004	}	\\	
Momentum calibration~(\ref{sec:p_scale})	&	0.0001	&	0.0014	&	0.0005	&	0.0002	&	0.0001	&	0.0003	&	\textbf{	0.0004	}	\\	
Max EMC acceptance~(\ref{sec:e_max})	&	0.0001	&	0.0011	&	0.0008	&	0.0001	&	0.0002	&	0.0005	&	\textbf{	0.0003	}	\\	
$\tau$ direction definition~(\ref{sec:tau_dir})	&	0.0003	&	0.0007	&	0.0008	&	0.0003	&	0.0001	&	0.0004	&	\textbf{	0.0003	}	\\	
Angular resolution~(\ref{sec:theta_res})	&	0.0003	&	0.0008	&	0.0003	&	0.0003	&	0.0002	&	0.0003	&	\textbf{	0.0003	}	\\	
Background modeling~(\ref{sec:backgrounds})	&	0.0005	&	0.0006	&	0.0010	&	0.0002	&	0.0003	&	0.0003	&	\textbf{	0.0003	}	\\	
Event transverse momentum~(\ref{sec:tot_pt})	&	0.0001	&	0.0013	&	0.0005	&	0.0002	&	0.0002	&	0.0004	&	\textbf{	0.0003	}	\\	
Momentum resolution~(\ref{sec:p_scale})	&	0.0001	&	0.0012	&	0.0004	&	0.0002	&	0.0001	&	0.0005	&	\textbf{	0.0003	}	\\	
$\rho$ mass acceptance~(\ref{sec:rho_mass})	&	0.0000	&	0.0011	&	0.0003	&	0.0001	&	0.0002	&	0.0005	&	\textbf{	0.0003	}	\\	
$\tau$ branching fraction~(\ref{tau_bf})	&	0.0001	&	0.0007	&	0.0004	&	0.0002	&	0.0002	&	0.0002	&	\textbf{	0.0002	}	\\	
$\cos\theta^\star$ acceptance~(\ref{sec:cts_ctp_angles})	&	0.0002	&	0.0006	&	0.0004	&	0.0001	&	0.0001	&	0.0004	&	\textbf{	0.0002	}	\\	
$\cos\psi$ acceptance~(\ref{sec:cts_ctp_angles}) &	0.0002	&	0.0003	&	0.0002	&	0.0002	&	0.0002	&	0.0003	&	\textbf{	0.0002	}	\\	\midrule
Total	&	0.0058	&	0.0062	&	0.0054	&	0.0030	&	0.0026	&	0.0038	&	\textbf{	0.0029	}	\\	\bottomrule\bottomrule
	\end{tabular}
	
\end{table*}
\section{Conclusions and Discussion}\label{sec:guide}
Using Tau Polarimetry to measure the average longitudinal polarization relies on the Standard Model, and so the existence of Beyond the Standard Model (BSM) physics could potentially appear as a non-zero value in the final result. At \pep2, no beam polarization is expected so a significant deviation from zero could indicate a BSM bias. Even though the measurement in this analysis is in good agreement with zero polarization a number of potential BSM effects were considered. Those related to the coupling of the electron and $\tau$ polarization through deviations from SM expectations in $g_{V,A}^\ell$ ($\ell=e,\mu,\tau$) are some of the most likely potential sources. The current world-average on $g_{V}^\tau$ suggests BSM effects could contribute a bias on the order of 0.0001, which is negligible for this analysis but could become a small fraction of the uncertainties for future experiments. A more substantial sector where BSM effects could arise is in the $\tau$ Michel parameter measurements, and specifically the chirality, $\xi$, of $\nu_\tau$. In the SM $\xi=1$ and has been experimentally constrained to $\xi=0.985\pm0.030$~\cite{pdg2022}. Any deviations from 1 in this parameter directly bias the average beam polarization measurement, and Tau Polarimetry would benefit from an improved measurement of $\xi$. 

While this analysis assumes the polarization is only present in the $e^-$ beam, Tau Polarimetry measures the average polarization of the mediator in the \epem collision. In the case that both beams are polarized, Tau Polarimetry cannot disentangle the individual beam polarizations without a secondary measurement of the $\epem\rightarrow\tautau$ cross-section.

In the development of this analysis, a number of key features were identified from which any future deployment of Tau Polarimetry at other $e^+e^-$ colliders would benefit. One of them is the systematic cancellation obtained from combining the results of the fits from the two electric charges. This is due to the effects of beam polarization on the kinematic observables being inverted with the sign of $\cos\theta$, or equivalently the electric charge. This means that any non-polarization sensitive biases will affect positively and negatively charged signals in opposite ways, and the biases will largely cancel out when averaged. Therefore, a large discrepancy between the polarization fits of the separate charges can indicate an uncontrolled source of bias. 

A major source of systematic uncertainty is related to the MC modeling of photon and \piz processes. Modeling issues were observed in three related variables: the angular separation of the final-state charged and neutral pions, the overall neutral pion efficiency, and the modeling of the calorimeter response to neutral particles in close proximity to charged particles. These potential sources of systematic uncertainties could be significantly reduced by the choice of a final state without a neutral pion, such as the \taupi~decay. However, we found that the dependence on PID modeling as well as the increased dilepton backgrounds introduce additional biases.

In a SuperKEKB upgraded with electron beam polarization, Belle II will benefit from having an existing unpolarized data set to compare the performance of Tau Polarimetry with and without polarization. Assuming the beam polarization is flipped in a controlled manner, Belle II will also be able to demonstrate the performance of Tau Polarimetry on arbitrary beam polarizations by using sub-sets of the polarized data. This should be considered as a necessary step in verifying the performance of Tau Polarimetry at non-zero polarizations.

The average longitudinal polarization of \pep2 has been measured to be $\langle P\rangle=\polresult \pm \polstat_{\textrm{stat}}\pm \polsys_{\textrm{sys}}$. This measurement demonstrates that a 0.3\% absolute systematic uncertainty can be achieved on the beam polarization measurement with approximately 500 \ifb~of data. 
\section{Acknowledgments}
We are grateful for the extraordinary contributions of our PEP-II colleagues in achieving the excellent luminosity and machine conditions that have made this work possible. The success of this project also relies critically on the expertise and dedication of the computing organizations that support \babar, including GridKa, UVic HEP-RC, CC-IN2P3, and CERN. The collaborating institutions wish to thank SLAC for its support and the kind hospitality extended to them. We also wish to acknowledge the important contributions of J.~Dorfan and our deceased colleagues E.~Gabathuler, W.~Innes, D.W.G.S.~Leith, A.~Onuchin, G.~Piredda, and R. F.~Schwitters.\\

\bibliography{pol_paper}
\clearpage
\onecolumngrid
\appendix
\section{Projection of the 3-dimensional event distributions onto each polarization observable} \label{app:plots}

\begin{figure*}[htb]
    \centering
    \begin{tabular}{cc}
        \begin{overpic}[width=0.4\textwidth]{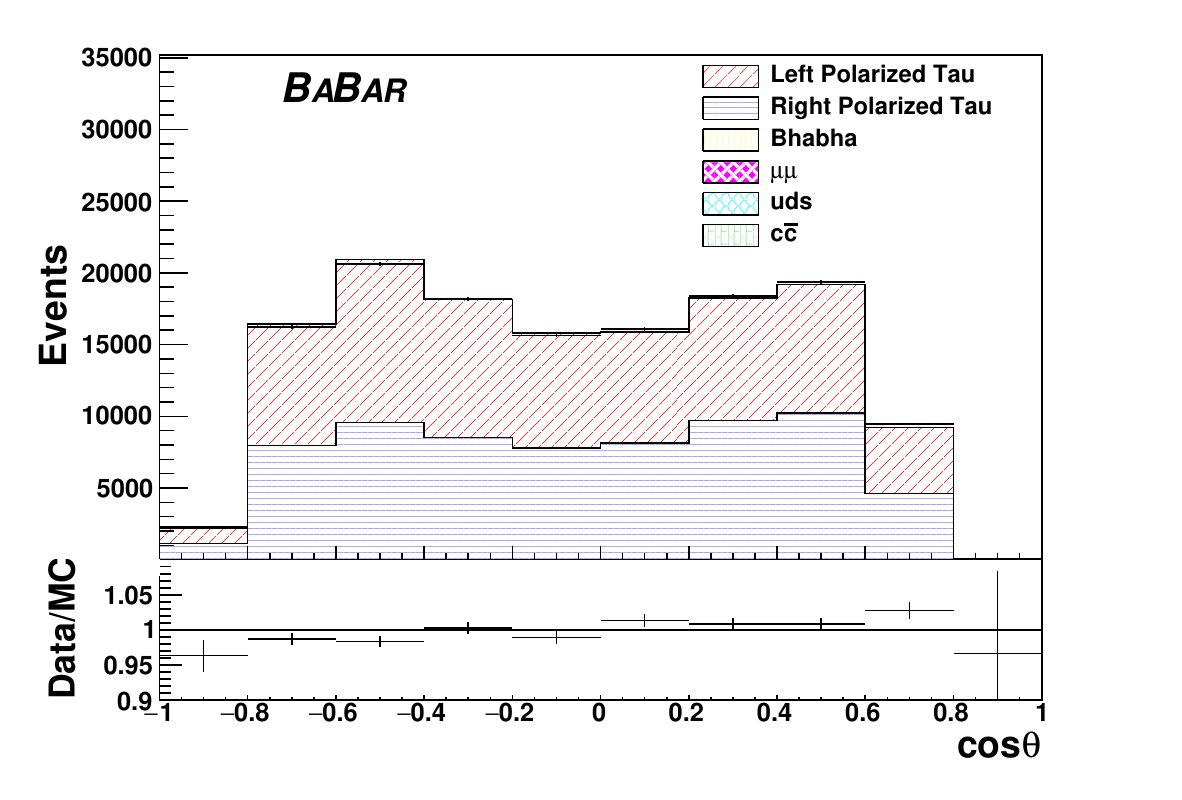}
            \put (20,55) {\textbf{a)}}
            \put (3,47) {\rotatebox{90}{\scriptsize\textbf{/0.2}}}
        \end{overpic} &
        \begin{overpic}[width=0.4\textwidth]{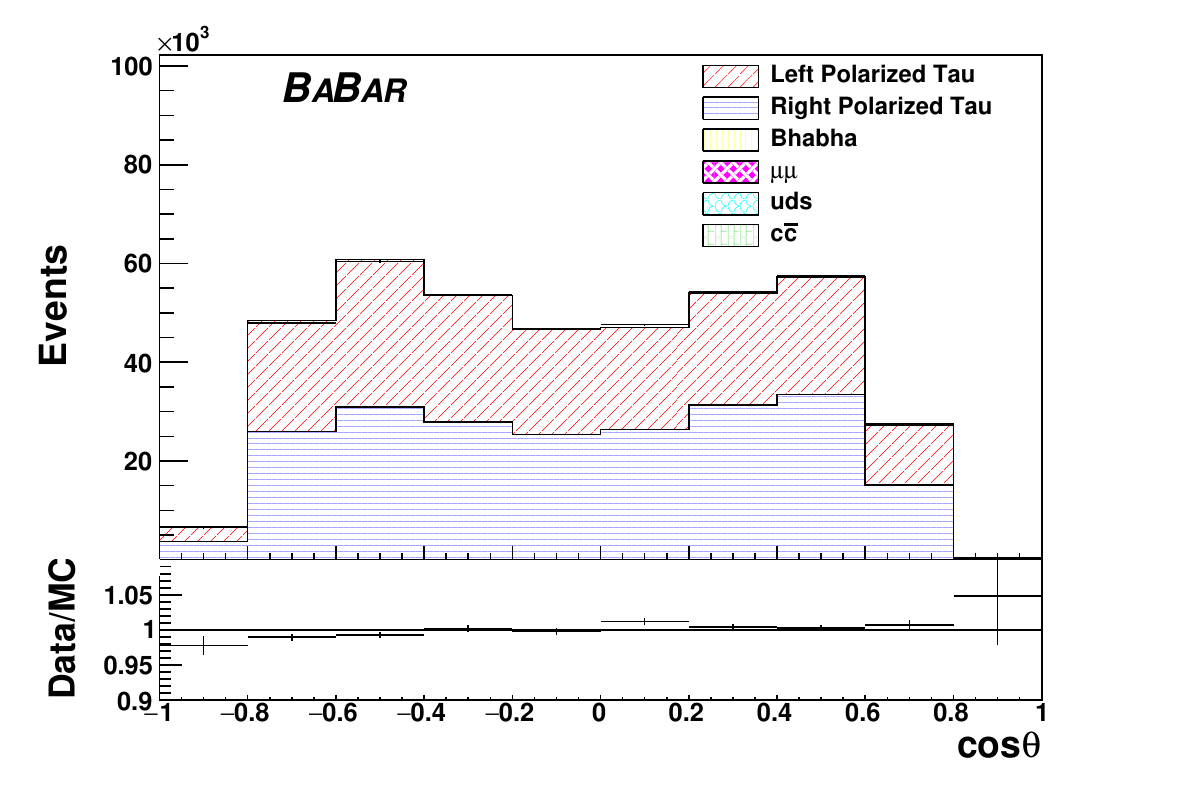}
            \put (20,55) {\textbf{b)}}
            \put (3,47) {\rotatebox{90}{\scriptsize\textbf{/0.2}}}
        \end{overpic} \\
        \begin{overpic}[width=0.4\textwidth]{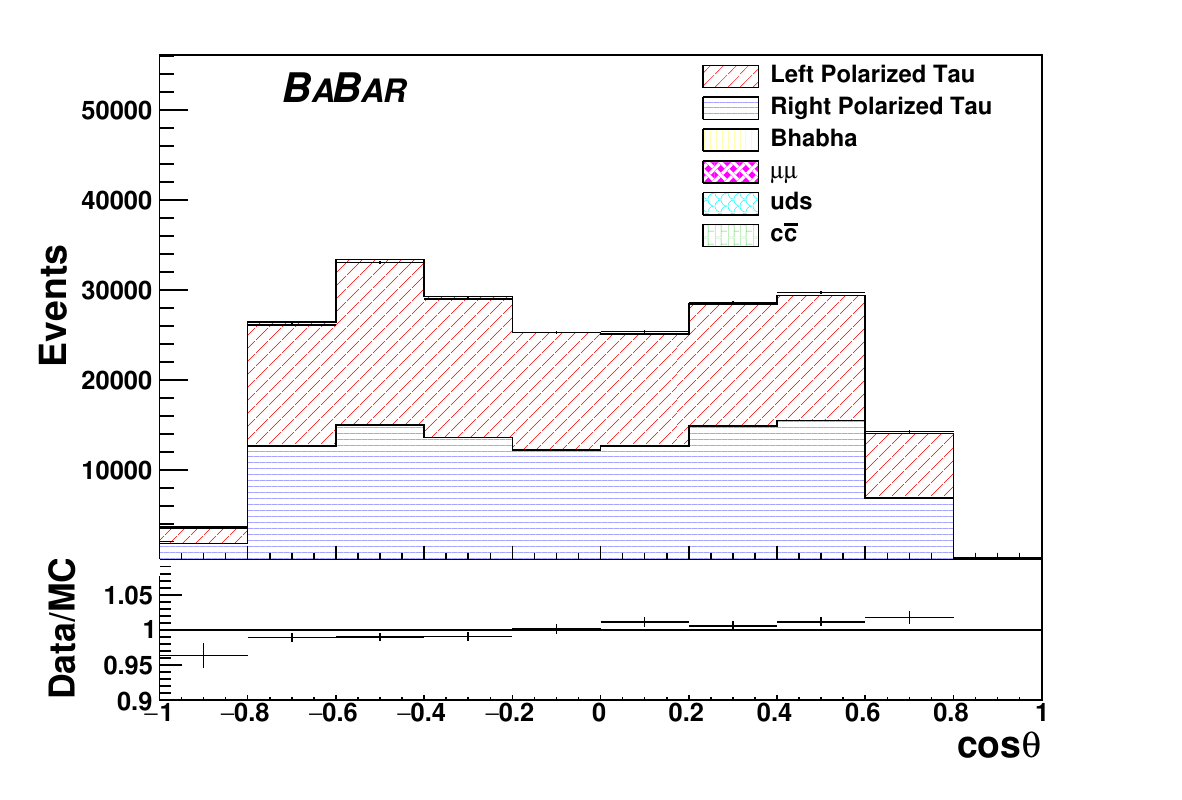}
            \put (20,55) {\textbf{c)}}
            \put (3,47) {\rotatebox{90}{\scriptsize\textbf{/0.2}}}
        \end{overpic} &
        \begin{overpic}[width=0.4\textwidth]{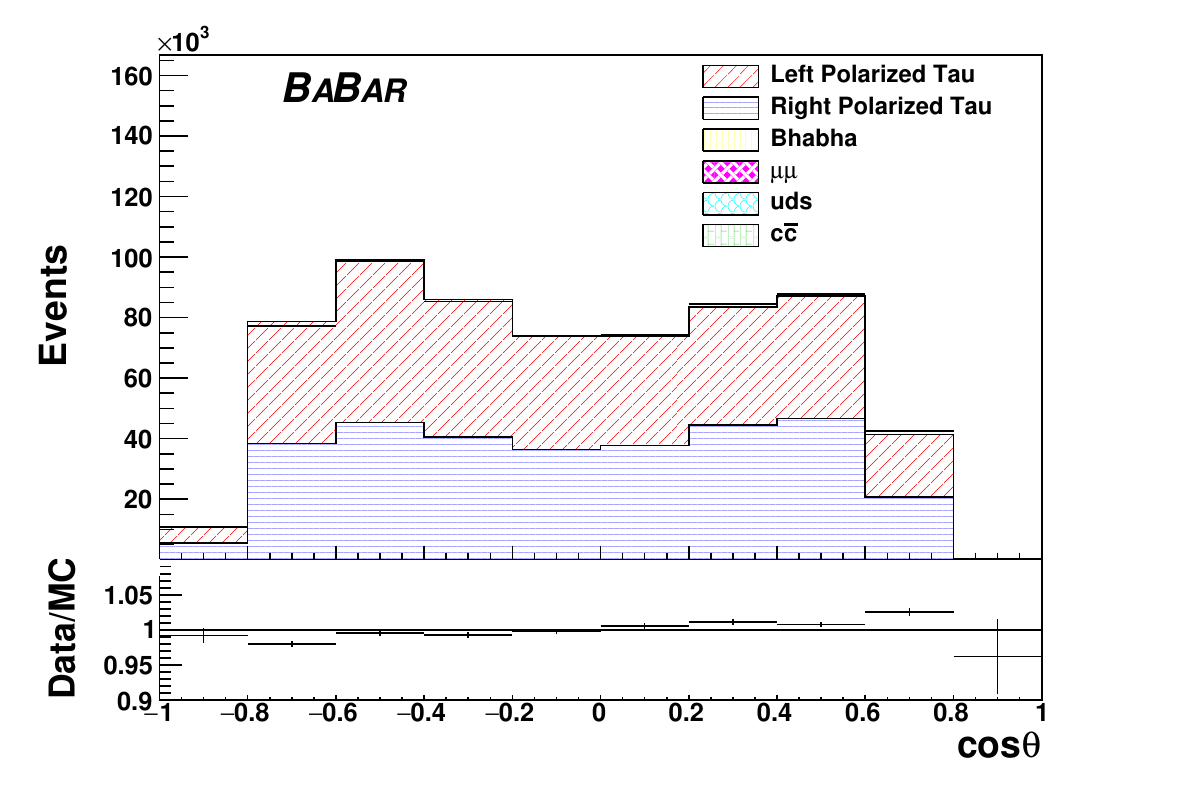}
            \put (20,55) {\textbf{d)}}
            \put (3,47) {\rotatebox{90}{\scriptsize\textbf{/0.2}}}
        \end{overpic} \\
        \begin{overpic}[width=0.4\textwidth]{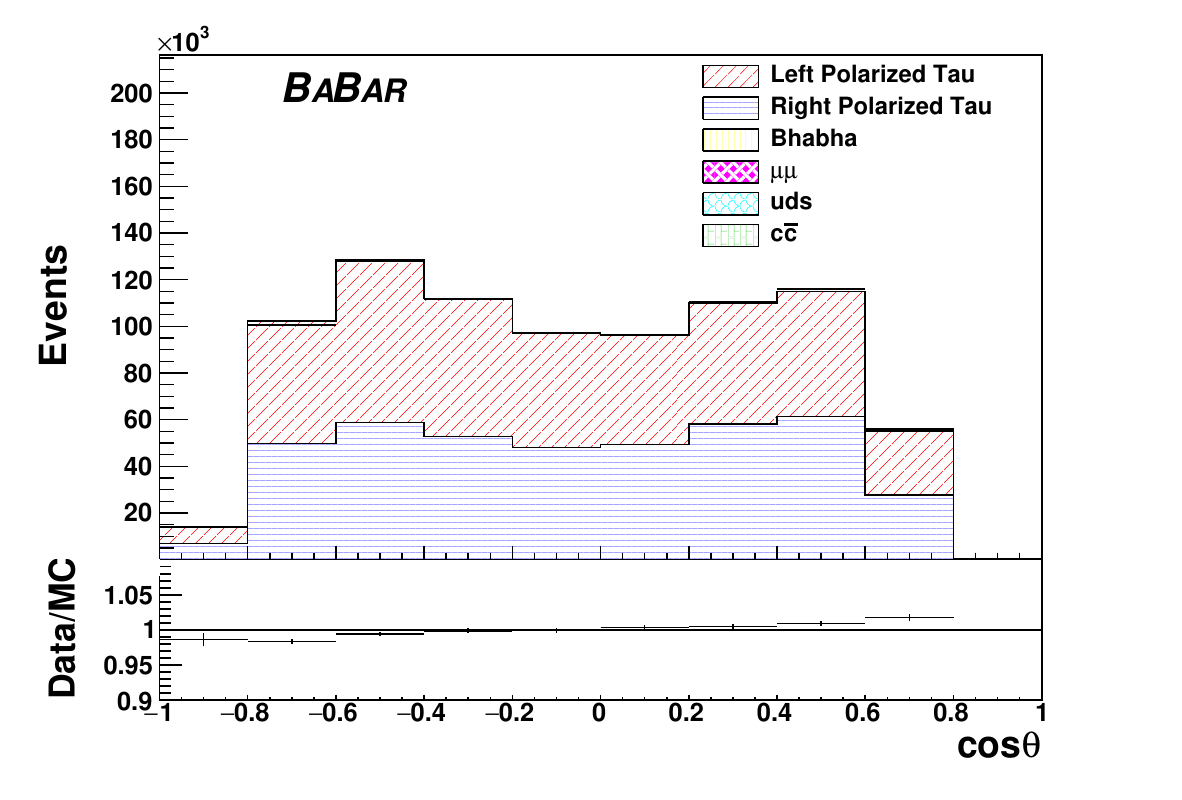}
            \put (20,55) {\textbf{e)}}
            \put (3,47) {\rotatebox{90}{\scriptsize\textbf{/0.2}}}
        \end{overpic} &
        \begin{overpic}[width=0.4\textwidth]{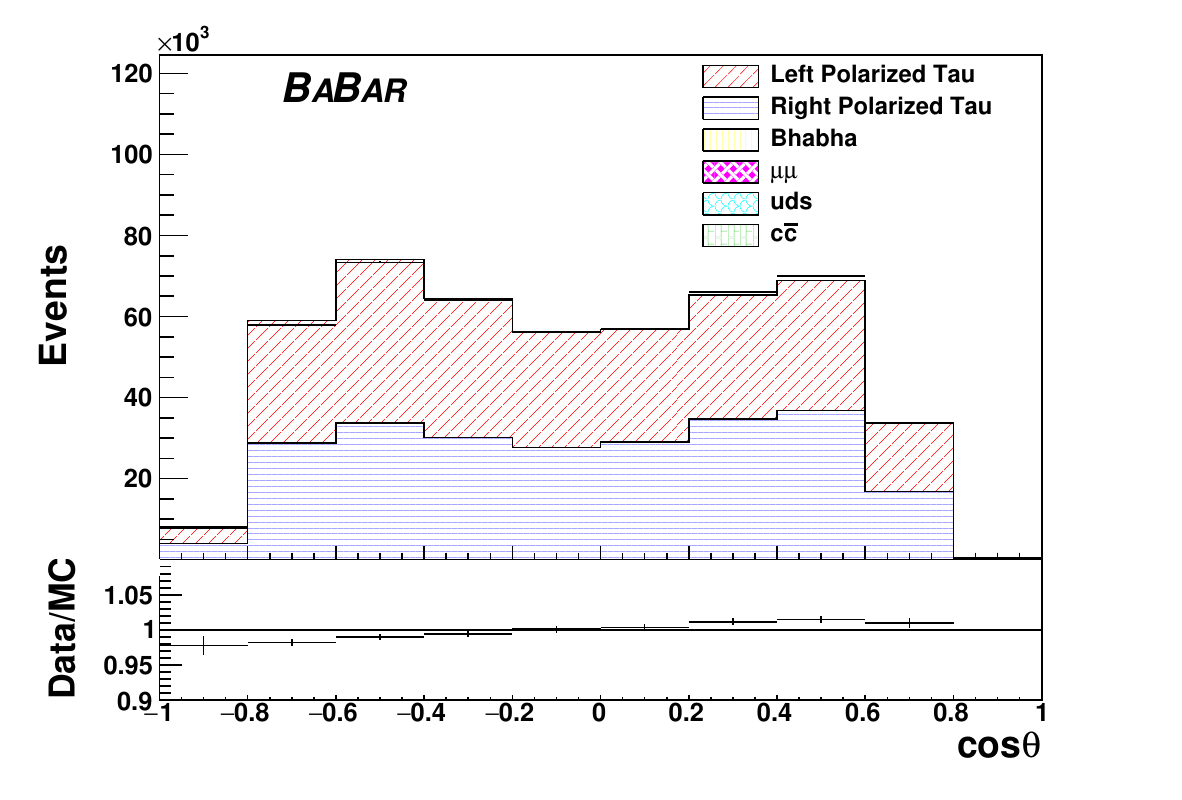}
            \put (20,55) {\textbf{f)}}
            \put (3,47) {\rotatebox{90}{\scriptsize\textbf{/0.2}}}
        \end{overpic} \\
    \end{tabular}
    \caption{One dimensional projection of $\cos\theta$ from $\rho^+$ fits for Runs 1-6, a) through f), respectively.}
    \label{fig:1d_ct_pfits}
\end{figure*}
\begin{figure*}
    \centering
    \begin{tabular}{cc}
        \begin{overpic}[width=0.4\textwidth]{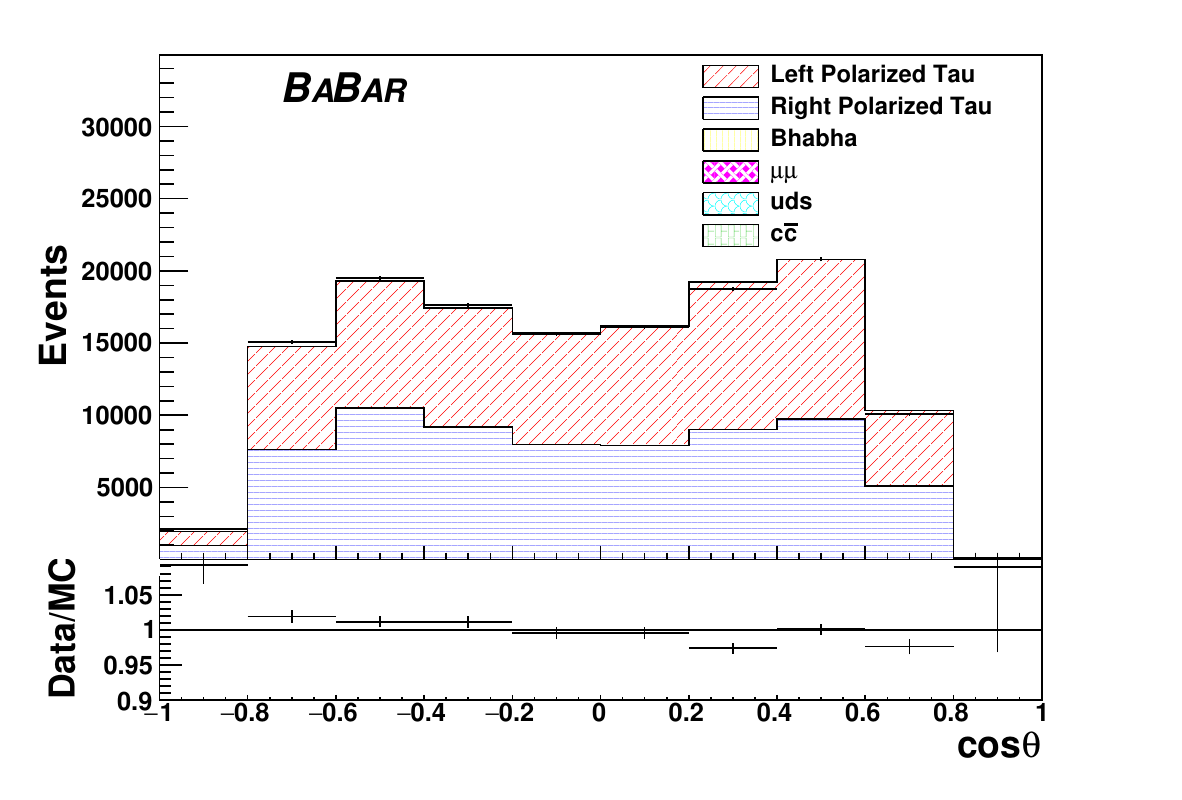}
            \put (20,55) {\textbf{a)}}
            \put (3,47) {\rotatebox{90}{\scriptsize\textbf{/0.2}}}
        \end{overpic} &
        \begin{overpic}[width=0.4\textwidth]{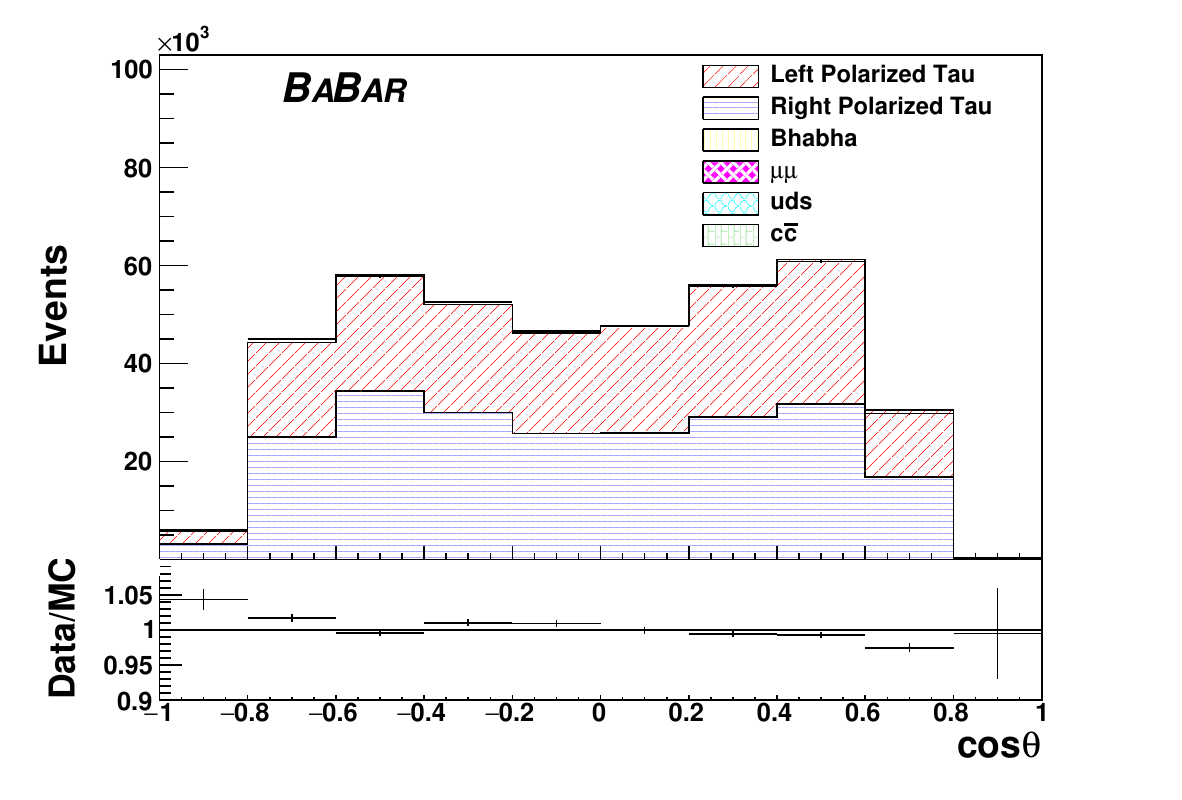}
            \put (20,55) {\textbf{b)}}
            \put (3,47) {\rotatebox{90}{\scriptsize\textbf{/0.2}}}
        \end{overpic} \\
        \begin{overpic}[width=0.4\textwidth]{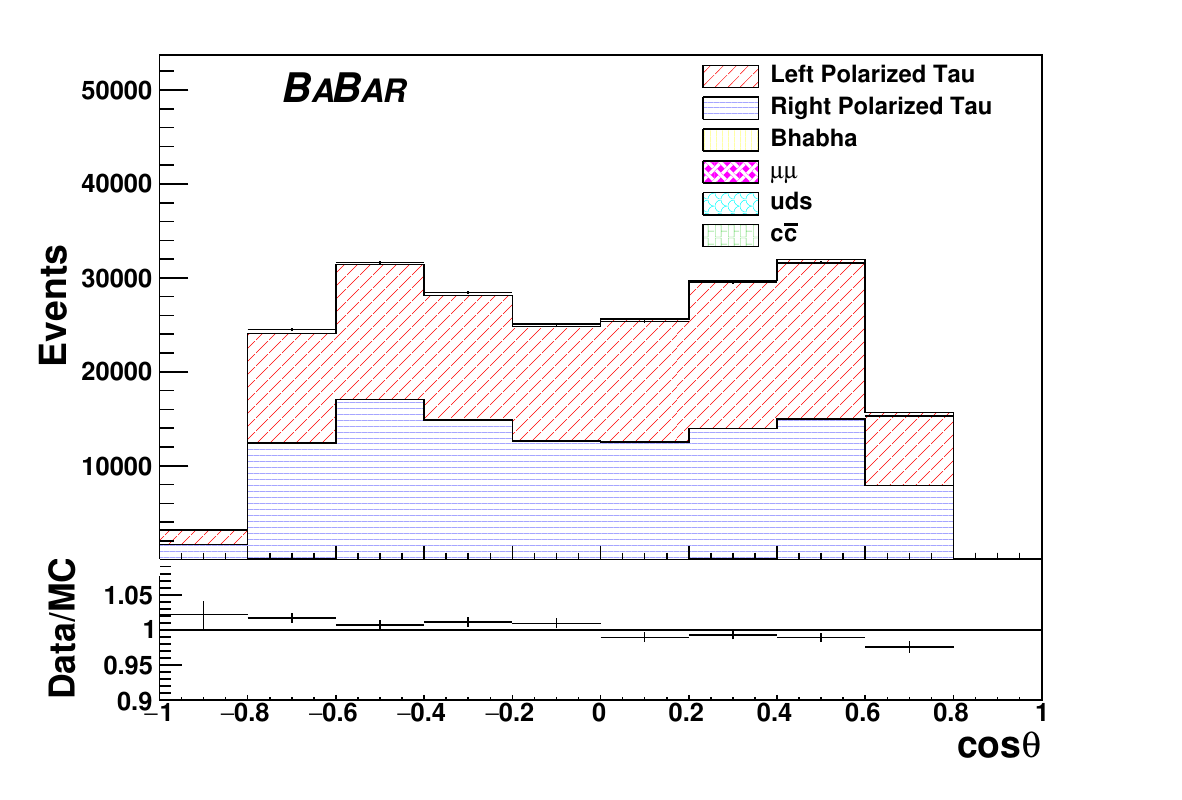}
            \put (20,55) {\textbf{c)}}
            \put (3,47) {\rotatebox{90}{\scriptsize\textbf{/0.2}}}
        \end{overpic} &
        \begin{overpic}[width=0.4\textwidth]{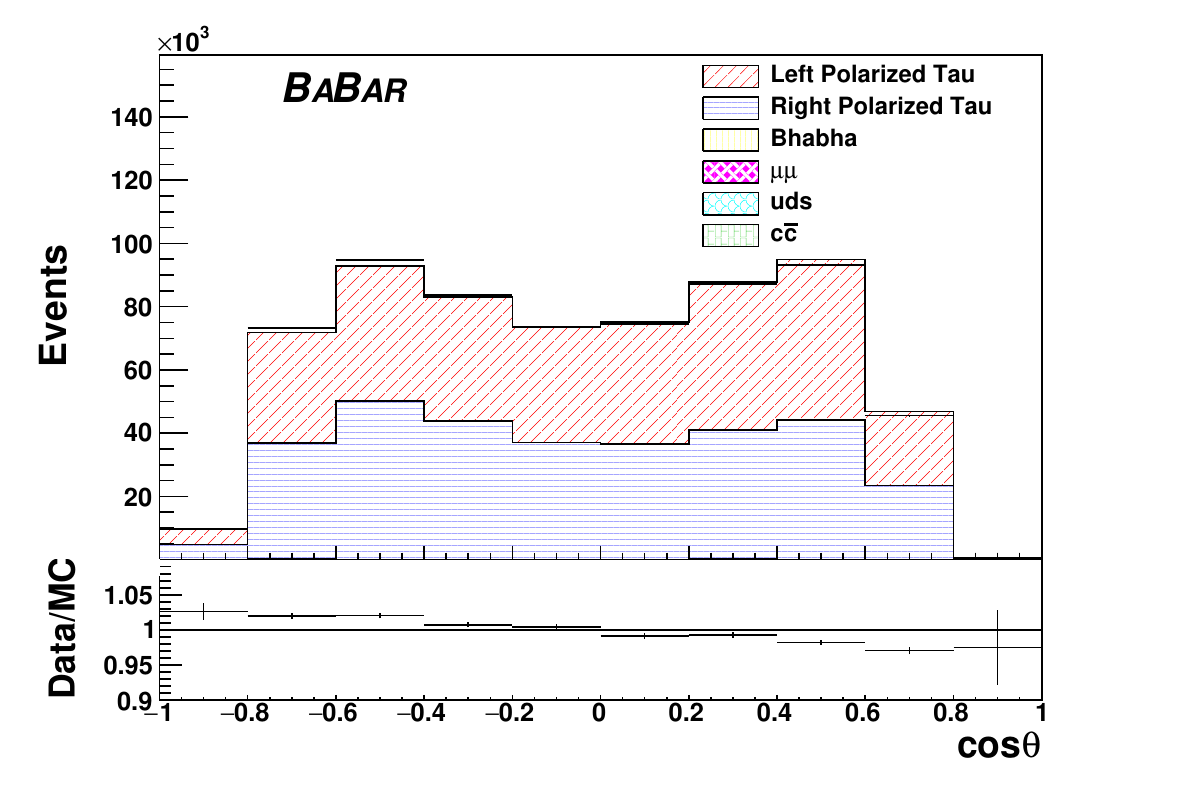}
            \put (20,55) {\textbf{d)}}
            \put (3,47) {\rotatebox{90}{\scriptsize\textbf{/0.2}}}
        \end{overpic} \\
        \begin{overpic}[width=0.4\textwidth]{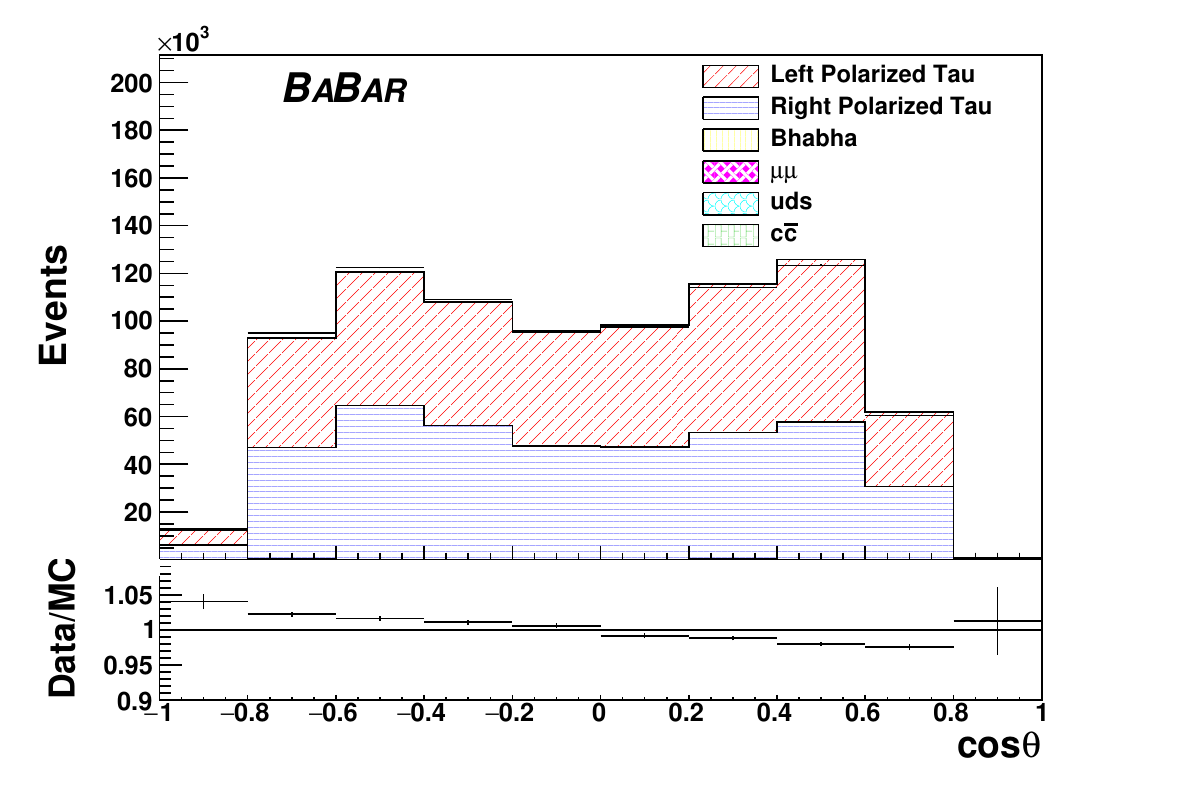}
            \put (20,55) {\textbf{e)}}
            \put (3,47) {\rotatebox{90}{\scriptsize\textbf{/0.2}}}
        \end{overpic} &
        \begin{overpic}[width=0.4\textwidth]{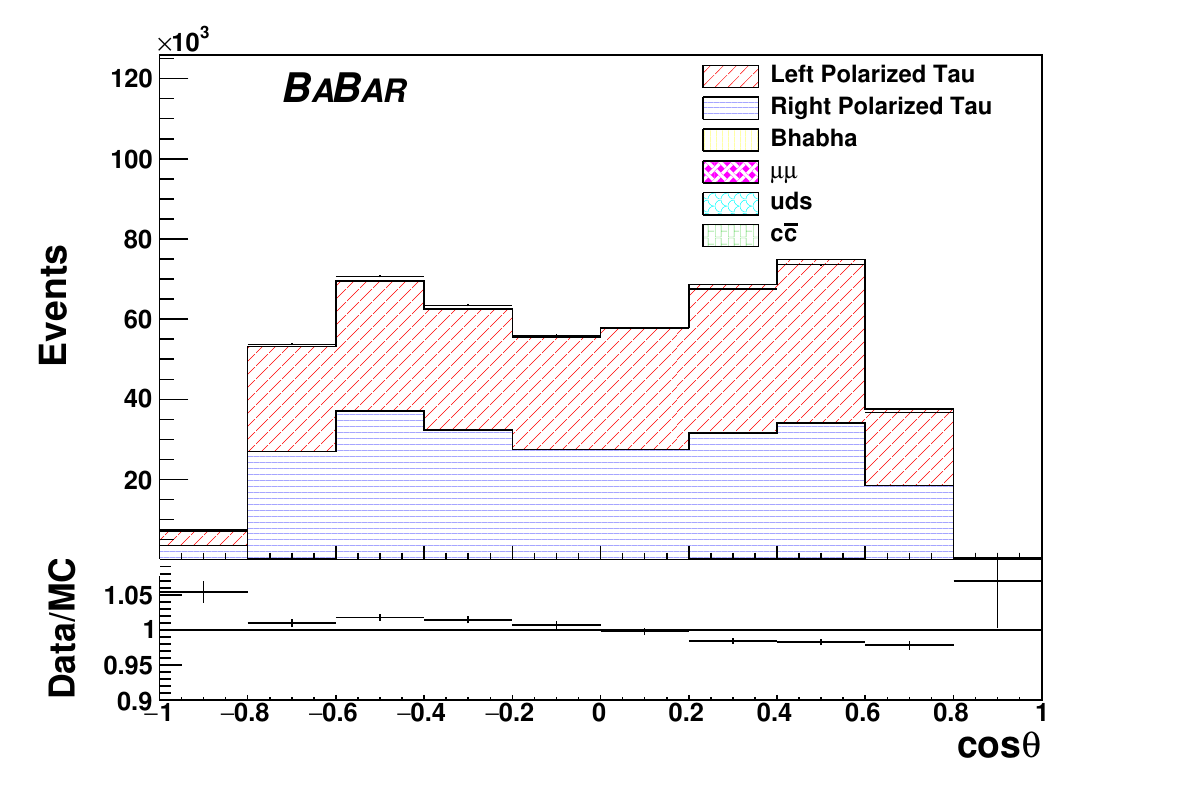}
            \put (20,55) {\textbf{f)}}
            \put (3,47) {\rotatebox{90}{\scriptsize\textbf{/0.2}}}
        \end{overpic} \\
    \end{tabular}
    \caption{One dimensional projection of $\cos\theta$ from $\rho^-$ fits for Runs 1-6, a) through f) respectively.}
    \label{fig:1d_ct_nfits}
\end{figure*}
\begin{figure*}
    \begin{tabular}{cc}
        \begin{overpic}[width=0.4\textwidth]{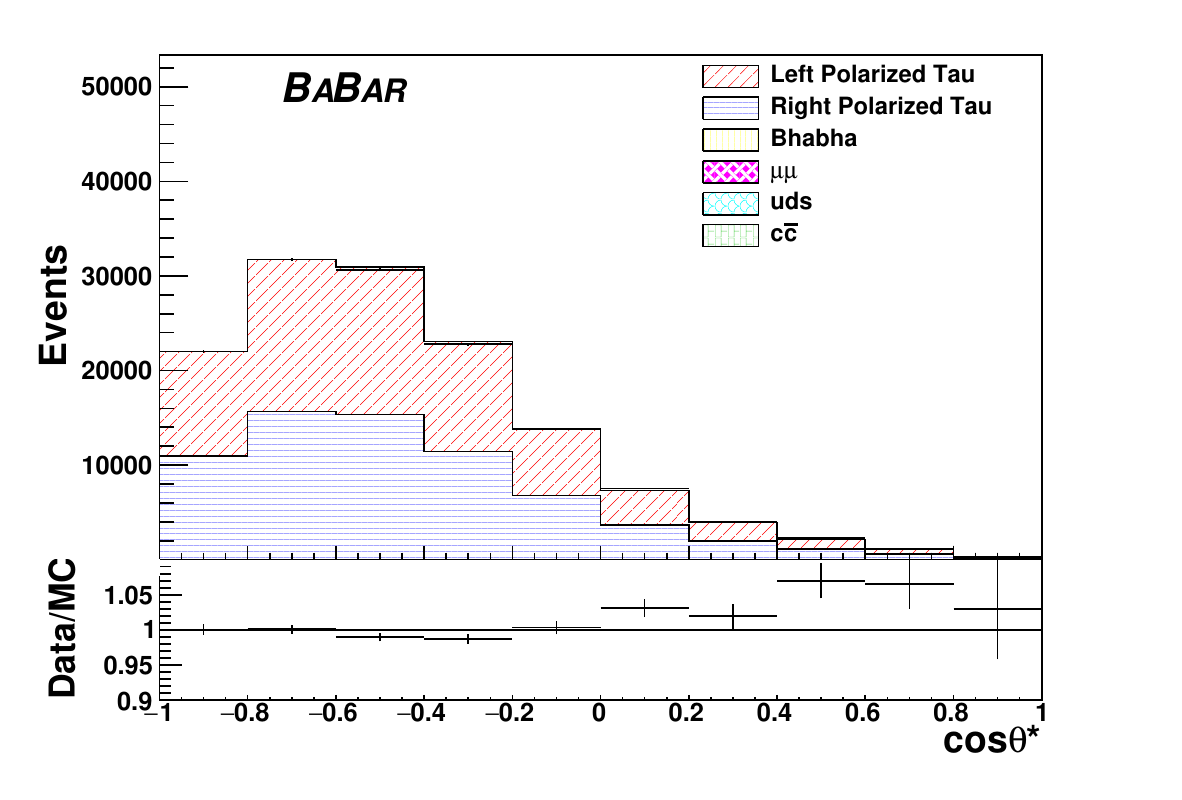}
            \put (20,55) {\textbf{a)}}
            \put (3,47) {\rotatebox{90}{\scriptsize\textbf{/0.2}}}
        \end{overpic} &
        \begin{overpic}[width=0.4\textwidth]{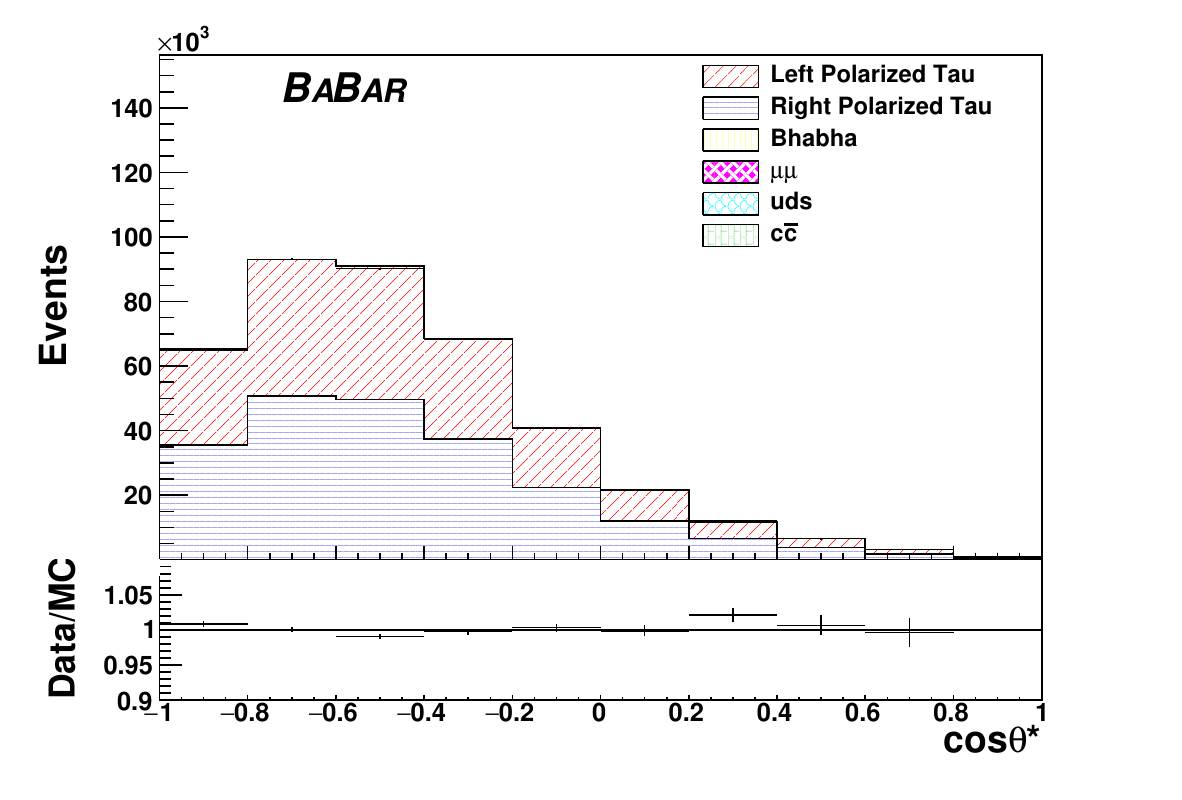}
            \put (20,55) {\textbf{b)}}
            \put (3,47) {\rotatebox{90}{\scriptsize\textbf{/0.2}}}
        \end{overpic} \\
        \begin{overpic}[width=0.4\textwidth]{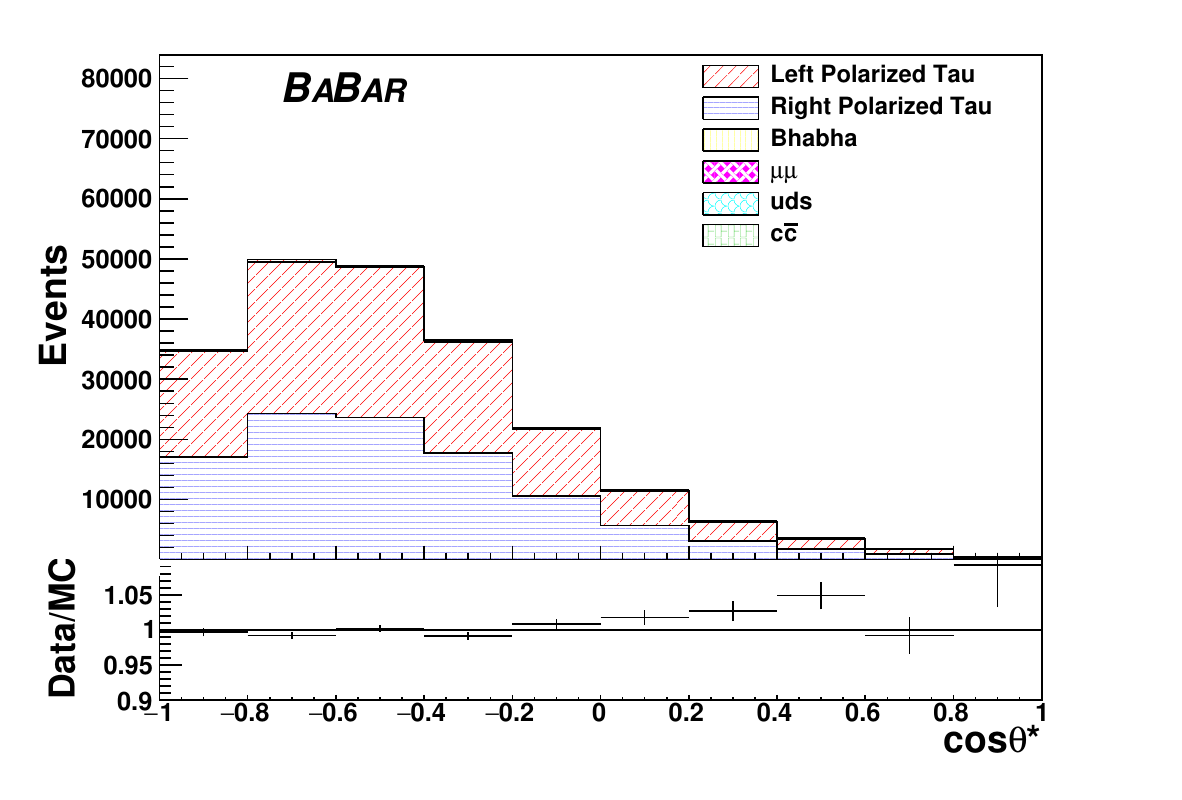}
            \put (20,55) {\textbf{c)}}
            \put (3,47) {\rotatebox{90}{\scriptsize\textbf{/0.2}}}
        \end{overpic} &
        \begin{overpic}[width=0.4\textwidth]{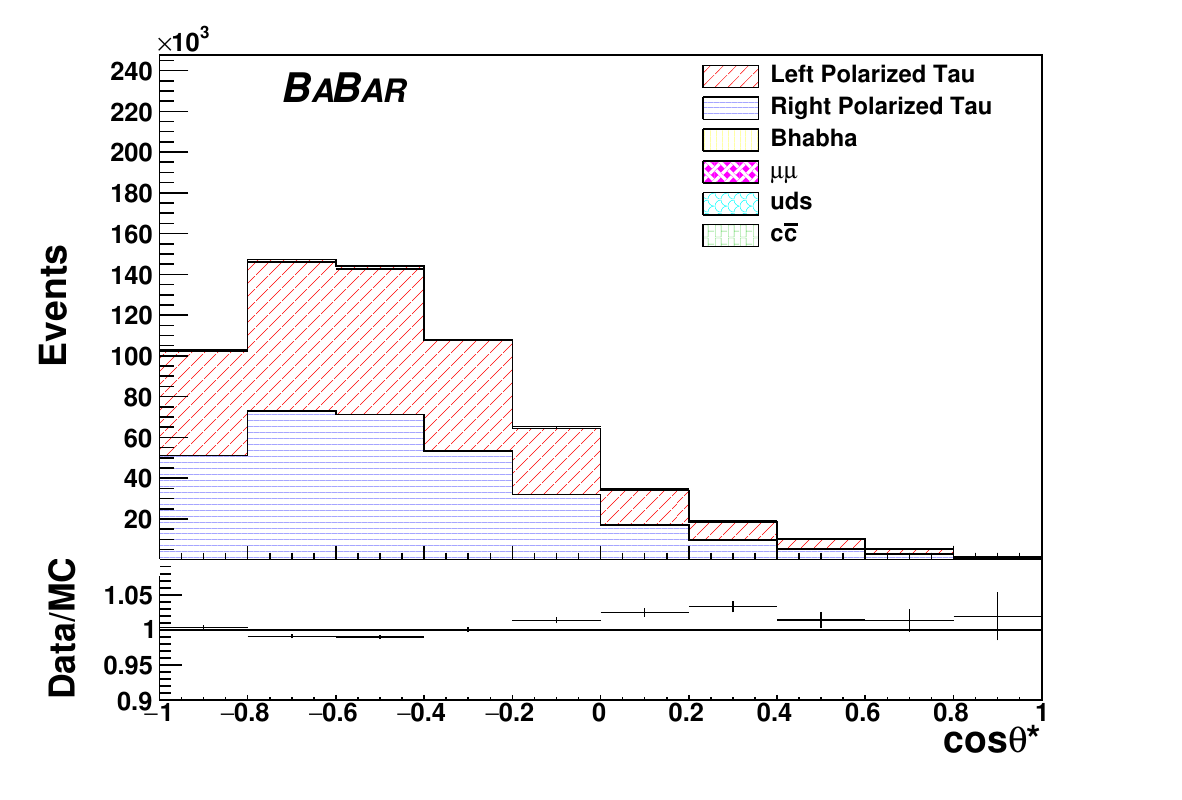}
            \put (20,55) {\textbf{d)}}
            \put (3,47) {\rotatebox{90}{\scriptsize\textbf{/0.2}}}
        \end{overpic} \\
        \begin{overpic}[width=0.4\textwidth]{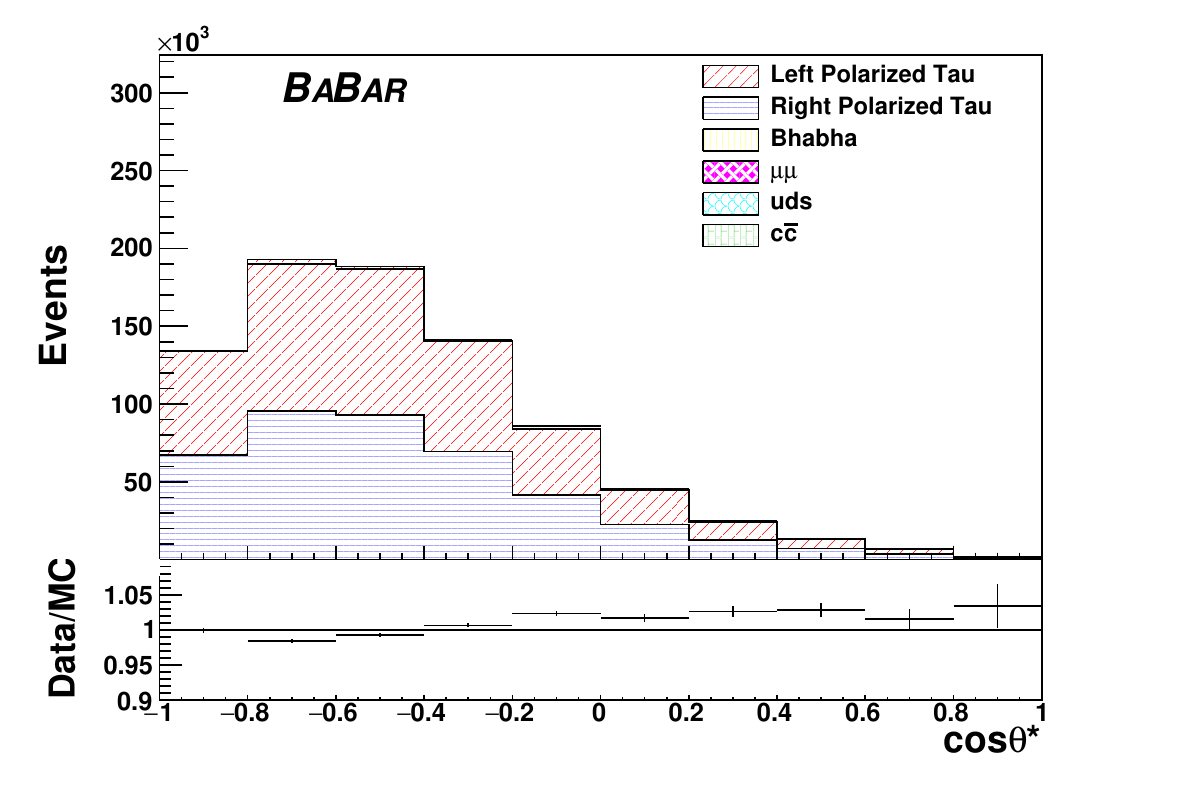}
            \put (20,55) {\textbf{e)}}
            \put (3,47) {\rotatebox{90}{\scriptsize\textbf{/0.2}}}
        \end{overpic} &
        \begin{overpic}[width=0.4\textwidth]{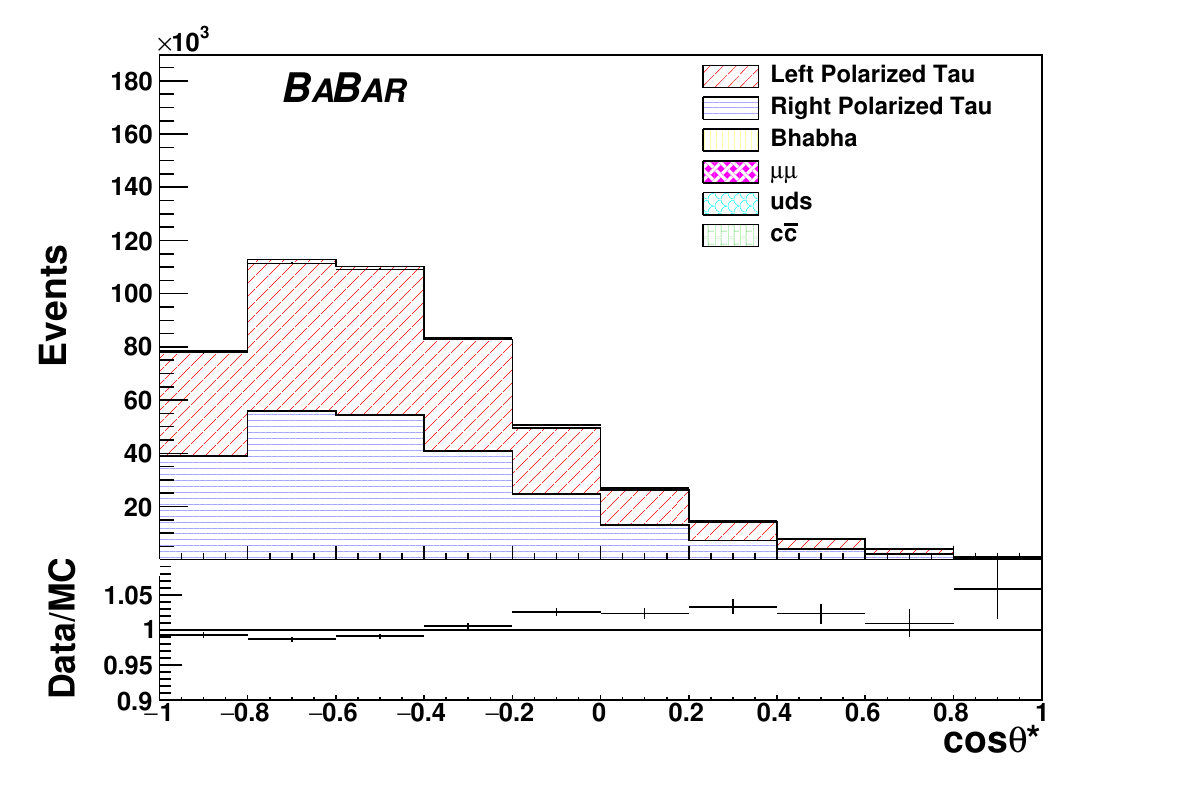}
            \put (20,55) {\textbf{f)}}
            \put (3,47) {\rotatebox{90}{\scriptsize\textbf{/0.2}}}
        \end{overpic} \\
    \end{tabular}
    \caption{One dimensional projection of $\cos\theta^*$ from $\rho^+$ fits for Runs 1-6, a) through f), respectively.}
    \label{fig:1d_zct_pfits}
\end{figure*}
\begin{figure*}
    \begin{tabular}{cc}
        \begin{overpic}[width=0.4\textwidth]{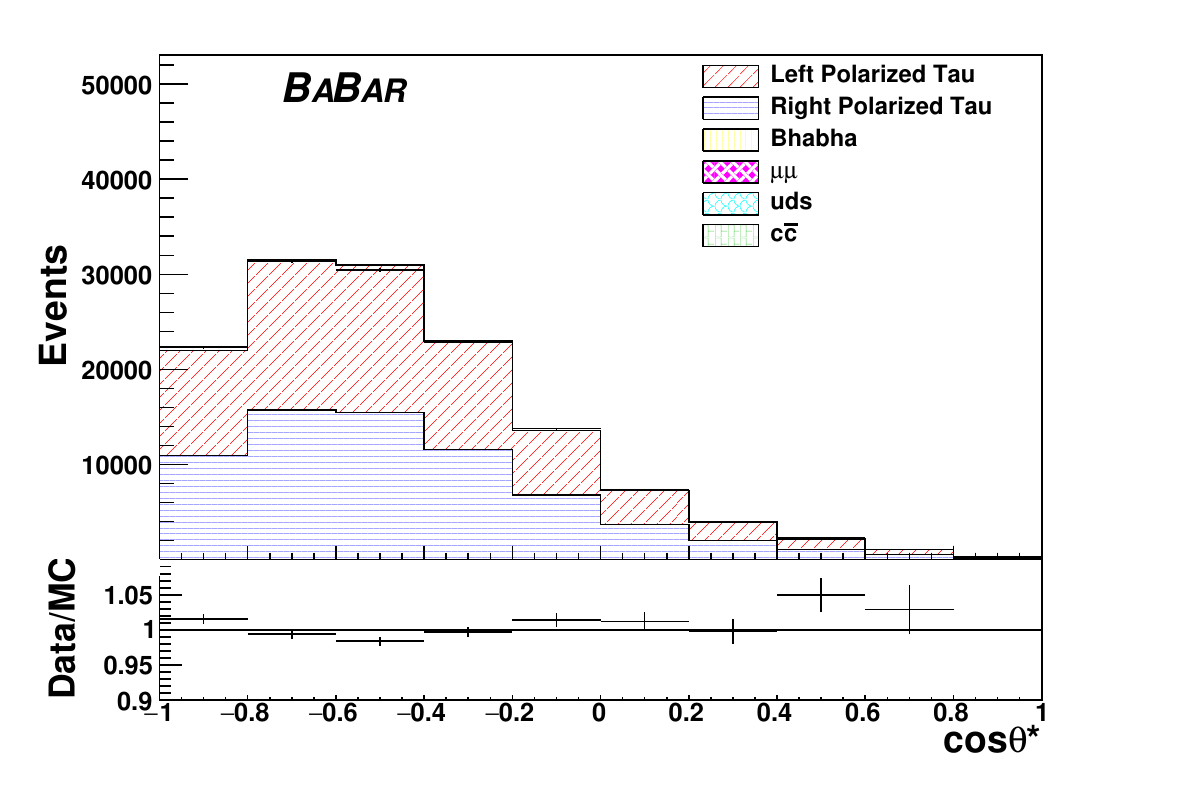}
            \put (20,55) {\textbf{a)}}
            \put (3,47) {\rotatebox{90}{\scriptsize\textbf{/0.2}}}
        \end{overpic} &
        \begin{overpic}[width=0.4\textwidth]{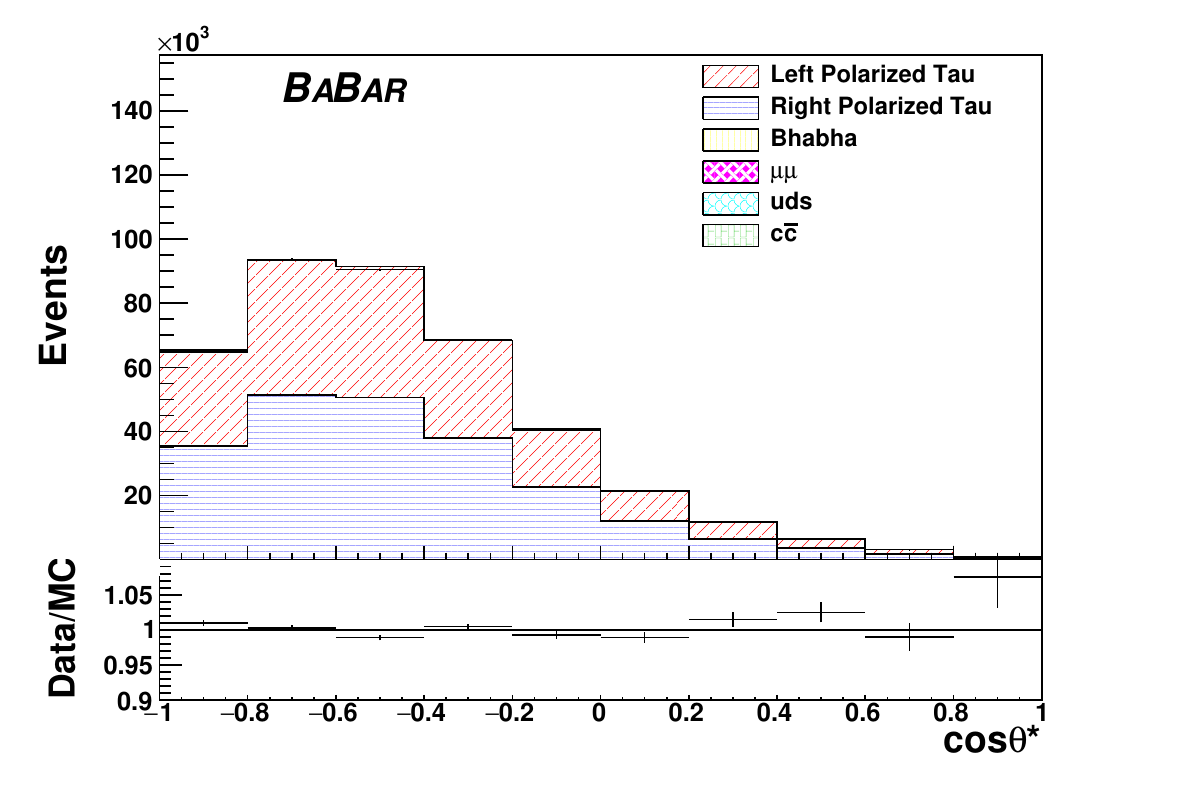}
            \put (20,55) {\textbf{b)}}
            \put (3,47) {\rotatebox{90}{\scriptsize\textbf{/0.2}}}
        \end{overpic} \\
        \begin{overpic}[width=0.4\textwidth]{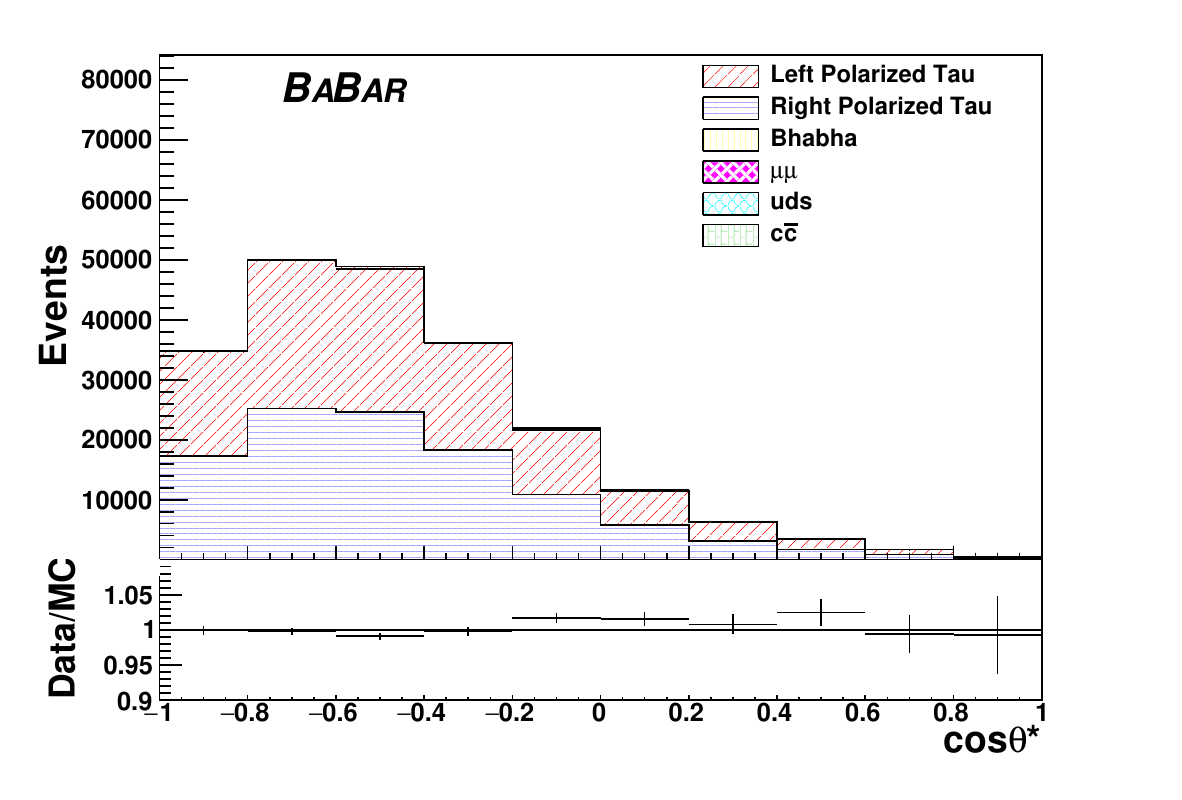}
            \put (20,55) {\textbf{c)}}
            \put (3,47) {\rotatebox{90}{\scriptsize\textbf{/0.2}}}
        \end{overpic} &
        \begin{overpic}[width=0.4\textwidth]{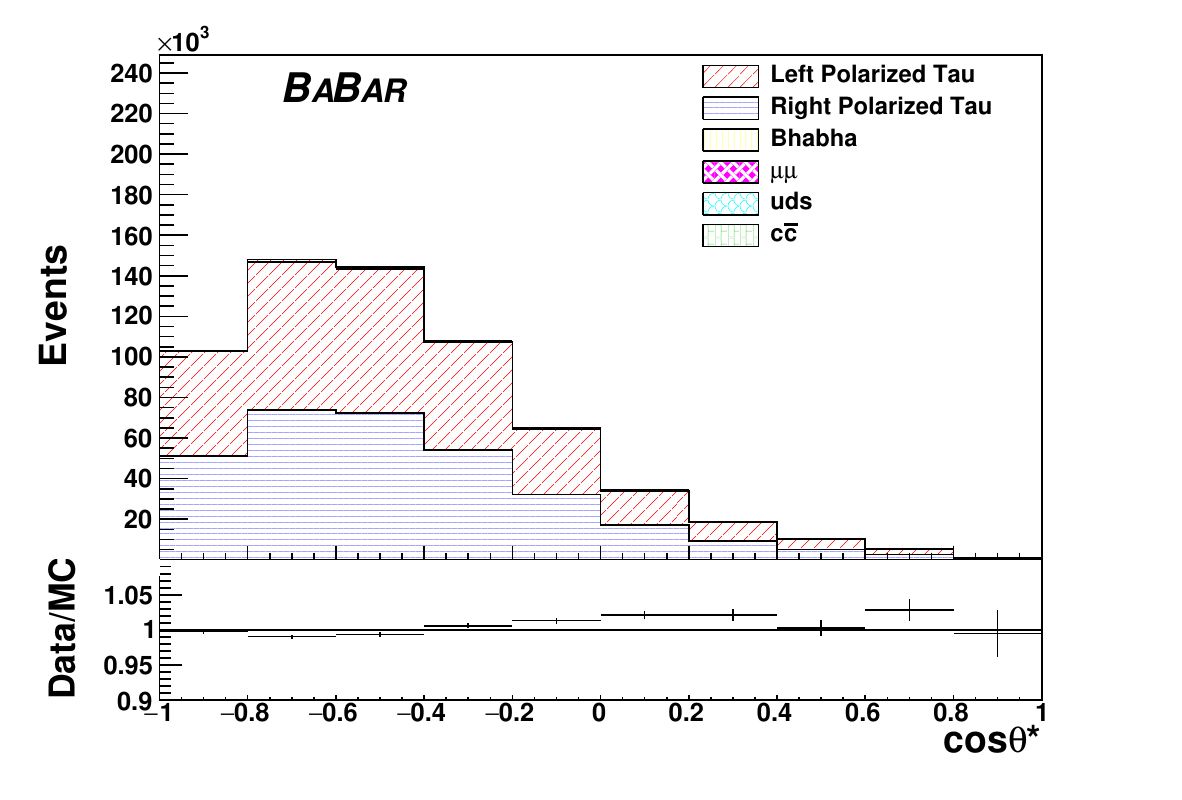}
            \put (20,55) {\textbf{d)}}
            \put (3,47) {\rotatebox{90}{\scriptsize\textbf{/0.2}}}
        \end{overpic} \\
        \begin{overpic}[width=0.4\textwidth]{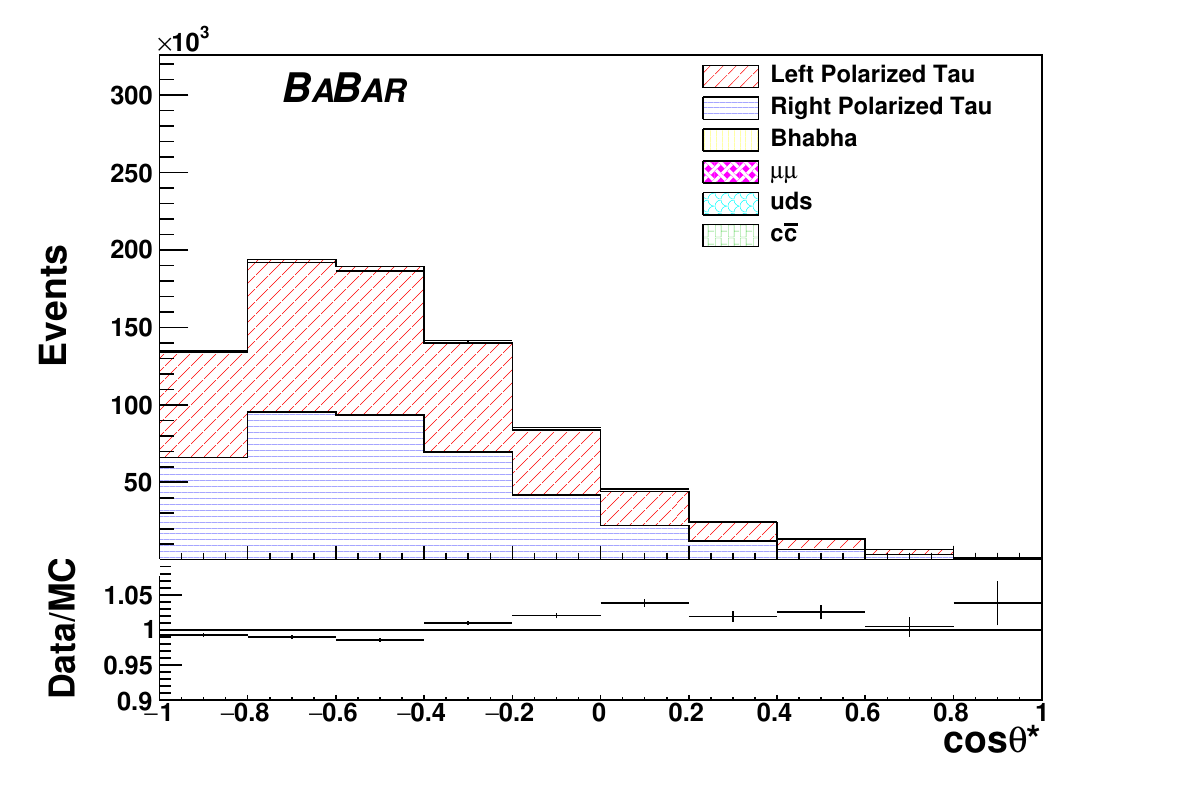}
            \put (20,55) {\textbf{e)}}
            \put (3,47) {\rotatebox{90}{\scriptsize\textbf{/0.2}}}
        \end{overpic} &
        \begin{overpic}[width=0.4\textwidth]{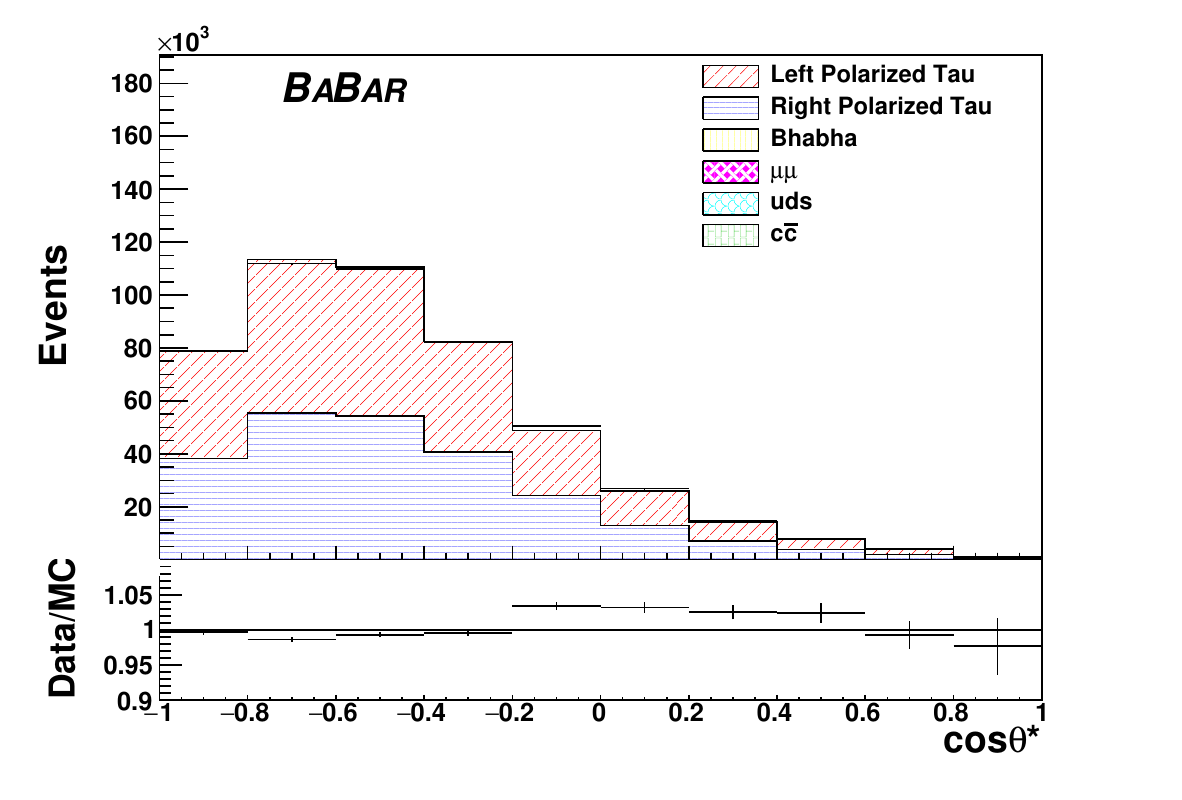}
            \put (20,55) {\textbf{f)}}
            \put (3,47) {\rotatebox{90}{\scriptsize\textbf{/0.2}}}
        \end{overpic} \\
    \end{tabular}
    \caption{One dimensional projection of $\cos\theta^*$ from $\rho^-$ fits for Runs 1-6, a) through f), respectively.}
    \label{fig:1d_zct_nfits}
\end{figure*}
\begin{figure*}
    \begin{tabular}{cc}
        \begin{overpic}[width=0.4\textwidth]{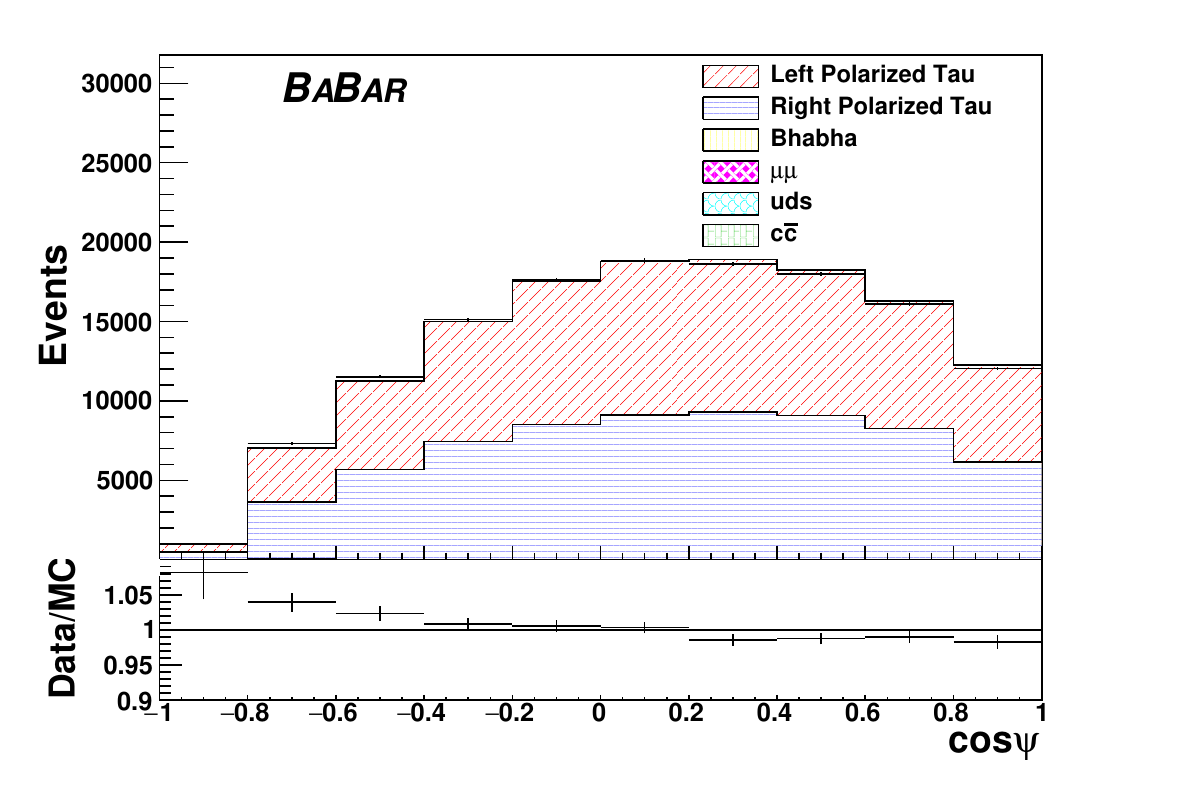}
            \put (20,55) {\textbf{a)}}
            \put (3,47) {\rotatebox{90}{\scriptsize\textbf{/0.2}}}
        \end{overpic} &
        \begin{overpic}[width=0.4\textwidth]{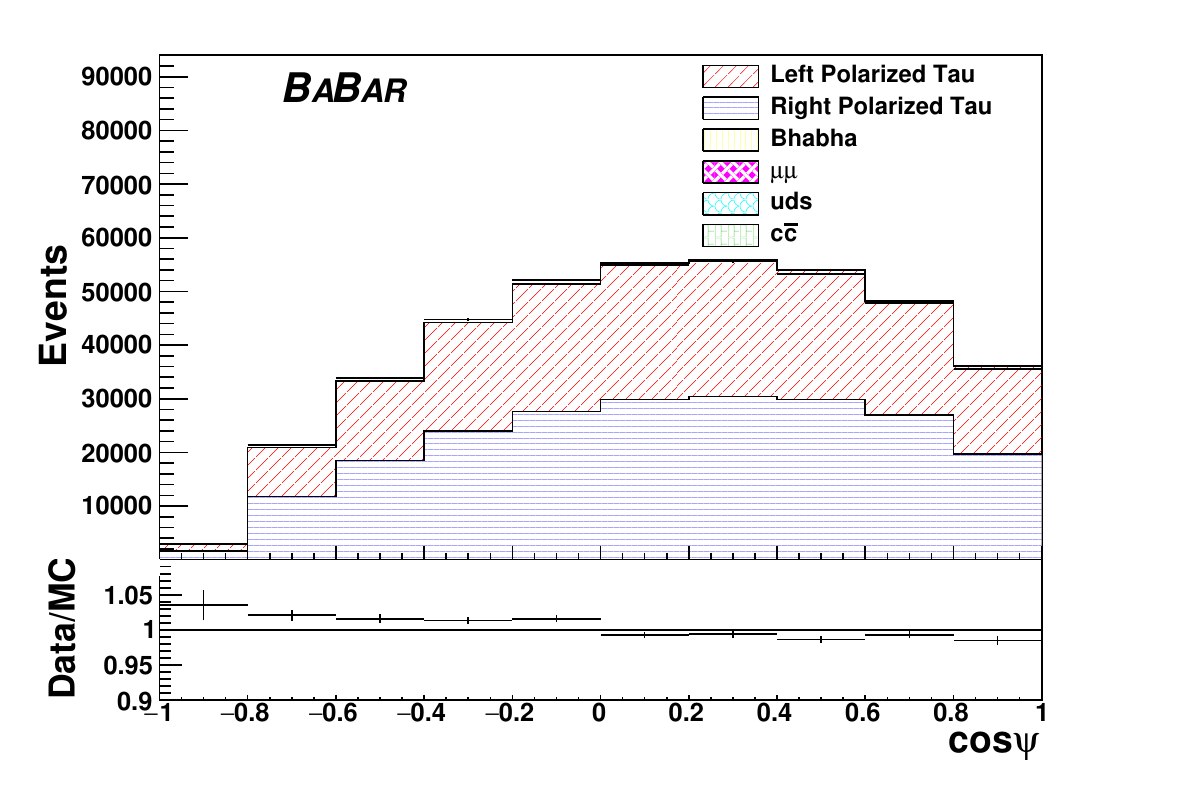}
            \put (20,55) {\textbf{b)}}
            \put (3,47) {\rotatebox{90}{\scriptsize\textbf{/0.2}}}
        \end{overpic} \\
        \begin{overpic}[width=0.4\textwidth]{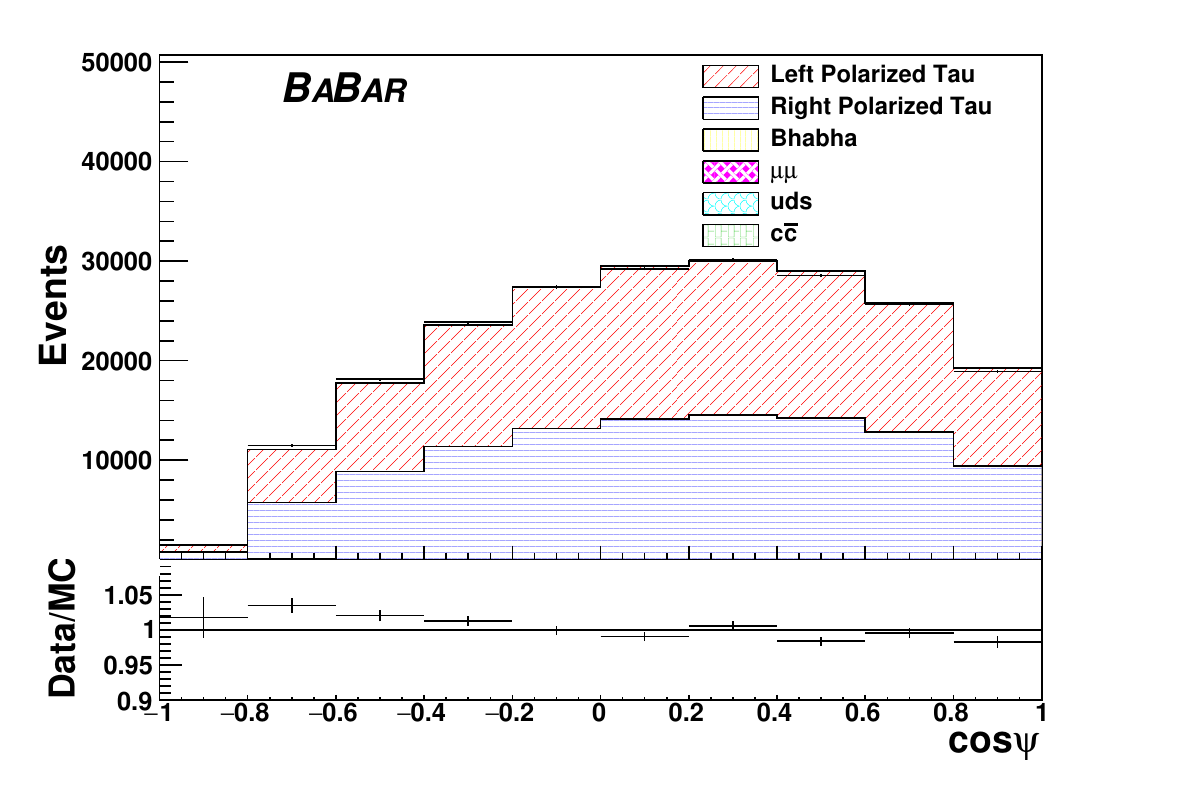}
            \put (20,55) {\textbf{c)}}
            \put (3,47) {\rotatebox{90}{\scriptsize\textbf{/0.2}}}
        \end{overpic} &
        \begin{overpic}[width=0.4\textwidth]{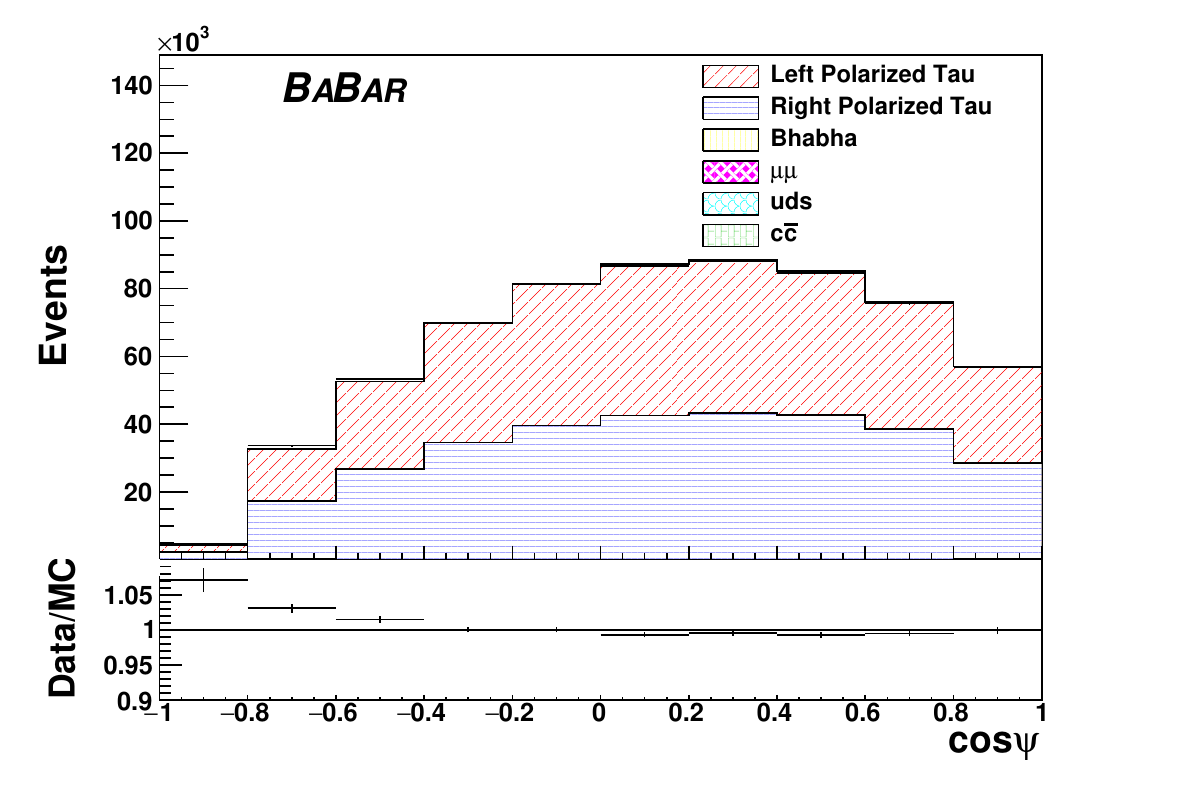}
            \put (20,55) {\textbf{d)}}
            \put (3,47) {\rotatebox{90}{\scriptsize\textbf{/0.2}}}
        \end{overpic} \\
        \begin{overpic}[width=0.4\textwidth]{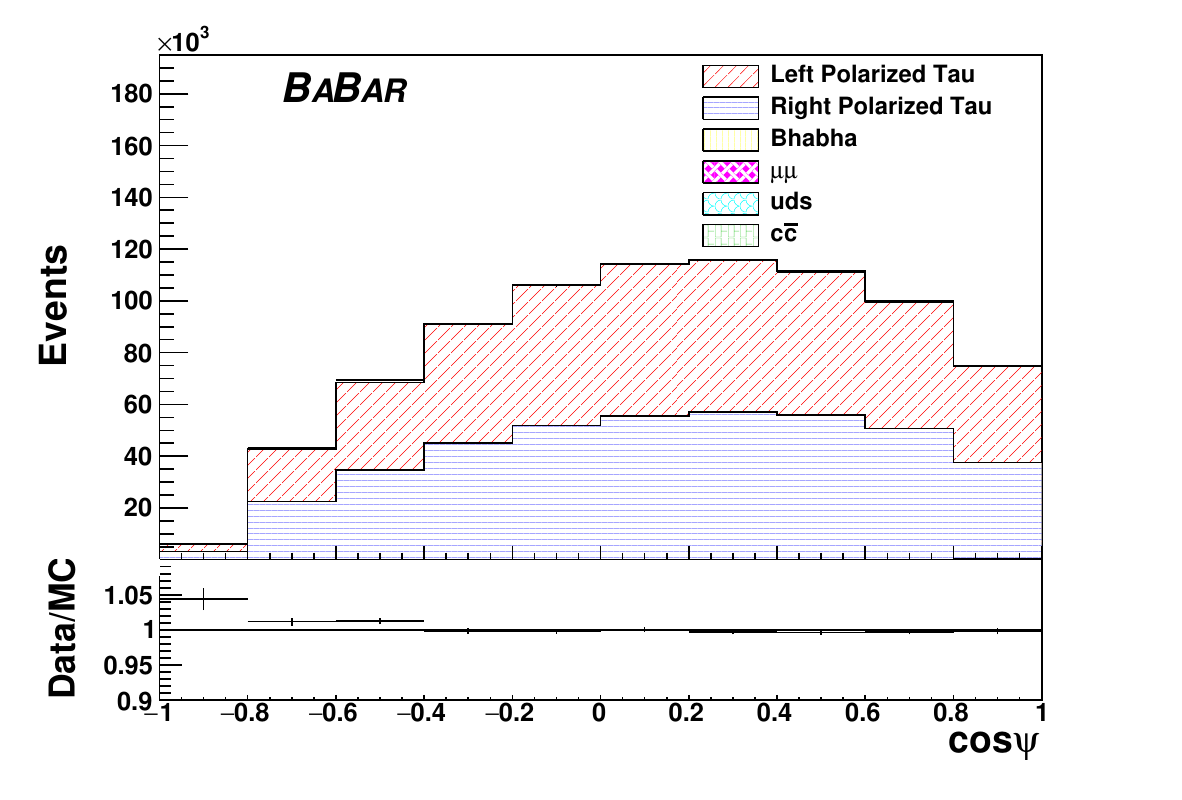}
            \put (20,55) {\textbf{e)}}
            \put (3,47) {\rotatebox{90}{\scriptsize\textbf{/0.2}}}
        \end{overpic} &
        \begin{overpic}[width=0.4\textwidth]{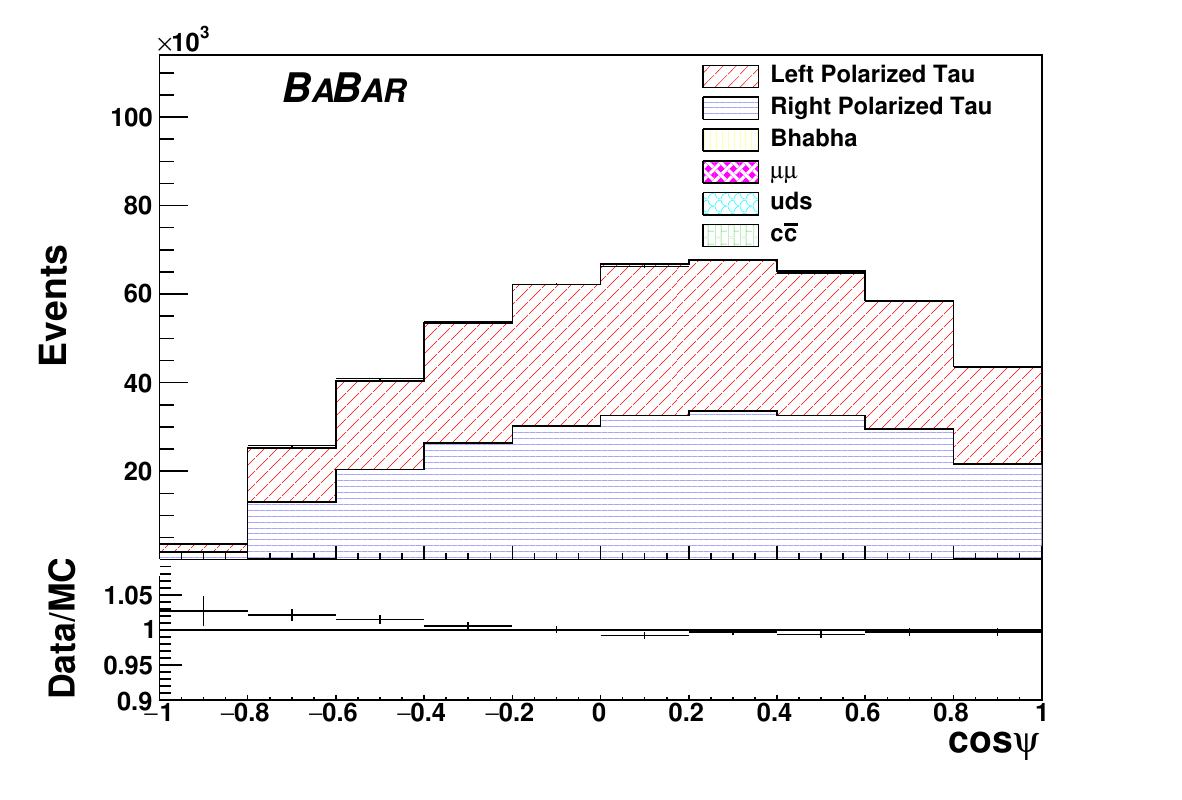}
            \put (20,55) {\textbf{f)}}
            \put (3,47) {\rotatebox{90}{\scriptsize\textbf{/0.2}}}
        \end{overpic} \\
    \end{tabular}
    \caption{One dimensional projection of $\cos\psi$ from $\rho^+$ fits for Runs 1-6, a) through f), respectively.}
    \label{fig:1d_xct_pfits}
\end{figure*}
\begin{figure*}
    \begin{tabular}{cc}
        \begin{overpic}[width=0.4\textwidth]{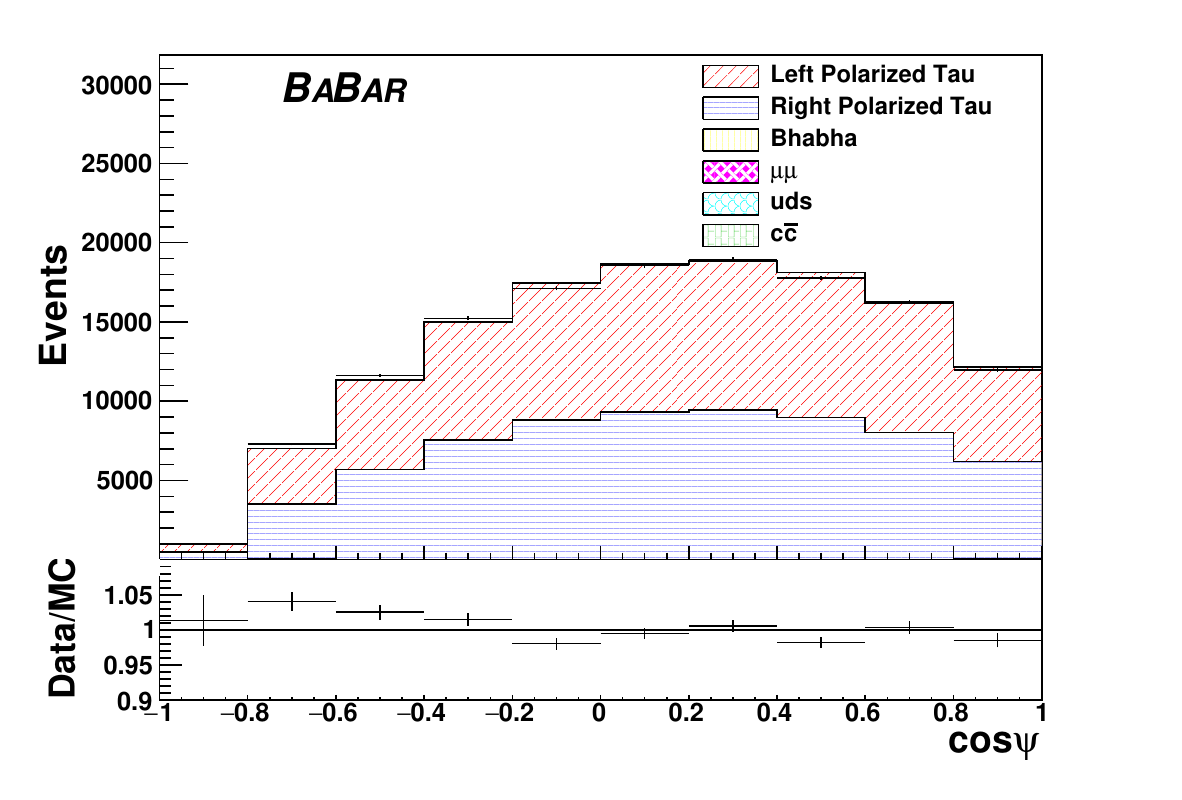}
            \put (20,55) {\textbf{a)}}
            \put (3,47) {\rotatebox{90}{\scriptsize\textbf{/0.2}}}
        \end{overpic} &
        \begin{overpic}[width=0.4\textwidth]{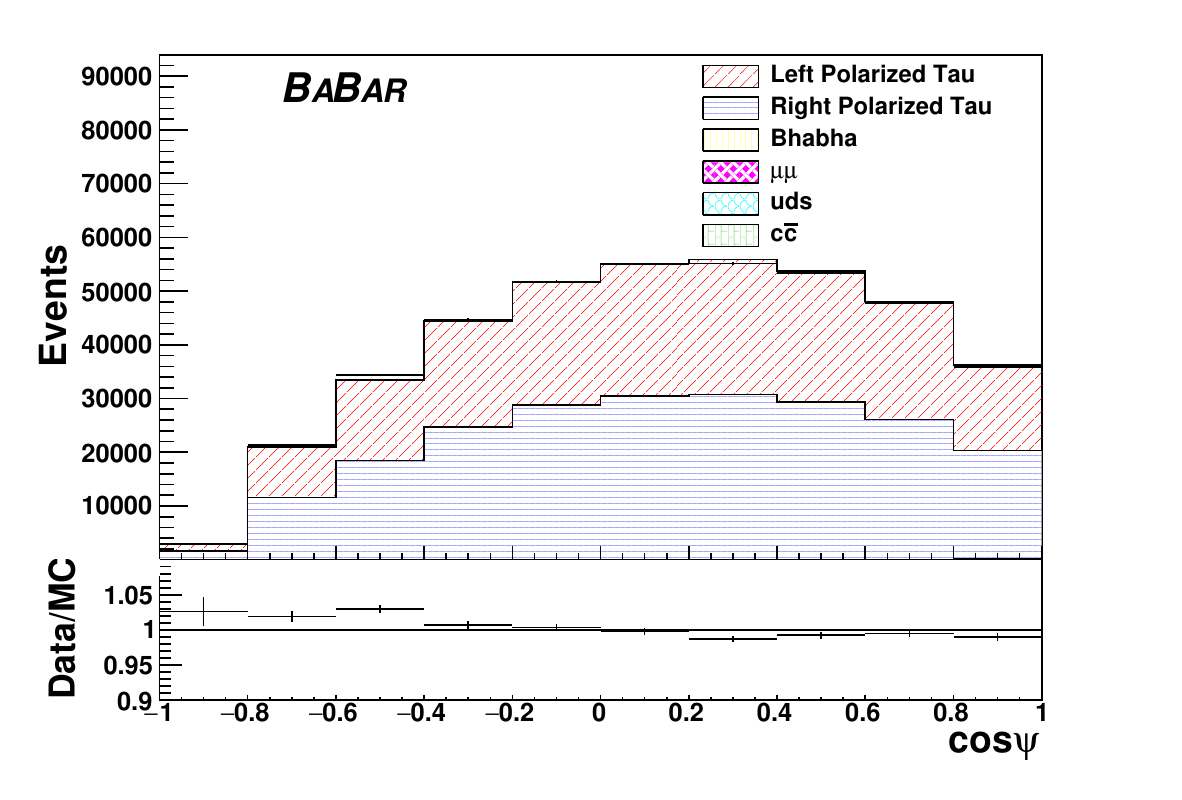}
            \put (20,55) {\textbf{b)}}
            \put (3,47) {\rotatebox{90}{\scriptsize\textbf{/0.2}}}
        \end{overpic} \\
        \begin{overpic}[width=0.4\textwidth]{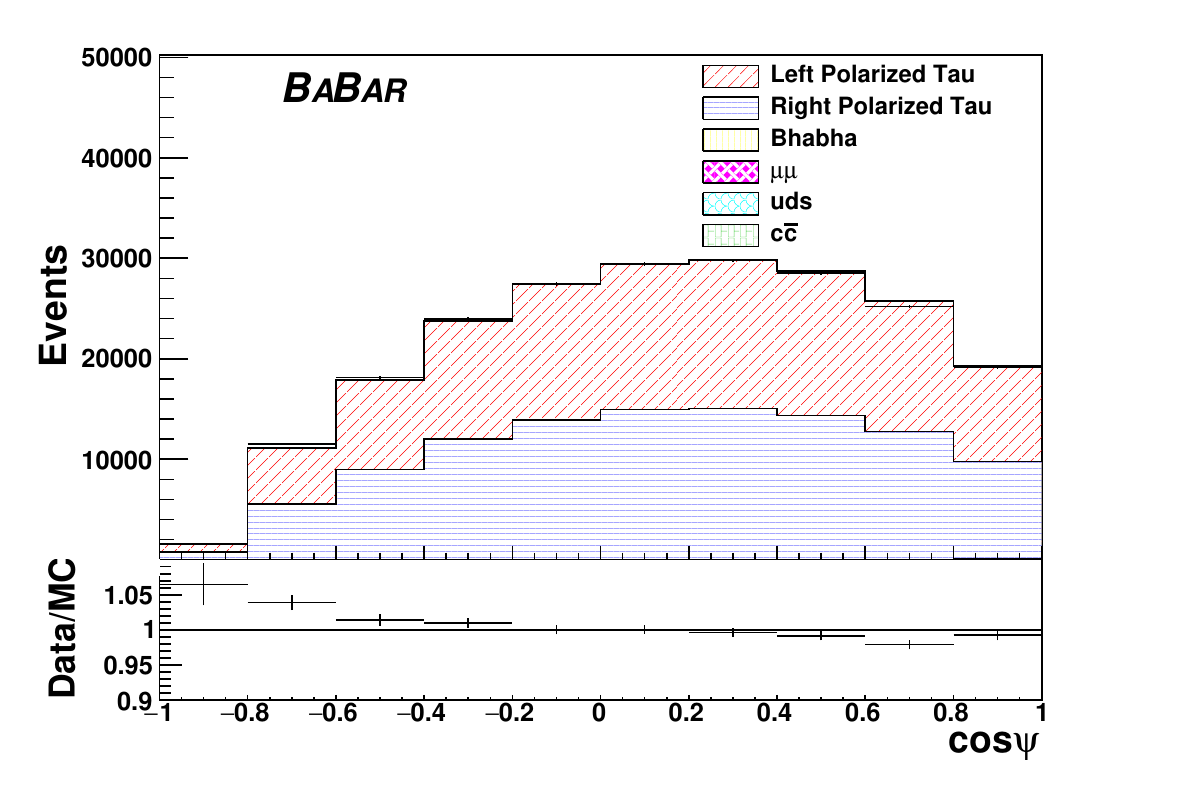}
            \put (20,55) {\textbf{c)}}
            \put (3,47) {\rotatebox{90}{\scriptsize\textbf{/0.2}}}
        \end{overpic} &
        \begin{overpic}[width=0.4\textwidth]{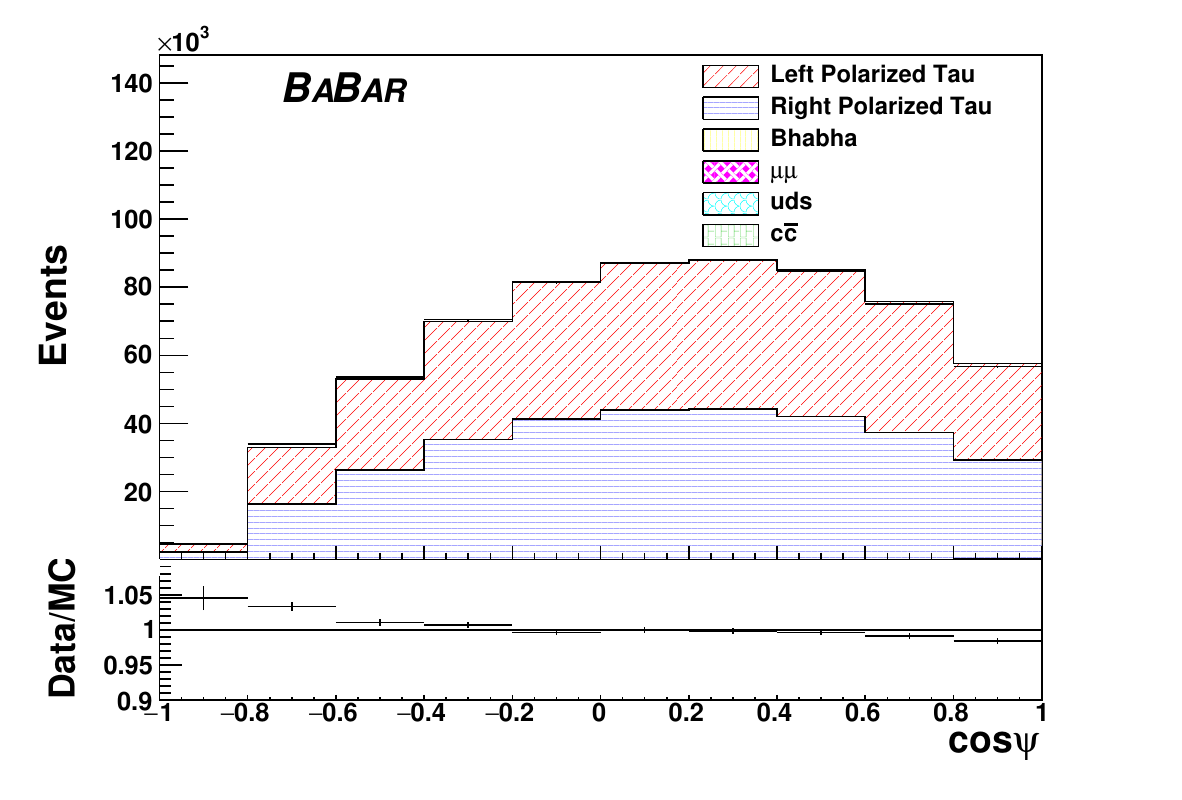}
            \put (20,55) {\textbf{d)}}
            \put (3,47) {\rotatebox{90}{\scriptsize\textbf{/0.2}}}
        \end{overpic} \\
        \begin{overpic}[width=0.4\textwidth]{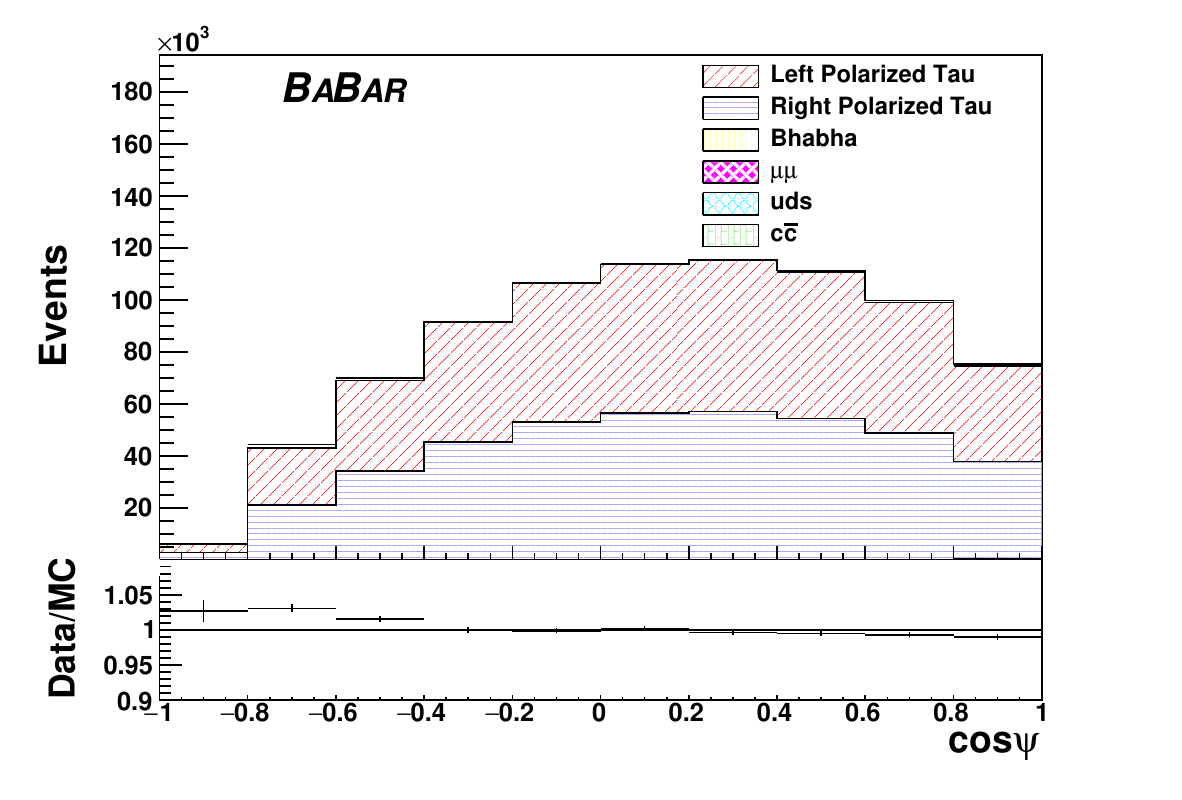}
            \put (20,55) {\textbf{e)}}
            \put (3,47) {\rotatebox{90}{\scriptsize\textbf{/0.2}}}
        \end{overpic} &
        \begin{overpic}[width=0.4\textwidth]{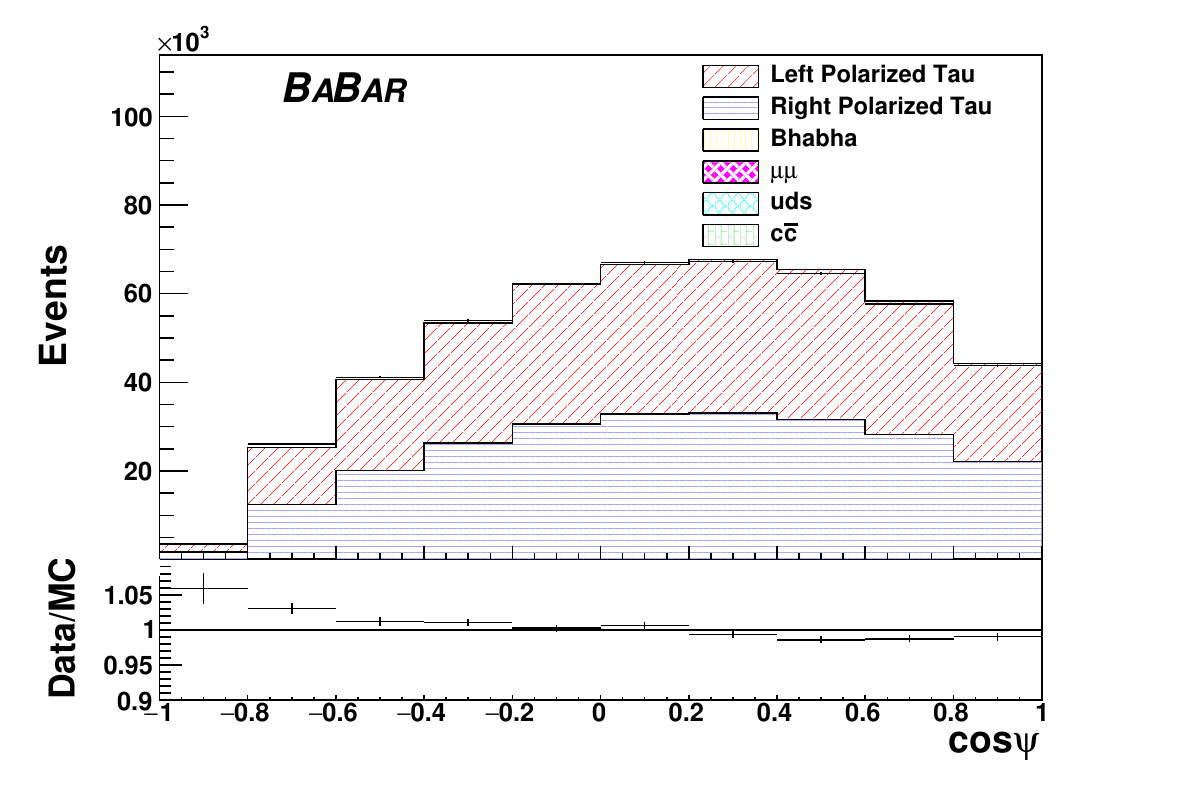}
            \put (20,55) {\textbf{f)}}
            \put (3,47) {\rotatebox{90}{\scriptsize\textbf{/0.2}}}
        \end{overpic} \\
    \end{tabular}
    \caption{One dimensional projection of $\cos\psi$ from $\rho^-$ fits for Runs 1-6, a) through f), respectively.}
    \label{fig:1d_xct_nfits}
\end{figure*}

\end{document}